\documentclass[AMdocument%  style
              ,optEnglish% language
              ,optBiber% 	bibliography tool
              ,optTikzExternalize
              ]{AMlatex}

\graphicspath{{figures/},{figures/introduction},{figures/theory},{figures/experimental}}%
\usetikzlibrary{pgfplots.polar}
\usepgfplotslibrary{groupplots}

\tikzset{external/system call={pdflatex \tikzexternalcheckshellescape -halt-on-error
		-interaction=batchmode -jobname "\image" "\texsource" --extra-mem-bot=100000000 --extra-mem-top=100000000 --pool-size=10000000 --buf-size=10000000}}
	
\newcommand{\IBSResultCSVdisp}{undefined} % csv file IBS
\newcommand{\IBSResultCSVvelo}{undefined} % csv file IBS
\newcommand{\IBSResultCSVacc}{undefined} % csv file IBS
\newcommand{\IBSlen}{undefined}

\newcommand{\yAmindisp}{undefined}
\newcommand{\yAmaxdisp}{undefined}
\newcommand{\yBmindisp}{undefined}
\newcommand{\yBmaxdisp}{undefined}
\newcommand{\LMmindisp}{undefined}
\newcommand{\LMmaxdisp}{undefined}

\newcommand{\yAminvelo}{undefined}
\newcommand{\yAmaxvelo}{undefined}
\newcommand{\yBminvelo}{undefined}
\newcommand{\yBmaxvelo}{undefined}
\newcommand{\LMminvelo}{undefined}
\newcommand{\LMmaxvelo}{undefined}

\newcommand{\yAminacc}{undefined}
\newcommand{\yAmaxacc}{undefined}
\newcommand{\yBminacc}{undefined}
\newcommand{\yBmaxacc}{undefined}
\newcommand{\LMminacc}{undefined}
\newcommand{\LMmaxacc}{undefined}

\newcommand{\IBSResultCSVAa}{undefined} % csv file IBS
\newcommand{\IBSResultCSVBa}{undefined} % csv file IBS
\newcommand{\IBSResultCSVAb}{undefined} % csv file IBS
\newcommand{\IBSResultCSVBb}{undefined} % csv file IBS
\newcommand{\IBSResultCSVAc}{undefined} % csv file IBS
\newcommand{\IBSResultCSVBc}{undefined} % csv file IBS
\newcommand{\IBSResultCSVAd}{undefined} % csv file IBS
\newcommand{\IBSResultCSVBd}{undefined} % csv file IBS
\newcommand{\IBSResultCSVAe}{undefined} % csv file IBS
\newcommand{\IBSResultCSVBe}{undefined} % csv file IBS
\newcommand{\IBSResultCSVAf}{undefined} % csv file IBS
\newcommand{\IBSResultCSVBf}{undefined} % csv file IBS
\newcommand{\IBSResultCSVAg}{undefined} % csv file IBS
\newcommand{\IBSResultCSVBg}{undefined} % csv file IBS

\newcommand{\yAmin}{undefined}
\newcommand{\yAmax}{undefined}
\newcommand{\LMmin}{undefined}
\newcommand{\LMmax}{undefined}

\newcommand{\yAminB}{undefined}
\newcommand{\yAmaxB}{undefined}
\newcommand{\LMminB}{undefined}
\newcommand{\LMmaxB}{undefined}

\newcommand{\tp}{^\mathrm{T}}

\usepackage[list=true,listformat=simple]{subcaption}

\newcommand{\legendary}{\includegraphics{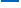}$\,$IBS Result, {\includegraphics{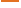}$\,$Measurement}}
\newcommand{\legendaryTwo}{\includegraphics{figures/line_blue.pdf}$\,$Time Domain IRF, {\includegraphics{figures/line_orange.pdf}$\,$Frequency Domain IRF}}

\usepackage{tabularx}

\newcolumntype{Y}{>{\centering\arraybackslash}X}

\usepackage{colortbl}
\usepackage{multirow}
\usepackage{hhline}

\newcommand{\trFailUnstable}{\cellcolor{lightgray!40!red}$\Large\boldsymbol{\infty}$}
\newcommand{\trFailCoupling}{\cellcolor{lightgray!40!red}$\Large\boldsymbol{\times}$}
\newcommand{\trPeakZero}{\cellcolor{lightgray!40!red}$\boldsymbol{\mathrm{O}}$}
\newcommand{\trPeakOne}{\cellcolor{lightgray!30!YellowOrange}$\boldsymbol{\checkmark}$}
\newcommand{\trPeakTwo}{\cellcolor{lightgray!30!yellow}$\boldsymbol{\checkmark\checkmark}$}
\newcommand{\trPeakThree}{\cellcolor{lightgray!60!green}$\boldsymbol{\checkmark\checkmark\checkmark}$}
\newcommand{\trPeakFour}{\cellcolor{lightgray!40!green}$\boldsymbol{\checkmark\checkmark\checkmark\checkmark}$}
\newcommand{\trPeakFive}{\cellcolor{lightgray!10!green}$\boldsymbol{\checkmark\checkmark\checkmark\checkmark +}$}

\begin{document}

% Titlepage
\vspace*{-0.4cm}\AMlayoutTitlePageDefault{%
	\fontsize{18.5}{14}\selectfont Enabling Experimental Impulse-Based Substructuring through Time Domain Deconvolution and Downsampling % Title
}{}{%
	Oliver Maximilian Zobel$^{*}$, Francesco Trainotti, Daniel J. Rixen\\
	Technical University of Munich, TUM School of Engineering and Design, Chair of Applied Mechanics, \\Boltzmannstr.~15, 85748 Garching, Germany\\
	$^*$ Correspondence: oliver.zobel@tum.de % Author
}%


% !TeX spellcheck = en_US
\section*{Abstract}%

Dynamic substructuring, especially the frequency-based variant (FBS) using frequency response functions (FRF), is gaining in popularity and importance, with countless successful applications, both numerically and experimentally. One drawback, however, is found when the responses to shocks are determined. Numerically, this might be especially expensive when a huge number of high-frequency modes have to be accounted for to correctly predict response amplitudes to shocks. In all cases, the initial response predicted using frequency-based substructuring might be erroneous, due to the forced periodization of the Fourier transform. This drawback can be eliminated by completely avoiding the frequency domain and remaining in the time domain, using the impulse-based substructuring method (IBS), which utilizes impulse response functions (IRF). While this method has already been utilized successfully for numerical test cases, none of the attempted experimental applications were successful. In this paper, an experimental application of IBS to rods considered as one-dimensional is tested in the context of shock analysis, with the goal of correctly predicting the maximum driving point response peak. The challenges related to experimental IBS applications are discussed and an improvement attempt is made by limiting the frequency content considered through low-pass filtering and downsampling. The combination of a purely time domain based estimation procedure for the IRFs and the application of low-pass filtering with downsampling to the measured responses enabled a correct prediction of the initial shock responses of the rods with IBS experimentally, using displacements, velocities and accelerations.
% !TeX spellcheck = en_US
\subsection*{Nomenclature}
\begin{minipage}[t]{0.5\textwidth}\vspace{0pt}
	\begin{tabular}{ll}
		dof(s) & Degree(s) of Freedom \\
		FBS	& Frequency-Based Substructuring \\
		FRF	& Frequency Response Function \\
		IBS	& Impulse-Based Substructuring \\
		IRF	& Impulse Response Function \\
		FFT	& Fast Fourier Transform \\
		IFFT & Inverse Fast Fourier Transform \\
		$\star$ & Placeholder symbol \\
		$\star^\ast$ & Complex conjugate of $\star$ \\
		$\star (t)$ & Time-continuous version of $\star$ \\
		$\star [k]$ & Time-discrete version of $\star$ \\
		$t, \tau$ & Time \\
		$k$ & Discrete time index \\
		$i$ & Arbitrary index \\
		$(s)$ & Substructure index \\
		$N_S$ & Number of substructures
	\end{tabular}
\end{minipage}
\begin{minipage}[t]{0.5\textwidth}\vspace{0pt}
	\begin{tabular}{ll}
		$y,\vy$ & General system response \\
		$d,\vd$ & Displacement-based response \\
		$v,\vv$ & Velocity-based response \\
		$a,\va$ & Acceleration-based response \\
		$h,\vH$ & General impulse response function \\
		$h_\text{D},\vH_\text{D}$ & Displacement impulse response function \\
		$h_\text{V},\vH_\text{V}$ & Velocity impulse response function \\
		$h_\text{A},\vH_\text{A}$ & Acceleration impulse response function \\
		$f,\vf$ & Externally applied force \\
		$\Delta t$ & Discrete time step size \\
		$\vB$ & Signed Boolean constraint matrix \\
		$\lambda,\vec{\lambda}$ & Lagrange multiplier(s) \\
		$g,\vg$ & Interface gap(s) \\
		$\vR$ & Auto-correlation matrix \\
		$f_\text{S}$ & Sampling frequency \\
		$M,N,Q$ & Counts/lengths of quantities
	\end{tabular}
\end{minipage}

\newpage

% Main Content
% !TeX spellcheck = en_US
\section{Introduction}
Dynamic substructuring is a method used to divide large structures into smaller parts, the substructures, which can then be analyzed independently. This is not only advantageous for very large and complex numerical system models with a large number of degrees of freedom (dofs), but also for experimental analysis of big structures, such as aircraft, that are impractical to handle as a whole. Also, dynamic substructuring allows to combine numeric models and experimentally identified components, for instance, a newly developed prototype with a hard-to-model environment and predict the behavior before manufacturing~\cite{Klerk2008}.

In 1988 substructuring using \textit{Frequency Response Functions} (FRF), the \textit{Frequency Based Substructuring} method (FBS), was formulated by \textsc{\citeauthor{Jetmundsen1988}}~\cite{Jetmundsen1988}. Over the years, the FBS method became popular and a multitude of successful applications, modifications and extensions can be found in literature~\cites{Allen2020}{Klerk2008}. However, one shortcoming of the method can be seen, when the time response to shocks or impacts should be calculated, as this requires a large frequency bandwidth and with that a large number of modes, making FBS very expensive and not very well suited~\cite{Rixen2011}.

Therefore, the idea to perform dynamic substructuring with the intent to determine shock responses directly in the time domain comes naturally. The time domain counterpart of FBS is the \textit{Impulse Based Substructuring} method (IBS), which uses \textit{Impulse Response Functions} (IRF) and convolution products. While it is no challenge to apply IBS numerically, where the required quantities can be obtained by direct time integration of the system, see for instance \cites{Gordis1995}{Rixen2011}{Rixen2013}, the experimental application is not as straightforward. So far, none of the trialed experimental applications of IBS, for instance by \textsc{\citeauthor{Rixen2010}} in \cite{Rixen2010} using displacement responses of thermoplastic rods or by \textsc{\citeauthor{VanDerSeijs2014}} in \cite{VanDerSeijs2014} using velocity responses of POM rods, were successful.

\begin{figure}[ht]
	\def\svgwidth{0.69\textwidth}
	\centering
	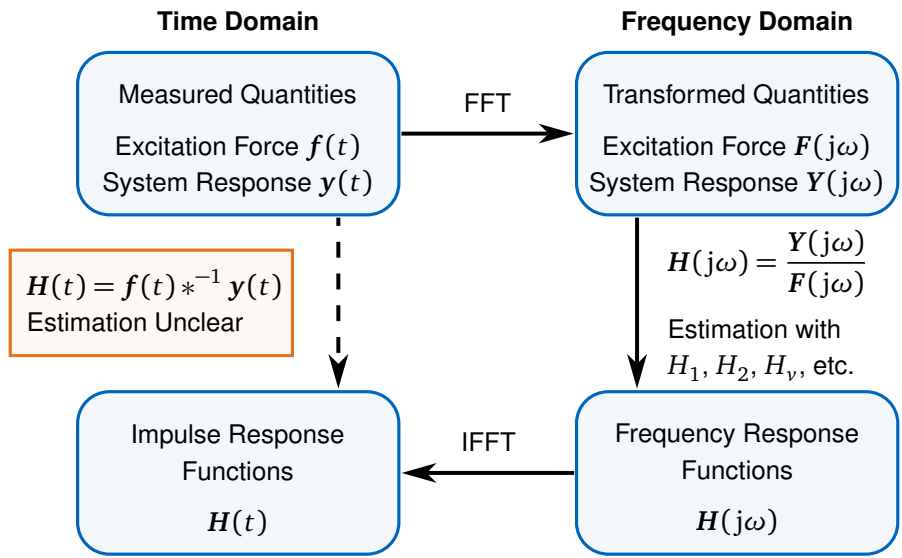
	\caption{Illustration of the relations between the measured system responses and excitation, the transfer characteristics in the time domain (Impulse Response Functions) and the ones in the frequency domain (Frequency Response Functions)}
	\label{fig:time_frequency_domain}
\end{figure}

One challenge of experimentally applying IBS lies in the correct identification of the IRFs $\vH(t)$, dual to FBS requiring accurate FRFs $\vH(\,\mathrm{j}\omega)$. For the estimation of FRFs from measured quantities, various estimators are established, e.g.\ the $H_1$, $H_2$ or $H_v$ estimator, which approximates the ratio between the Fourier-transformed responses $\vY(\,\mathrm{j}\omega)$ and the applied force $\vF(\,\mathrm{j}\omega)$, while also reducing the influence of noise and other variations between individual impacts, see also \cref{fig:time_frequency_domain}. In the time domain, the deconvolution of the responses $\vy(t)$ with the applied force $\vf(t)$ has to be calculated, but ideally also estimated such that the influence of noise is reduced. For this, an estimator is desired that enables averaging of multiple impacts, but none has been established so far.

Looking at the connections between the aforementioned quantities in \cref{fig:time_frequency_domain}, one idea might be to calculate the IRFs by applying an \textit{Inverse Fast Fourier Transform (IFFT)} to FRFs calculated using the established estimators. The disadvantage of this approach is that then all limitations and drawbacks of the frequency domain would apply to the IRFs as well, e.g.\ leakage effects due to the forced periodization, which would normally be avoided with IBS. This implies that if a sufficient estimator for IRFs in the time domain could be found, there will be some advantages, e.g.\ not requiring the response to decay within the measurement window.

\newpage

As mentioned previously, IRFs are especially suited to describe shock responses, i.e.\ transients with high-frequency content, because such transients can be characterized by a relatively short time series that can have a high modal density. For such short, transient events, the periodic assumption of the frequency domain is not very well suited. At the same time, IBS is not suited for long-term calculations because the computation costs of the convolution products per time step grow linearly, making it impractical to compute.

Another advantage is the possibility to add non-linear elements in the IBS algorithm~\cite{Rixen2013}, which is not possible with FBS due to the assumed linearity for the frequency domain transformation. While some methods for the addition of non-linear elements exist in literature for purely numerical applications~\cite{VanderValk2012}, no (hybrid) applications have been presented to couple experimental IRFs with a numerical non-linear counterpart. In addition, a successful experimental IBS application would open the door to a more sophisticated generalization to non-linear response functions assembly~\cite{Tawfiq2004}.

The goal of this paper is to successfully apply IBS experimentally to determine shock responses, for which mostly the highest response peak is of interest, e.g.\ the maximum acceleration of a component. For realistic impacts, the shock is not instantaneous, but a force is applied over time, denoted as an imperfect impact. The first peak will therefore not be the highest one, requiring an accurate estimation of at least the second response peak to an externally applied force. Since IBS could not be applied successfully in an experimental context so far, the test cases are limited to systems considered one-dimensional, here rods made out of POM and aluminum.

In this paper, first, the theory of IBS is quickly summarized and then two possibilities for estimating IRFs from measurements, one in the time domain and one in the frequency domain, are presented. Due to limitations of both, the excitation and sensors, not all of the measured frequency bandwidth might be useful. For this reason, methods for limiting the frequency content considered within the IBS scheme are briefly discussed, followed by the description of the test cases and the utilized experimental setups. Then, the results of experimental IBS are discussed for unmodified responses and then for responses with limited frequency content. Lastly, the observed issues of the experimental IBS application are investigated and the results summarized.\vspace{-0.1cm}
% !TeX spellcheck = en_US
\section{Theory}\label{sec:Theory}\vspace{-0.1cm}

The \textit{Impulse Based Substructuring} technique was first proposed by \textsc{Gordis} in 1995 in \cite{Gordis1995} using Volterra Integral equations. Later on, the dually assembled form was proposed by \textsc{Rixen} in 2010 in \cite{Rixen2011}, where the substructures are coupled together using reaction forces, i.e.\ the Lagrange multipliers. The latter will be used in this paper.

\subsection{System Responses using Impulse Response Functions}

\begin{figure}[ht!]
	\centering
	\def\svgwidth{0.78\textwidth}
	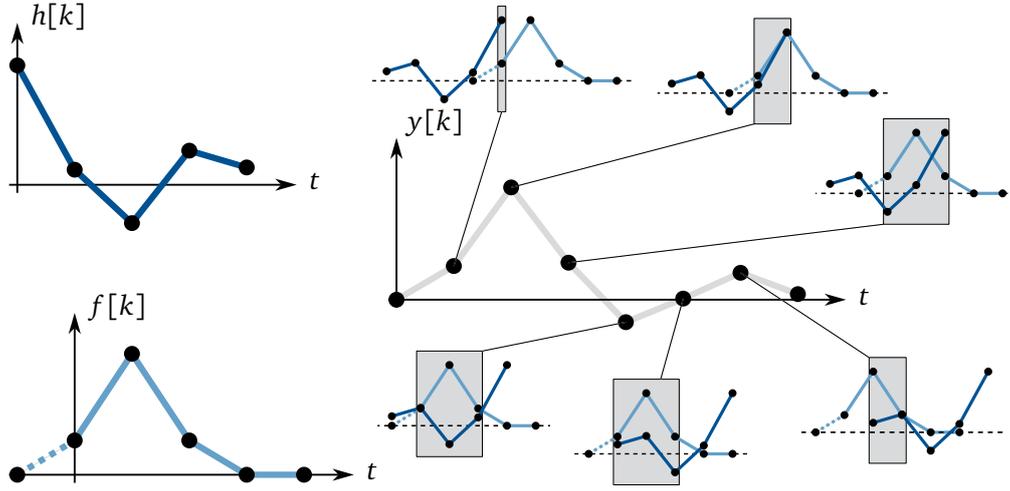
	\caption{Illustration of the Duhamel convolution integral discretization using the Cauchy product}
	\label{fig:conv_disc}
\end{figure}

The dynamic response $y(t)$ of a linear system to an arbitrary external force $f(t)$ is given by the convolution of the \textit{Impulse Response Function (IRF)} $h(t)$ with the applied force, denoted as the Duhamel integral, see e.g.\ \cite{Geradin2014}:
\begin{equation}
	y(t)=\int h(t-\tau)f(\tau)\dd{\tau}\qquad\text{or}\qquad\vy(t)=\int \vH(t-\tau)\vf(\tau)\dd{\tau}\label{eq:Duhamel}
\end{equation}
Since all measured quantities are time-discrete, the convolution integral also has to be discretized in time. For this, commonly the Cauchy product is used, e.g.\ by \textit{MATLAB} or \textit{SciPy}, where the time-discrete system response $y[k]$, with the discrete time step size $\Delta t$, is then governed by:
\begin{equation}
	y[k]=\sum_{i=0}^{k-1}h[k-(i+1)]\;f[i]\;\Delta t\qquad\forall k>0\label{eq:conv_disc}
\end{equation}
In this equation, the indices are arranged such that the response at $k=0$ is given by the initial condition $y_0$, which is always equal to zero in this paper, and then follows as:
\begin{equation}
	\begin{aligned}
	y[0]&=y_0=0 \\
	y[1]&=h[0]f[0]\;\Delta t \\
	y[2]&=h[1]f[0]\;\Delta t+h[0]f[1]\;\Delta t \\
	y[3]&=h[2]f[0]\;\Delta t+h[1]f[1]\;\Delta t+h[0]f[2]\;\Delta t \\[-0.2cm]
	&\;\;\vdots
	\end{aligned}
\end{equation}
Using simulations, this discretization scheme was shown to be stable within the context of IBS. The discrete convolution procedure of force $f[k]$ and IRF $h[k]$ is also illustrated in \cref{fig:conv_disc}.
\newpage

\subsection{Impulse-Based Substructuring}

The basis for the formulation of the \textit{Impulse Based Substructuring} method is the solution of the system's response due to external forces, namely Duhamel's integral (\cref{eq:Duhamel}), stating that the response $\vy(t)$ of a system is determined by the convolution product of the impulse response function matrix $\vH(t)$ with the vector of applied forces $\vf(t)$. This is written for every substructure $s$, denoted by an upper right index $(s)$.

In order to couple two substructures, an interface has to be defined between them. Then, two conditions have to be enforced: \textbf{Interface equilibrium}, requiring the sum of the interface forces to be zero for the assembled system, and \textbf{interface compatibility} requiring that the interface degrees of freedom exhibit the same response on both substructures, i.e.\ that no gap between the two substructures exist at any time. This can be described using signed Boolean constraint matrices $\vB^{(s)}$, that match the interface degrees of freedom to each other~\cite{Allen2020}.

Using the same matrices $\vB^{(s)}$, the interface forces $\vec{\lambda}(t)$, the Lagrange multipliers, can be assigned to the correct degrees of freedom. The Lagrange multipliers are used to enforce the interface equilibrium, which couples the substructures dynamically. They are calculated based on the interface gap $\vg[k]$ using IRFs, such that the application of the force generates a response that exactly closes the gap. Putting all this together yields the integral formulation that describes the \textit{Impulse Based Substructuring} method~\cite{Rixen2011}:
\begin{equation}
	\left\lbrace\begin{array}{ll}\displaystyle
		\vy^{(s)}(t)=\int_{0}^{t}\vH^{(s)}(t-\tau)\;\left(\vf^{(s)}(\tau)+{\vB^{(s)}}\tp\vec{\lambda}(\tau)\right)\dd{\tau} & \qquad\text{Equations of Motion} \\\displaystyle
		\sum_{s=1}^{N_S}\vB^{(s)}\vy^{(s)}(t)=\vg[k]=\vzero & \qquad\text{Compatibility}
	\end{array}\right. \label{eq:IBS-Int}
\end{equation}
Note that the IBS formulation is valid for displacements, velocities and accelerations. Nonetheless, the quantity chosen determines also the quantity with which the interface compatibility is evaluated. When velocities or accelerations are chosen, a small drift over time could lead to a displacement gap between the substructures. Since in this paper, only short times are considered for the IBS evaluation, the influence of this is negligible. 

The convolution integral in the \textit{Equations of Motion} of \cref{eq:IBS-Int} are discretized using the Cauchy product from \cref{eq:conv_disc} and then the last summand is extracted from the sum:
\begin{equation}
	\begin{aligned}
		\vy^{(s)}[k]&=\sum_{i=0}^{k-1}\vH^{(s)}[k-(i+1)]\;\left(\vf^{(s)}[i]+{\vB^{(s)}}\tp\vec{\lambda}[i]\right)\Delta t\qquad\forall k>0 \\
		\vy^{(s)}[k]&=\left(\sum_{i=0}^{k-2}\vH^{(s)}[k-(i+1)]\;\left(\vf^{(s)}[i]+{\vB^{(s)}}\tp\vec{\lambda}[i]\right)\Delta t\right)+ \\
		&\qquad\qquad\qquad\qquad+\vH^{(s)}[0]\;\left(\vf^{(s)}[k-1]+{\vB^{(s)}}\tp\vec{\lambda}[k-1]\right)\Delta t\qquad\forall k>0
	\end{aligned}
\end{equation}

As can be seen, the response $\vy^{(s)}[k]$ at the discrete time step $k$ depends at most on the Lagrange multiplier $\vec{\lambda}[k-1]$ from the previous time step, not on $\vec{\lambda}[k]$, meaning that the compatibility at time step $k$ has to be enforced by $\vec{\lambda}[k-1]$. As this quantity is not known a priori, the calculation of the assembled system response $\vy^{(s)}[k]$ has to be split up. For this, first, the response of the system at time step $k$ is predicted without $\vec{\lambda}[k-1]$:\vspace{-0.4cm}

\paragraph{Predictor Step}

\begin{equation}
	\tilde{\vy}^{(s)}[k]=\left(\sum_{i=0}^{k-2}\vH^{(s)}[k-(i+1)]\;\left(\vf^{(s)}[i]+{\vB^{(s)}}\tp\vec{\lambda}[i]\right)\Delta t\right)+\vH^{(s)}[0]\;\vf^{(s)}[k-1]\;\Delta t\qquad\forall k>0
\end{equation}
Then, based on the interface gap $\vg[k]$ as determined through the \textit{Compatibility} equation of \cref{eq:IBS-Int}, the required Lagrange multiplier $\vec{\lambda}[k-1]$ is calculated based on the first IRF values $\vH^{(s)}[0]$:\vspace{-0.4cm}

\paragraph{Calculation of Lagrange Multipliers}

\begin{equation}
	\left(\sum_{s=1}^{N_S}\vB^{(s)}\vH^{(s)}[0]{\vB^{(s)}}\tp\right)\vec{\lambda}[k-1]=-\sum_{s=1}^{N_S}\vB^{(s)}\tilde{\vy}^{(s)}[k]\cdot\dfrac{1}{\Delta t}=-\vg[k]\cdot\dfrac{1}{\Delta t}\qquad\forall k>0 \label{eq:IBS_LM_from_predictor_D1}
\end{equation}
With this equation, the Lagrange multipliers $\vec{\lambda}[k-1]$ are calculated such that the interface responses at the time step $k$ are compatible. The right-hand side describes the resulting gap $\vg[k]$ between the interface dofs and the left-hand side describes the initial response of the interface dofs to the Lagrange multipliers $\vec{\lambda}[k-1]$ yet to be determined. By evaluating the equality between the two sides, also note the minus sign on the right-hand side, the Lagrange multipliers $\vec{\lambda}[k-1]$ are found such that the impulse response to them exactly closes the gap. For linear systems, it is sufficient to premultiply the right-hand side by the inverse of the term in brackets on the left-hand side and evaluate the found equation for $\vec{\lambda}[k-1]$, while for non-linear systems an iterative solution is required, see e.g.\ \cite{VanderValk2012}.

It is to be noted that this calculation solely relies on the IRFs at $k=0$. In case of IRF errors that result in these values being equal or less than zero, either no Lagrange multiplier can be found or the sign is erroneous, leading to fully unstable IBS results. A special case are displacement IRFs, where $\vH[0]$ is always equal to zero~\cite{Rixen2010}. To apply IBS in the cases where the first value $\vH[0]$ of a driving point IRF is negative or zero, i.e.\ a force would generate a response in the wrong direction or no response at all, all IRFs are shifted backward in time by one sample.

Lastly, the predicted system response $\tilde{\vy}^{(s)}[k]$ is updated to include the response to the Lagrange multipliers $\vec{\lambda}[k-1]$ just determined, that ideally close any interface gaps $\vg[k]$:\vspace{-0.4cm}

\paragraph{Corrector Step}

\begin{equation}
	\vy_{\lambda}^{(s)}[k]=\vH^{(s)}[0]{\vB^{(s)}}\tp\vec{\lambda}[k-1]\;\Delta t\qquad\forall k>0 \label{eq:CorrectorD1}
\end{equation}

\begin{figure}[ht!]
	\small
	\centering
	\def\svgwidth{0.98\textwidth}
	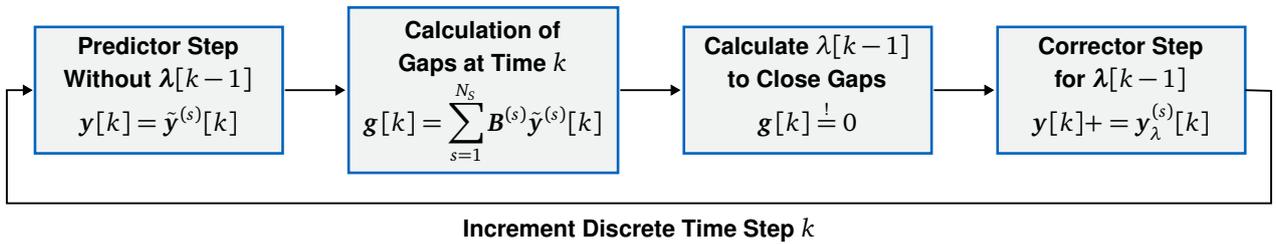
	\caption{Flowchart of IBS scheme calculations within each time step}
	\label{fig:IBS_flowchart}
\end{figure}

The required procedure during each discrete time step $k$ is summarized and illustrated in \cref{fig:IBS_flowchart}.

\vspace{-0.2cm}

To generally assemble the substructures, the IBS scheme could be evaluated with a unit impulse at $k=0$ applied as an external force $\vf^{(s)}$. In this paper, for simplicity's and comparison's sake, the force applied in the reference measurements is used for the IBS scheme. Then, the response of the physically assembled, measured system and the response of the system virtually assembled using IBS should be fully identical, enabling validation of the IBS method's performance.

For an experimental application, besides setting up the required singed Boolean constraint matrices $\vB^{(s)}$, the last quantity to identify are the IRFs $\vH^{(s)}$ of each substructure. As mentioned previously, impacting a structure with an impact hammer represents an imperfect impulse excitation and yields an impulse response $\vy$, not an impulse response function $\vH$. Besides any IRF being required to be normalized in absolute magnitude with respect to the applied force used to identify it, i.e.\ normalizing the impulse to \qty{1}{\newton\second}, the spectral content of the excitation also has to be normalized, because the excited frequencies in the responses $\vy$ are biased by the imperfect impulse shape. As, therefore, responses $\vy$ to an (imperfect) impulse do not generally represent the correct frequency content of the system dynamics, IRFs $\vH$ have to be identified with suitable procedures. In the following, one possibility to calculate IRFs through the frequency domain and one in the time domain is introduced.

\subsection{Frequency Domain IRF Identification}

Looking at the overview of the involved quantities in \cref{fig:time_frequency_domain}, the procedure to calculate IRFs from measured excitation forces $\vf$ and responses $\vy$ might seem straightforward: Transform the quantities into the frequency domain, estimate an FRF, e.g.\ using the $H_1$ estimator, and apply an IFFT to retrieve the IRFs.

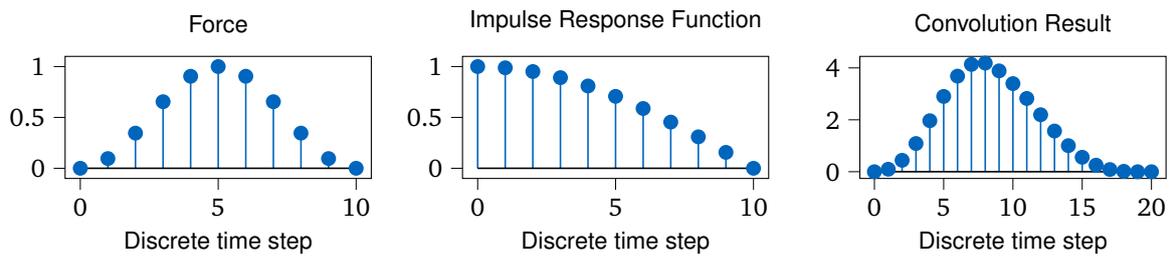
\begin{figure}[ht!]
	\small
	\centering
	% This file was created with tikzplotlib v0.10.1.
\begin{tikzpicture}

\definecolor{darkcyan0101189}{RGB}{0,101,189}
\definecolor{darkgray176}{RGB}{176,176,176}

\begin{groupplot}[group style={group size=3 by 1, horizontal sep=1.2cm}]
\nextgroupplot[
height=3.2cm,
tick align=outside,
tick pos=left,
title=Force,
width=0.33\textwidth,
x grid style={darkgray176},
xlabel={Discrete time step},
xmin=-0.55, xmax=10.55,
xtick style={color=black},
y grid style={darkgray176},
ymin=-0.1, ymax=1.1,
ytick style={color=black}
]
\path [draw=darkcyan0101189, semithick]
(axis cs:0,0)
--(axis cs:0,0);

\path [draw=darkcyan0101189, semithick]
(axis cs:1,0)
--(axis cs:1,0.095492);

\path [draw=darkcyan0101189, semithick]
(axis cs:2,0)
--(axis cs:2,0.34549);

\path [draw=darkcyan0101189, semithick]
(axis cs:3,0)
--(axis cs:3,0.65451);

\path [draw=darkcyan0101189, semithick]
(axis cs:4,0)
--(axis cs:4,0.90451);

\path [draw=darkcyan0101189, semithick]
(axis cs:5,0)
--(axis cs:5,1);

\path [draw=darkcyan0101189, semithick]
(axis cs:6,0)
--(axis cs:6,0.90451);

\path [draw=darkcyan0101189, semithick]
(axis cs:7,0)
--(axis cs:7,0.65451);

\path [draw=darkcyan0101189, semithick]
(axis cs:8,0)
--(axis cs:8,0.34549);

\path [draw=darkcyan0101189, semithick]
(axis cs:9,0)
--(axis cs:9,0.095492);

\path [draw=darkcyan0101189, semithick]
(axis cs:10,0)
--(axis cs:10,0);

\addplot [semithick, darkcyan0101189, mark=*, mark size=2.5, mark options={solid}, only marks]
table {%
0 0
1 0.095492
2 0.34549
3 0.65451
4 0.90451
5 1
6 0.90451
7 0.65451
8 0.34549
9 0.095492
10 0
};
\addplot [semithick, black]
table {%
0 0
10 0
};

\nextgroupplot[
height=3.2cm,
tick align=outside,
tick pos=left,
title=Impulse Response Function,
width=0.33\textwidth,
x grid style={darkgray176},
xlabel={Discrete time step},
xmin=-0.55, xmax=10.55,
xtick style={color=black},
y grid style={darkgray176},
ymin=-0.1, ymax=1.1,
ytick style={color=black}
]
\path [draw=darkcyan0101189, semithick]
(axis cs:0,0)
--(axis cs:0,1);

\path [draw=darkcyan0101189, semithick]
(axis cs:1,0)
--(axis cs:1,0.98769);

\path [draw=darkcyan0101189, semithick]
(axis cs:2,0)
--(axis cs:2,0.95106);

\path [draw=darkcyan0101189, semithick]
(axis cs:3,0)
--(axis cs:3,0.89101);

\path [draw=darkcyan0101189, semithick]
(axis cs:4,0)
--(axis cs:4,0.80902);

\path [draw=darkcyan0101189, semithick]
(axis cs:5,0)
--(axis cs:5,0.70711);

\path [draw=darkcyan0101189, semithick]
(axis cs:6,0)
--(axis cs:6,0.58779);

\path [draw=darkcyan0101189, semithick]
(axis cs:7,0)
--(axis cs:7,0.45399);

\path [draw=darkcyan0101189, semithick]
(axis cs:8,0)
--(axis cs:8,0.30902);

\path [draw=darkcyan0101189, semithick]
(axis cs:9,0)
--(axis cs:9,0.15643);

\path [draw=darkcyan0101189, semithick]
(axis cs:10,0)
--(axis cs:10,6.1232e-17);

\addplot [semithick, darkcyan0101189, mark=*, mark size=2.5, mark options={solid}, only marks]
table {%
0 1
1 0.98769
2 0.95106
3 0.89101
4 0.80902
5 0.70711
6 0.58779
7 0.45399
8 0.30902
9 0.15643
10 0
};
\addplot [semithick, black]
table {%
0 0
10 0
};

\nextgroupplot[
height=3.2cm,
tick align=outside,
tick pos=left,
title=Convolution Result,
width=0.33\textwidth,
x grid style={darkgray176},
xlabel={Discrete time step},
xmin=-1.05, xmax=21.05,
xtick style={color=black},
y grid style={darkgray176},
ymin=-0.25921, ymax=4.4434,
ytick style={color=black}
]
\path [draw=darkcyan0101189, semithick]
(axis cs:0,0)
--(axis cs:0,0);

\path [draw=darkcyan0101189, semithick]
(axis cs:1,0)
--(axis cs:1,0.095492);

\path [draw=darkcyan0101189, semithick]
(axis cs:2,0)
--(axis cs:2,0.43981);

\path [draw=darkcyan0101189, semithick]
(axis cs:3,0)
--(axis cs:3,1.0866);

\path [draw=darkcyan0101189, semithick]
(axis cs:4,0)
--(axis cs:4,1.9646);

\path [draw=darkcyan0101189, semithick]
(axis cs:5,0)
--(axis cs:5,2.9009);

\path [draw=darkcyan0101189, semithick]
(axis cs:6,0)
--(axis cs:6,3.6826);

\path [draw=darkcyan0101189, semithick]
(axis cs:7,0)
--(axis cs:7,4.1348);

\path [draw=darkcyan0101189, semithick]
(axis cs:8,0)
--(axis cs:8,4.1842);

\path [draw=darkcyan0101189, semithick]
(axis cs:9,0)
--(axis cs:9,3.8848);

\path [draw=darkcyan0101189, semithick]
(axis cs:10,0)
--(axis cs:10,3.3954);

\path [draw=darkcyan0101189, semithick]
(axis cs:11,0)
--(axis cs:11,2.8225);

\path [draw=darkcyan0101189, semithick]
(axis cs:12,0)
--(axis cs:12,2.1949);

\path [draw=darkcyan0101189, semithick]
(axis cs:13,0)
--(axis cs:13,1.5674);

\path [draw=darkcyan0101189, semithick]
(axis cs:14,0)
--(axis cs:14,1.0037);

\path [draw=darkcyan0101189, semithick]
(axis cs:15,0)
--(axis cs:15,0.55673);

\path [draw=darkcyan0101189, semithick]
(axis cs:16,0)
--(axis cs:16,0.2525);

\path [draw=darkcyan0101189, semithick]
(axis cs:17,0)
--(axis cs:17,0.083555);

\path [draw=darkcyan0101189, semithick]
(axis cs:18,0)
--(axis cs:18,0.014938);

\path [draw=darkcyan0101189, semithick]
(axis cs:19,0)
--(axis cs:19,5.8472e-18);

\path [draw=darkcyan0101189, semithick]
(axis cs:20,0)
--(axis cs:20,0);

\addplot [semithick, darkcyan0101189, mark=*, mark size=2.5, mark options={solid}, only marks]
table {%
0 0
1 0.095492
2 0.43981
3 1.0866
4 1.9646
5 2.9009
6 3.6826
7 4.1348
8 4.1842
9 3.8848
10 3.3954
11 2.8225
12 2.1949
13 1.5674
14 1.0037
15 0.55673
16 0.2525
17 0.083555
18 0.014938
19 0
20 0
};
\addplot [semithick, black]
table {%
0 0
20 0
};
\end{groupplot}

\end{tikzpicture}\vspace{-0.2cm}
	\caption{\textbf{Linear Convolution} of \textit{Force} and \textit{Impulse Response Function} to the \textit{Convolution Result}}
	\label{fig:lin_vs_circ_lin}
\end{figure}

Special care has to be taken when transforming the time signals to the frequency domain because a multiplication or division in the frequency domain is not necessarily identical to the convolution or deconvolution in the time domain. The desired operation is a \textit{Linear Convolution} or \textit{Linear Deconvolution} as depicted in \cref{fig:lin_vs_circ_lin}.

\begin{figure}[ht!]
	\small
	\centering
	\input{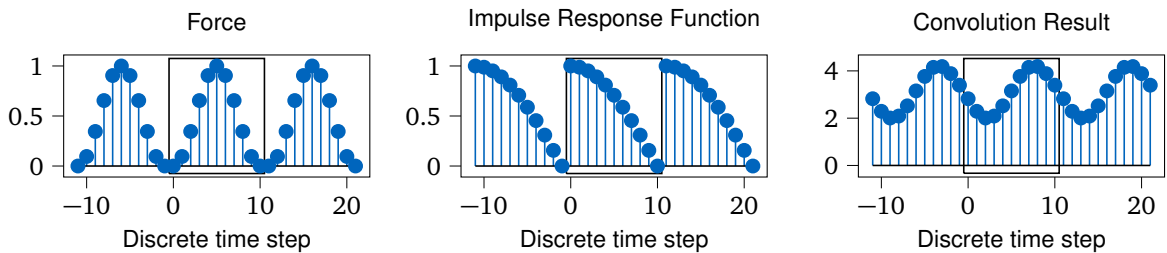}\vspace{-0.2cm}
	\caption{\textbf{Circular Convolution} of \textit{Force} and \textit{Impulse Response Function} to the \textit{Convolution Result} using FFT}
	\label{fig:lin_vs_circ_FFT}
\end{figure}

Here, the \textit{Force} is convoluted with the \textit{Impulse Response Function} to the \textit{Convolution Result}. Both signals are convoluted over their entire length or amount of samples, denoted $N_1$ and $N_2$, where the length of the convolution result is $N_1+N_2-1$. When the same time signals are transformed into the frequency domain and then multiplied, the result in \cref{fig:lin_vs_circ_FFT} is found, where the original signals reside within the boxes drawn in the figure.

\begin{figure}[ht!]
	\small
	\centering
	% This file was created with tikzplotlib v0.10.1.
\begin{tikzpicture}

\definecolor{darkcyan0101189}{RGB}{0,101,189}
\definecolor{darkgray176}{RGB}{176,176,176}

\begin{groupplot}[group style={group size=3 by 1, horizontal sep=1.2cm}]
\nextgroupplot[
height=3.2cm,
tick align=outside,
tick pos=left,
title=Force,
width=0.33\textwidth,
x grid style={darkgray176},
xlabel={Discrete time step},
xmin=-1.1, xmax=21.1,
xtick style={color=black},
y grid style={darkgray176},
ymin=-0.1, ymax=1.1,
ytick style={color=black}
]
\path [draw=darkcyan0101189, semithick]
(axis cs:0,0)
--(axis cs:0,0);

\path [draw=darkcyan0101189, semithick]
(axis cs:1,0)
--(axis cs:1,0.095492);

\path [draw=darkcyan0101189, semithick]
(axis cs:2,0)
--(axis cs:2,0.34549);

\path [draw=darkcyan0101189, semithick]
(axis cs:3,0)
--(axis cs:3,0.65451);

\path [draw=darkcyan0101189, semithick]
(axis cs:4,0)
--(axis cs:4,0.90451);

\path [draw=darkcyan0101189, semithick]
(axis cs:5,0)
--(axis cs:5,1);

\path [draw=darkcyan0101189, semithick]
(axis cs:6,0)
--(axis cs:6,0.90451);

\path [draw=darkcyan0101189, semithick]
(axis cs:7,0)
--(axis cs:7,0.65451);

\path [draw=darkcyan0101189, semithick]
(axis cs:8,0)
--(axis cs:8,0.34549);

\path [draw=darkcyan0101189, semithick]
(axis cs:9,0)
--(axis cs:9,0.095492);

\path [draw=darkcyan0101189, semithick]
(axis cs:10,0)
--(axis cs:10,0);

\path [draw=darkcyan0101189, semithick]
(axis cs:11,0)
--(axis cs:11,0);

\path [draw=darkcyan0101189, semithick]
(axis cs:12,0)
--(axis cs:12,0);

\path [draw=darkcyan0101189, semithick]
(axis cs:13,0)
--(axis cs:13,0);

\path [draw=darkcyan0101189, semithick]
(axis cs:14,0)
--(axis cs:14,0);

\path [draw=darkcyan0101189, semithick]
(axis cs:15,0)
--(axis cs:15,0);

\path [draw=darkcyan0101189, semithick]
(axis cs:16,0)
--(axis cs:16,0);

\path [draw=darkcyan0101189, semithick]
(axis cs:17,0)
--(axis cs:17,0);

\path [draw=darkcyan0101189, semithick]
(axis cs:18,0)
--(axis cs:18,0);

\path [draw=darkcyan0101189, semithick]
(axis cs:19,0)
--(axis cs:19,0);

\path [draw=darkcyan0101189, semithick]
(axis cs:20,0)
--(axis cs:20,0);

\addplot [semithick, darkcyan0101189, mark=*, mark size=2.5, mark options={solid}, only marks]
table {%
0 0
1 0.095492
2 0.34549
3 0.65451
4 0.90451
5 1
6 0.90451
7 0.65451
8 0.34549
9 0.095492
10 0
11 0
12 0
13 0
14 0
15 0
16 0
17 0
18 0
19 0
20 0
};
\addplot [semithick, black]
table {%
0 0
20 0
};
\addplot [semithick, black]
table {%
-0.5 -0.075
10.5 -0.075
10.5 1.075
-0.5 1.075
-0.5 -0.075
};

\nextgroupplot[
height=3.2cm,
tick align=outside,
tick pos=left,
title=Impulse Response Function,
width=0.33\textwidth,
x grid style={darkgray176},
xlabel={Discrete time step},
xmin=-1.1, xmax=21.1,
xtick style={color=black},
y grid style={darkgray176},
ymin=-0.1, ymax=1.1,
ytick style={color=black}
]
\path [draw=darkcyan0101189, semithick]
(axis cs:0,0)
--(axis cs:0,1);

\path [draw=darkcyan0101189, semithick]
(axis cs:1,0)
--(axis cs:1,0.98769);

\path [draw=darkcyan0101189, semithick]
(axis cs:2,0)
--(axis cs:2,0.95106);

\path [draw=darkcyan0101189, semithick]
(axis cs:3,0)
--(axis cs:3,0.89101);

\path [draw=darkcyan0101189, semithick]
(axis cs:4,0)
--(axis cs:4,0.80902);

\path [draw=darkcyan0101189, semithick]
(axis cs:5,0)
--(axis cs:5,0.70711);

\path [draw=darkcyan0101189, semithick]
(axis cs:6,0)
--(axis cs:6,0.58779);

\path [draw=darkcyan0101189, semithick]
(axis cs:7,0)
--(axis cs:7,0.45399);

\path [draw=darkcyan0101189, semithick]
(axis cs:8,0)
--(axis cs:8,0.30902);

\path [draw=darkcyan0101189, semithick]
(axis cs:9,0)
--(axis cs:9,0.15643);

\path [draw=darkcyan0101189, semithick]
(axis cs:10,0)
--(axis cs:10,6.1232e-17);

\path [draw=darkcyan0101189, semithick]
(axis cs:11,0)
--(axis cs:11,0);

\path [draw=darkcyan0101189, semithick]
(axis cs:12,0)
--(axis cs:12,0);

\path [draw=darkcyan0101189, semithick]
(axis cs:13,0)
--(axis cs:13,0);

\path [draw=darkcyan0101189, semithick]
(axis cs:14,0)
--(axis cs:14,0);

\path [draw=darkcyan0101189, semithick]
(axis cs:15,0)
--(axis cs:15,0);

\path [draw=darkcyan0101189, semithick]
(axis cs:16,0)
--(axis cs:16,0);

\path [draw=darkcyan0101189, semithick]
(axis cs:17,0)
--(axis cs:17,0);

\path [draw=darkcyan0101189, semithick]
(axis cs:18,0)
--(axis cs:18,0);

\path [draw=darkcyan0101189, semithick]
(axis cs:19,0)
--(axis cs:19,0);

\path [draw=darkcyan0101189, semithick]
(axis cs:20,0)
--(axis cs:20,0);

\addplot [semithick, darkcyan0101189, mark=*, mark size=2.5, mark options={solid}, only marks]
table {%
0 1
1 0.98769
2 0.95106
3 0.89101
4 0.80902
5 0.70711
6 0.58779
7 0.45399
8 0.30902
9 0.15643
10 0
11 0
12 0
13 0
14 0
15 0
16 0
17 0
18 0
19 0
20 0
};
\addplot [semithick, black]
table {%
0 0
20 0
};
\addplot [semithick, black]
table {%
-0.5 -0.075
10.5 -0.075
10.5 1.075
-0.5 1.075
-0.5 -0.075
};

\nextgroupplot[
height=3.2cm,
tick align=outside,
tick pos=left,
title=Convolution Result,
width=0.33\textwidth,
x grid style={darkgray176},
xlabel={Discrete time step},
xmin=-1.1, xmax=21.1,
xtick style={color=black},
y grid style={darkgray176},
ymin=-0.25921, ymax=4.4434,
ytick style={color=black}
]
\path [draw=darkcyan0101189, semithick]
(axis cs:0,0)
--(axis cs:0,-9.0206e-17);

\path [draw=darkcyan0101189, semithick]
(axis cs:1,0)
--(axis cs:1,0.095492);

\path [draw=darkcyan0101189, semithick]
(axis cs:2,0)
--(axis cs:2,0.43981);

\path [draw=darkcyan0101189, semithick]
(axis cs:3,0)
--(axis cs:3,1.0866);

\path [draw=darkcyan0101189, semithick]
(axis cs:4,0)
--(axis cs:4,1.9646);

\path [draw=darkcyan0101189, semithick]
(axis cs:5,0)
--(axis cs:5,2.9009);

\path [draw=darkcyan0101189, semithick]
(axis cs:6,0)
--(axis cs:6,3.6826);

\path [draw=darkcyan0101189, semithick]
(axis cs:7,0)
--(axis cs:7,4.1348);

\path [draw=darkcyan0101189, semithick]
(axis cs:8,0)
--(axis cs:8,4.1842);

\path [draw=darkcyan0101189, semithick]
(axis cs:9,0)
--(axis cs:9,3.8848);

\path [draw=darkcyan0101189, semithick]
(axis cs:10,0)
--(axis cs:10,3.3954);

\path [draw=darkcyan0101189, semithick]
(axis cs:11,0)
--(axis cs:11,2.8225);

\path [draw=darkcyan0101189, semithick]
(axis cs:12,0)
--(axis cs:12,2.1949);

\path [draw=darkcyan0101189, semithick]
(axis cs:13,0)
--(axis cs:13,1.5674);

\path [draw=darkcyan0101189, semithick]
(axis cs:14,0)
--(axis cs:14,1.0037);

\path [draw=darkcyan0101189, semithick]
(axis cs:15,0)
--(axis cs:15,0.55673);

\path [draw=darkcyan0101189, semithick]
(axis cs:16,0)
--(axis cs:16,0.2525);

\path [draw=darkcyan0101189, semithick]
(axis cs:17,0)
--(axis cs:17,0.083555);

\path [draw=darkcyan0101189, semithick]
(axis cs:18,0)
--(axis cs:18,0.014938);

\path [draw=darkcyan0101189, semithick]
(axis cs:19,0)
--(axis cs:19,-4.8446e-16);

\path [draw=darkcyan0101189, semithick]
(axis cs:20,0)
--(axis cs:20,0);

\addplot [semithick, darkcyan0101189, mark=*, mark size=2.5, mark options={solid}, only marks]
table {%
0 -0
1 0.095492
2 0.43981
3 1.0866
4 1.9646
5 2.9009
6 3.6826
7 4.1348
8 4.1842
9 3.8848
10 3.3954
11 2.8225
12 2.1949
13 1.5674
14 1.0037
15 0.55673
16 0.2525
17 0.083555
18 0.014938
19 -0
20 0
};
\addplot [semithick, black]
table {%
0 0
20 0
};
\end{groupplot}

\end{tikzpicture}\vspace{-0.2cm}
	\caption{\textbf{Linear Convolution} using \textbf{Zero-Padded FFT} of \textit{Force} and \textit{Impulse Response Function} to \textit{Convolution Result}}
	\label{fig:lin_vs_circ_ZP}
\end{figure}
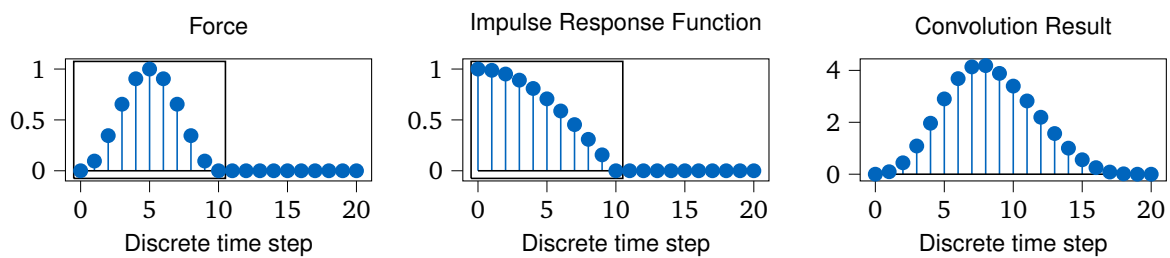

Due to the properties of the Fourier transform, every time signal fed to the transformation is continued periodically. Therefore, the result of the convolution, in this case called a \textit{Circular Convolution}, is different from the one using a \textit{Linear Convolution}. This is due to the periodic continuation being also involved in the convolution, resulting in non-zero values, where they would have been equal to zero in the case of a linear convolution. Correct results can be achieved when the time signals are padded with zeros, such that their length is equal to the length of the resulting linear convolution, i.e.\ $N_1+N_2-1$. Then, in theory, the results of the convolution in the time domain and in the frequency domain using zero-padded FFT/IFFT are identical. The results are shown in \cref{fig:lin_vs_circ_ZP} for one period.

While the signals are still periodically continued, there is no overlap with this continuation within the convolution interval. Nevertheless, the forced periodization can still introduce jumps that lead to leakage, which has to be handled using the traditional procedures, e.g.\ by applying window functions to the time signals before transformation.

Denoting the FFT of a zero-padded response as $Y_{\text{ZP}}(\,\mathrm{j}\omega)$ with corresponding force $F_{\text{ZP}}(\,\mathrm{j}\omega)$, the auto-power spectrum $S_{FF}(\,\mathrm{j}\omega)$ and cross-power spectrum $S_{FY}(\,\mathrm{j}\omega)$, respectively their averaged versions, are defined as:
\begin{align}
	S_{FF}(\,\mathrm{j}\omega) &= F_{\text{ZP}}^\ast(\,\mathrm{j}\omega)F_{\text{ZP}}(\,\mathrm{j}\omega)\qquad\qquad\text{with}\qquad\qquad S_{FF}^{\text{avg}}(\,\mathrm{j}\omega) = \sum_{i=1}^n\dfrac{1}{n}S_{FF,i}(\,\mathrm{j}\omega) \\
	S_{FY}(\,\mathrm{j}\omega) &= F_{\text{ZP}}^\ast(\,\mathrm{j}\omega)Y_{\text{ZP}}(\,\mathrm{j}\omega)\qquad\qquad\text{with}\qquad\qquad S_{FY}^{\text{avg}}(\,\mathrm{j}\omega) = \sum_{i=1}^n\dfrac{1}{n}S_{FY,i}(\,\mathrm{j}\omega)
\end{align}
where the asterisk superscript denotes the complex conjugate. Using these spectra, the $H_1$ FRF estimator can be calculated~\cite{RandallAllemang2020} and the experimental impulse response function is then found by:
\begin{equation}
	h[k]=\text{IFFT}\lbrace H_1(\,\mathrm{j}\omega)\rbrace\qquad\text{where}\qquad H_1(\,\mathrm{j}\omega) = \dfrac{S_{FY}^{\text{avg}}(\,\mathrm{j}\omega)}{S_{FF}^{\text{avg}}(\,\mathrm{j}\omega)}\label{eq:H1F}
\end{equation}

\subsection{Time Domain IRF Identification}

A calculation procedure for IRFs in the time domain can be found by rewriting the discretized Duhamel integral from \cref{eq:conv_disc} as an equivalent matrix-vector product, assuming an excitation by an imperfect impulse which is only non-zero for $M$ samples and set to zero using a force window for every other value:
\begin{equation}
	\underbrace{\left[\begin{matrix}
			y[1] \\
			\\[-0.1cm]
			y[2] \\
			\\
			\vdots \\
			\\
			y[N-1] \\
			\\[-0.1cm]
			y[N]
		\end{matrix}\right]}_{\displaystyle\underset{(N\times 1)}{\bar{\vy}}}=\underbrace{\left[\begin{matrix}
			f[0] & 0 & \dots & 0 \\
			f[1] & f[0] & \ddots & \vdots \\
			\vdots & f[1] & \ddots & 0 \\
			f[M-1] & \vdots & \ddots & f[0] \\
			0 & f[M-1] & \vdots & f[1] \\
			\vdots & \ddots & \ddots & \vdots \\
			0 & \dots & \;0\; & f[M-1]
		\end{matrix}\right]}_{\displaystyle\underset{(N\times Q)}{\vF}}\underbrace{\left[\begin{matrix}
			h[0] \\[0.1cm]
			h[1] \\[0.1cm]
			\vdots \\[0.1cm]
			h[Q-1]
		\end{matrix}\right]}_{\displaystyle\underset{(Q\times 1)}{\bar{\vh}}}\Delta t \label{eq:ConvVandS}
\end{equation}
where $N$ is the length of one measured response $y[k]$, stored in the vector $\bar{\vy}$, where the overline denotes that the elements of a single time series are stored in a vector and that $\bar{\vy}$ does not contain multiple dofs. $M$ is the length of the experimentally applied force $f[k]$, arranged in the convolution matrix $\vF$, and $Q$ is the length of the impulse response function $h[k]$ to be identified. The IRF length is prescribed by $Q=N-M+1$, assuming it to be zero after $Q$, because the length of the full convolution product is $N=Q+M-1$.

Since the force convolution matrix $\vF$ is not square, the IRF $\bar{\vh}$ has to be solved for in a least-squares sense, which is achieved by premultiplying with $\vF\tp$, yielding the pseudo-inverse~\cite{VanDerSeijs2014}:
\begin{equation}
	\bar{\vh}=\dfrac{1}{\Delta t}\left(\vF\tp\vF\right)^{-1}\vF\tp\bar{\vy}=\dfrac{1}{\Delta t}\vR^{-1}\vF\tp\bar{\vy}\label{eq:hexp_inverse_force_filter}
\end{equation}
where $\vR$ is the auto-correlation matrix of the experimental force $f[k]$ and $\vF\tp\bar{\vy}$ is the cross-correlation between the experimental force $f[k]$ and response $y[k]$. From \cref{eq:hexp_inverse_force_filter} the IRF $\bar{\vh}$ can then be retrieved using a sparse linear solver, since $Q\gg M$ for the cases considered here. Multiple impacts $i$ can be averaged by stacking the respective $\vF_i$ matrices vertically and calculating an averaged auto-correlation matrix $\vR^{\text{avg}}=\vF_{\text{stacked}}\tp\vF_{\text{stacked}}^{\vphantom{\text{T}}}$. Then, summing up the cross-correlation vectors $\vF_i\tp\bar{\vy}_i$ yields the averaged IRF, calculated exclusively in the time domain~\cite{VanDerSeijs2014}:
\begin{equation}
	\bar{\vh}^{\text{avg}}=\dfrac{1}{\Delta t}{\left(\vR^{\text{avg}}\right)}^{-1}\sum_{i=1}^{n_i}\left(\vF_i\tp\bar{\vy}_i\right)\label{eq:E4_est}
\end{equation}
Looking closer at \cref{eq:E4_est}, this equation is the time domain version of the $H_1$ FRF estimator (\cref{eq:H1F}), building the ratio of the cross-correlation, between the excitation $f_\text{exp}[k]$ and the response $y_\text{exp}[k]$, and the auto-correlation of the force $f_\text{exp}[k]$. Note that none of the used quantities are transformed into the frequency domain.

\subsection{Discussion of Impulse Response Function Identification Methods}

Both IRF identification methods use the $H_1$ estimator, once in the frequency domain and once in the time domain. The quality of the calculated IRFs can potentially be improved by substituting the $H_1$ estimator with a different one, for example, the $H_v$ estimator, respectively the equivalent time domain version.

While processing the IRFs through the frequency domain might seem like a drawback due to the assumed linearity of the Fourier transform, this does not matter as IRFs themselves are a linear concept. The clear advantage is the cheap computation of FRFs and their IFFT as well as the possibility to use any of the many established FRF estimators. However, a significant drawback is the required response length, which must be sufficiently long for a good frequency resolution. More importantly, the response of the system is required to decay within the measurement window, because otherwise, the forced periodization of the Fourier transform will introduce leakage.

The drawback of the time domain method is the very expensive computation, especially the inversion of the (averaged) auto-correlation matrix $\vR$, which depends on the sparsity of the matrix, determined by the length of the experimentally applied force $f[k]$, and the overall size of the matrix, determined by the sample rate and the chosen IRF length, i.e.\ the number of samples. When multiple impacts are averaged, this does not change the size of the auto-correlation matrix $\vR^{\text{avg}}$, but only the number of multiplications required for building $\vR^{\text{avg}}$ and the final IRF $\bar{\vh}^{\text{avg}}$, the computationally most expensive inversion is unaffected.

Nevertheless, the computational costs can be reduced significantly by the fact that the responses do not have to decay fully within the measurement window, because there is no forced periodization in the time domain. While there are also cut-off effects with IRFs, i.e.\ an impulse at the end that forces the response to zero, this does not influence the responses generated by IBS as long as the IRF is calculated for a longer time than the IBS procedure is evaluated. However, calculating the IRFs in the time domain for only short times, e.g.\ in this paper for \SI{50}{\milli\second}, takes about an order of magnitude longer than calculating the IRF in the frequency domain for an length of approximately \SI{1.2}{\second}. A short summary of this comparison can be found in \cref{tab:IRF_FvT}.\vspace{-0.1cm}

\begin{table}[ht]
	\centering
	\begin{tabularx}{0.9\textwidth}{c|X|X}
		 & \centering\arraybackslash\textbf{Frequency Domain} & \centering\arraybackslash\textbf{Time Domain} \\\hline
		 \textbf{Computational costs} & $\oplus$ Cheap, only multiplication and division of complex numbers & $\ominus$ Expensive, inversion of (large) auto-correlation matrix $\vR$ \\\hline
		 \textbf{Required response length} & $\ominus$ Longer response required for sufficient frequency resolution & $\oplus$ Only few milliseconds required. e.g.\ used here: \qty{50}{\milli\second} \\\hline
		 \textbf{Decay of responses} & $\ominus$ Must decay within measurement window, otherwise leakage (due to forced periodicity) & $\oplus$ Decay within measurement window not required, but minimally $t_\text{measure}>t_\text{IBS}$
	\end{tabularx}
	\caption{Comparison of IRF identification in the frequency domain versus in the time domain}
	\label{tab:IRF_FvT}
\end{table}\vspace{-0.7cm}

\subsection{Limiting the Frequency Content by Downsampling with Low-Pass Filter}

In addition to the selection of the IRF calculation method, the frequency bandwidth used for the IRFs and with that for the IBS scheme has to be chosen as well. For FBS this can be achieved by truncating the FRFs at a certain frequency, disregarding all higher frequencies. Since the span of the frequency axis depends on the used sampling frequency $f_\text{S}$, truncation of the frequency content in the time domain can be achieved by downsampling.

In order to avoid aliasing, a suitable low-pass filter has to be applied to the time series prior to downsampling. For this paper, the downsampling implementation of \textit{SciPy} (\textit{signal.decimate}) is used, which applies an 8th order Chebyshev type I low-pass filter prior to removing sample points according to the chosen whole number downsampling factor. The corner frequency of the filter is set at \SI{80}{\percent} of the new sampling frequency after downsampling. Any phase shift introduced by the filter is compensated by filtering twice, once forward and once backward.
%\Cref{fig:Filter} shows the amplitude response of the filter with respect to the corner frequency.
%
%\begin{figure}[ht]
%	\centering
%	\small
%	\input{figures/theory/filter.tikz}
%	\caption{Amplitude response of the utilized 8th order Chebyshev type I low-pass filter}
%	\label{fig:Filter}
%\end{figure}

While it would also be possible to only apply a low-pass filter and retain the original sampling frequency, the removal of samples yields computational cost improvements. For a fixed IBS computation length, a downsampling factor of two halves the amount of response values $y[k]$ that have to be calculated. Since the computation cost of the (discretized) convolution integrals per time step increases linearly, the computation time reduction is greater than \SI{50}{\percent}. Further, if downsampling is applied to the response and excitation time series prior to an IRF calculation in the time domain, the size of the auto-correlation matrix is reduced by the downsampling factor squared. Because the calculations required to solve the linear system for the IRF identification do not linearly depend on the matrix size, the computation time can be reduced significantly. For this reason, the measurements are downsampled prior to the IRF calculations, i.e.\ the IRFs are not computed from the fully sampled raw response signals.
% !TeX spellcheck = en_US
\section{Experimental Test Cases}\label{sec:Exp}

The experimental test cases considered here are rods made out of aluminum (EN AW 7075) and Polyoxymethylene (POM), which are assumed to behave like one-dimensional bars. For reference measurements, the rods are not assembled physically and instead, longer rods pose as a perfectly assembled system. While this test case has limited relevance for practical applications, it serves as a good validation setup because the number of variables is as limited as possible, e.g., there is no variation of the interface assembly.

\subsection{Description of Test Cases}

\begin{figure}[ht!]
	\centering
	\def\svgwidth{0.7\textwidth}
	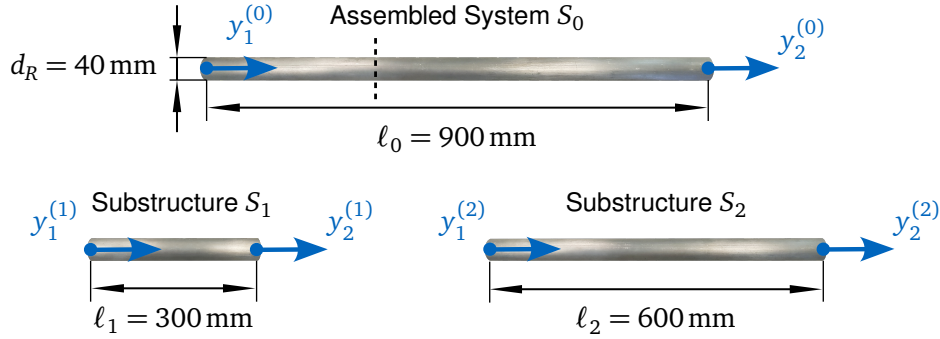
	\caption{Overview of geometry and coordinates of the 1D rods considered as test pieces, shown here: aluminum rods}
	\label{fig:1D_bar_overview}
\end{figure}

For both materials, three test pieces were manufactured, once two rods considered as the substructure $S_1$ and $S_2$ with a length of $\ell_1=\SI{300}{\milli\meter}$ and $\ell_2=\SI{600}{\milli\meter}$ respectively, and once the physical reference of the assembled system $S_0$ with a length of $\ell_0=\SI{900}{\milli\meter}$, see \cref{fig:1D_bar_overview}. The rods are assumed to only have one translational dof at each end. Within the IBS scheme, the response $y_2^{(1)}$ is coupled with $y_1^{(2)}$, using the signed Boolean constraint matrices $\vB^{(1)}=\left[\begin{matrix} 0 & 1 \end{matrix}\right]$ and $\vB^{(2)}=\left[\begin{matrix} -1 & 0 \end{matrix}\right]$. The properties of both materials are taken from the relevant datasheets and summarized below in \cref{tab:1D_prop_material}. From this, the longitudinal wave propagation speed ${c_{\text{wave}} = \sqrt{E/\rho}}$~\cite{Rixen2010} is calculated, which describes the fundamental dynamics observable in the longitudinal direction.\vspace{-0.2cm}

\vspace{-0.2cm}

\begin{table}[ht!]
	\centering
	\small
	\begin{tabular}{c|c|c}
		\textbf{Property} & \textbf{EN AW 7075}~\cite{ENAW7075} & \textbf{Sustarin C${}^\copyright$}~\cite{SusDS}  \\\hline
		Common name & Aluminum & Polyoxymethylene \\\hline
		Young's modulus $E$ & \SI{71}{\giga\pascal} & \SI{2.76}{\giga\pascal} \\\hline
		Density $\rho$ &  \SI[per-mode=symbol]{2800}{\kilogram\per\cubic\meter} & \SI[per-mode=symbol]{1400}{\kilogram\per\cubic\meter} \\\hline
		Wave propagation speed $c_{\text{wave}}$ & \SI[per-mode=symbol]{5036}{\meter\per\second} & \SI[per-mode=symbol]{1404}{\meter\per\second}
	\end{tabular}
	\captionof{table}{Material properties of aluminum and POM rod test pieces}
	\label{tab:1D_prop_material}
\end{table}\vspace{-0.2cm}

Using the length $\ell$ of the respective rod, the time it takes for the shock wave created by an impact to travel through the rod to the other side can be calculated as ${t_\text{travel}=\ell/c_{\text{wave}}}$. More importantly, the time the shock wave takes to return to the driving point is given by ${t_{\text{return}} = 2\ell/c_{\text{wave}}}$, which determines the frequency of the first observable eigenfrequency in the longitudinal direction as ${f_1=1/t_{\text{return}}}$. An overview of the aforementioned quantities for all considered test structures is given in \cref{tab:Rod_dyn}.

\begin{table}[ht!]
	\centering
	\small
	\begin{tabular}{c|c|c|c|c}
		\multicolumn{2}{c|}{\textbf{Property}} & \textbf{Substructure 1} & \textbf{Substructure 2} & \textbf{Assembled System} \\\hline
		\multicolumn{2}{c|}{Length $\ell$} & \SI{300}{\milli\meter} & \SI{600}{\milli\meter} & \SI{900}{\milli\meter} \\\hline
		\multirow{2}{*}{Wave travel time $t_{\text{travel}}$} & POM & \SI{0.214}{\milli\second} & \SI{0.427}{\milli\second} & \SI{0.641}{\milli\second} \\\cline{2-5}
		& Alu & \SI{0.060}{\milli\second} & \SI{0.119}{\milli\second} & \SI{0.179}{\milli\second} \\\hline
		\multirow{2}{*}{Wave return time $t_{\text{return}}$} & POM & \SI{0.427}{\milli\second} & \SI{0.855}{\milli\second} & \SI{1.28}{\milli\second} \\\cline{2-5}
		& Alu & \SI{0.119}{\milli\second} & \SI{0.238}{\milli\second} & \SI{0.357}{\milli\second} \\\hline
		\multirow{2}{*}{First eigenfrequency $f_1$} & POM & \SI{2.34}{\kilo\hertz} & \SI{1.17}{\kilo\hertz} & \SI{0.781}{\kilo\hertz} \\\cline{2-5}
		& Alu & \SI{8.39}{\kilo\hertz} & \SI{4.20}{\kilo\hertz} & \SI{2.80}{\kilo\hertz}
	\end{tabular}
	\caption{Summary of expected dynamics of aluminum and POM rod test pieces}
	\label{tab:Rod_dyn}
\end{table}

To capture the minimally required first eigenfrequency of substructure $S_1$, for the aluminum rods a sampling frequency of $f_\text{S}\approx\SI{20}{\kilo\hertz}$ is required, for the POM rods $f_\text{S}\approx\SI{6}{\kilo\hertz}$.

\subsection{Experimental Setup}

\begin{figure}[ht!]
	\centering
	\begin{subfigure}{0.375\textwidth}
		\centering
		\includegraphics[width=\columnwidth]{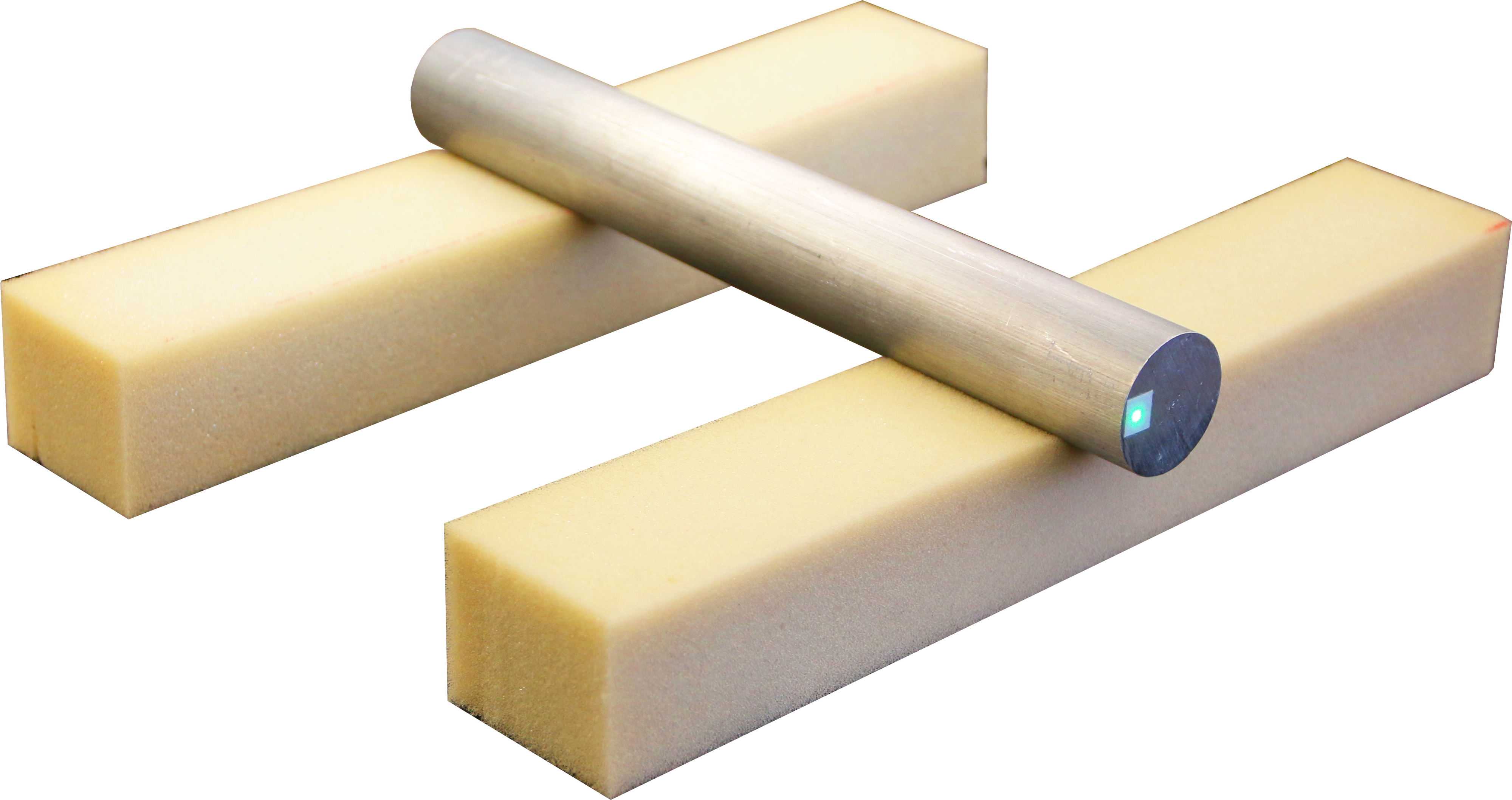}
		\caption{Measurement of displacements \& velocities with Laservibrometer on aluminum rod}
	\end{subfigure}\hspace{0.4cm}
	\begin{subfigure}{0.375\textwidth}
		\centering
		\includegraphics[width=\columnwidth]{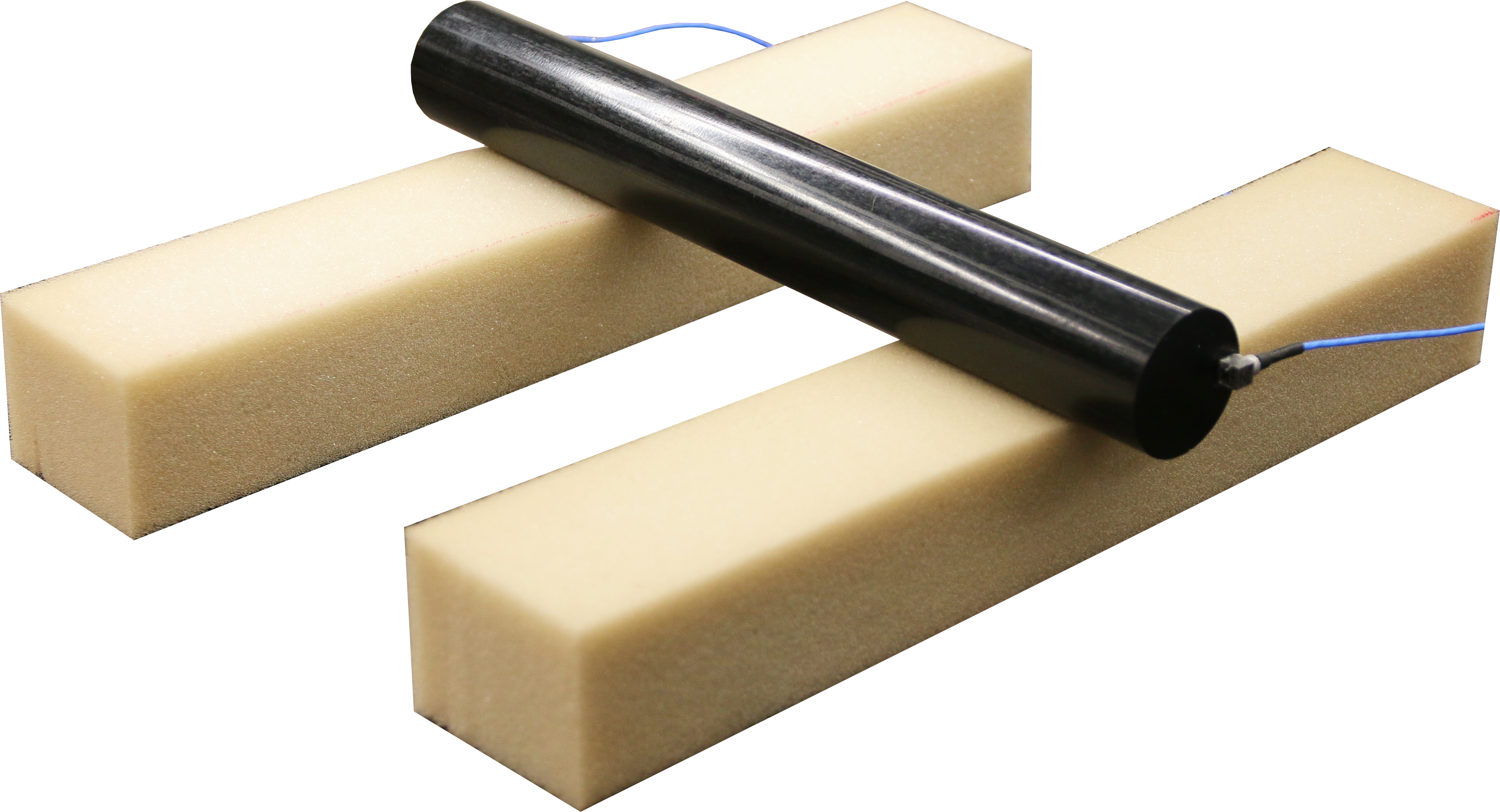}
		\caption{Measurement of accelerations using triaxial accelerometer on POM rod}
	\end{subfigure}
	\caption{Approximate 'free-free' setup of test rods using foam}
	\label{fig:foam_setup}
\end{figure}

\begin{figure}[ht!]
	\centering
	\includegraphics[width=\textwidth]{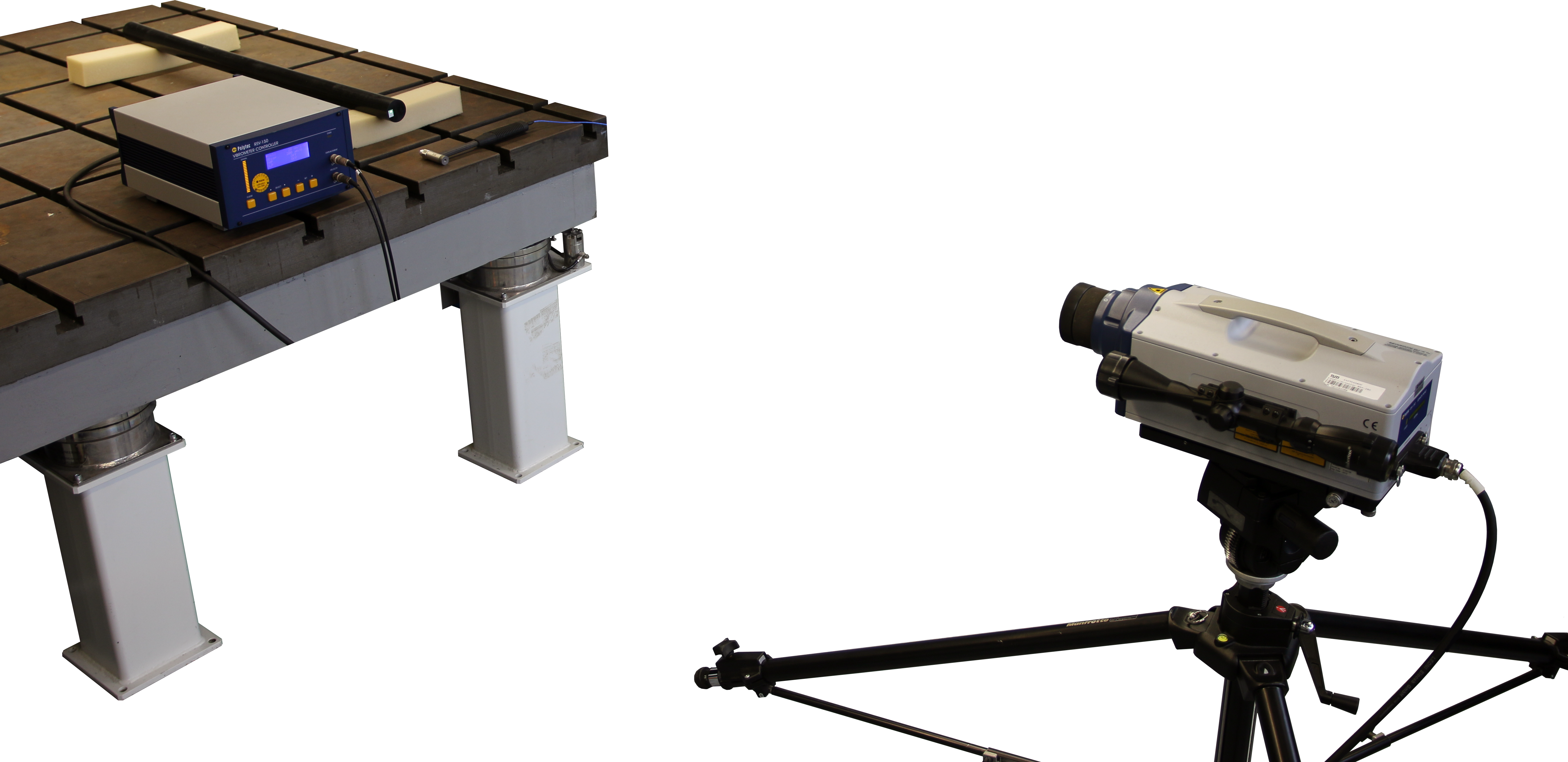}
	\caption{Experimental setup for the measurement of displacements and velocities using a Laservibrometer}
	\label{fig:exp_laser_setup}
\end{figure}

To capture the time series of excitation and responses, the \textit{LMS SCADAS mobile measurement system} is used with the highest available sampling rate of $f_\text{S}=\SI{102.4}{\kilo\hertz}$ and a measurement duration of approximately \SI{1.2}{\second}. For the IRF calculation in the frequency domain, the full measurement duration is used, while for the calculation in the time domain only \SI{50}{\milli\second} is used. The rods are approximately set up in a 'free-free' configuration by placing them on two foam pieces as shown in \cref{fig:foam_setup}, atop of an isolated table, see \cref{fig:exp_laser_setup}. 

For the aluminum rods, the rigid body translation is at roughly $7$-$\SI{10}{\hertz}$, for the POM rods at roughly $8$-$\SI{12}{\hertz}$. Displacement, velocity and acceleration responses to hammer impacts are measured, where the former two quantities are obtained with a \textit{PolyTec RSV-150 Laservibrometer} and the latter one with \textit{PCB Piezotronics Model 356A03 Triaxial ICP® accelerometers}. The responses are measured off-center to allow for a impact at the center.

Accelerations can be measured on both sides simultaneously, while for displacements and velocities separate measurements have to be performed for each side. For IBS, a full $2\times 2$ IRF matrix is required for each structure, containing the response at the driving point edge surface and the edge at the opposite end, and this for impacts on both sides. Here, the rods are assumed to be symmetric with respect to which side is impacted, hence only half of the required responses for the IRF matrix are measured, and the other values are derived from symmetry.

%\begin{figure}[ht]
%	\centering
%	\begin{subfigure}{0.55\textwidth}
%	\begin{subfigure}{\textwidth}
%		\centering
%		\includegraphics[width=0.8\columnwidth]{figures/experimental/LaserVib.png}
%		\caption{PolyTec RSV-150 Laservibrometer with controller, image from PolyTec}
%		\label{fig:LaserVib}
%	\end{subfigure}\vspace{0.4cm}\\
%	\begin{subfigure}{\textwidth}
%		\centering
%		\includegraphics[width=0.4\columnwidth]{figures/experimental/sens.png}
%		\caption{PCB Piezotronics Model 356A03 Triaxial ICP® accelerometer}
%		\label{fig:sens}
%	\end{subfigure}
%	\end{subfigure}\hspace{0.4cm}
%	\begin{subfigure}{0.35\textwidth}
%		\centering
%		\includegraphics[width=0.9\columnwidth]{figures/experimental/hammer_rot.png}
%		\caption{PCB Piezotronics 086B03 manual hammer with steel tip (also used with vinyl tip)}
%		\label{fig:hammer}
%	\end{subfigure}
%	\caption{Overview of utilized sensors for displacement \& velocity responses (Laservibrometer), acceleration responses (Accelerometer) and system excitation (manual impact hammer)}
%\end{figure}

\begin{figure}[ht!]
	\small
	\centering
	\begin{subfigure}[t]{0.15\textwidth}\vspace{0.4cm}
		\begin{subfigure}{0.8\textwidth}
			\centering
			\includegraphics[width=0.85\columnwidth]{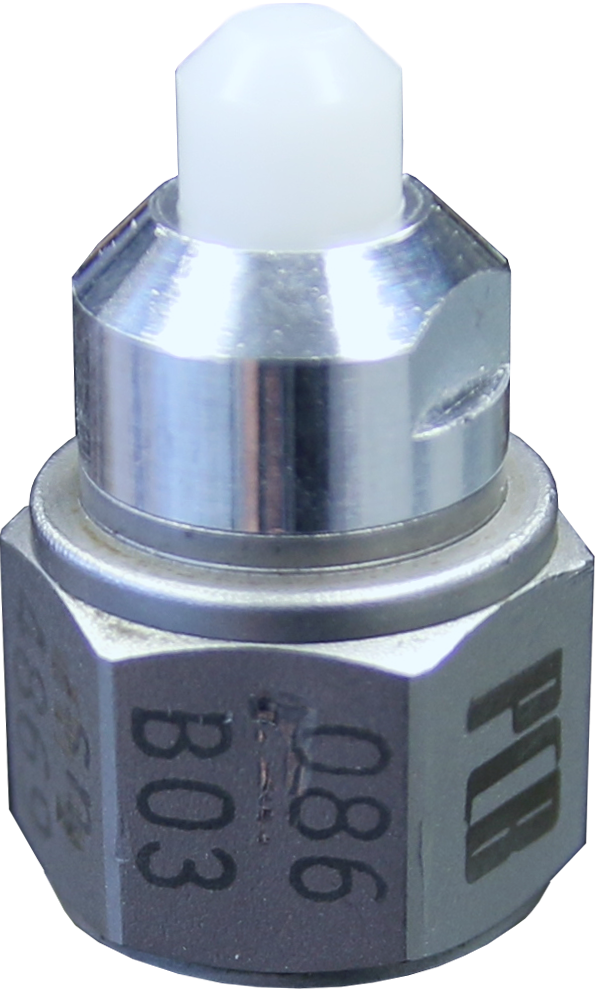}\vspace{0.2cm}
			{\includegraphics{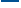}$\,$Vinyl tip}
		\end{subfigure}\vspace{0.6cm}\\
		\begin{subfigure}{0.8\textwidth}
			\centering
			\includegraphics[width=0.85\columnwidth]{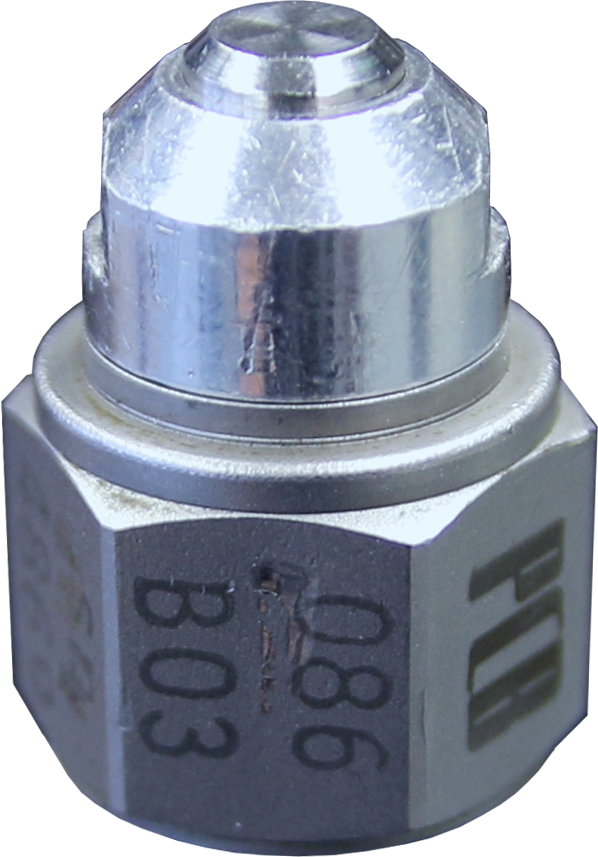}\vspace{0.2cm}
			{\includegraphics{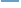}$\,$Steel tip}
		\end{subfigure}
	\end{subfigure}
	\begin{subfigure}[t]{0.8\textwidth}\vspace{0pt}
		\input{figures/experimental/impact_shape.tikz}
	\end{subfigure}\vspace{-0.4cm}
	\caption{Excitation forces on aluminum and POM rods of the available hammer tips in time and frequency domain}
	\label{fig:impacts_compare}
\end{figure}

The rod test pieces are excited using a \textit{PCB Piezotronics 086B03 manual hammer}, once using a vinyl tip and once using a steel tip. Depending on the material pairing, different imperfect impact lengths, and with that, varying excitation bandwidths, can be achieved. One exemplary impact, out of the 20 performed impacts with each tip on each rod material, is shown in \cref{fig:impacts_compare}, once in the time domain and once in the frequency domain.
\newpage

As can be seen, the steel tip has a better excitation bandwidth, i.e.\ the drop-off occurs significantly later, especially for the aluminum rods, where the excitation up to \SI{10}{\kilo\hertz} is more uniform than with the vinyl tip and higher magnitudes for almost all frequencies can be seen. With the POM rods, the difference between both tips is less pronounced, but the steel tip still performs slightly better. While it therefore would be recommended to always use a steel tip, its higher tendency for double impacts means that this is not always possible.

Before the IRFs are calculated using the procedures outlined previously, the measured data is temporally aligned, i.e.\ the time points within the pre-trigger that do not contain system dynamics are removed. Additionally, a force window is applied to the measured excitation, extracting the applied impulse with unaltered magnitudes.

\subsection{Experimental Results Using Full Measurement Bandwidth}

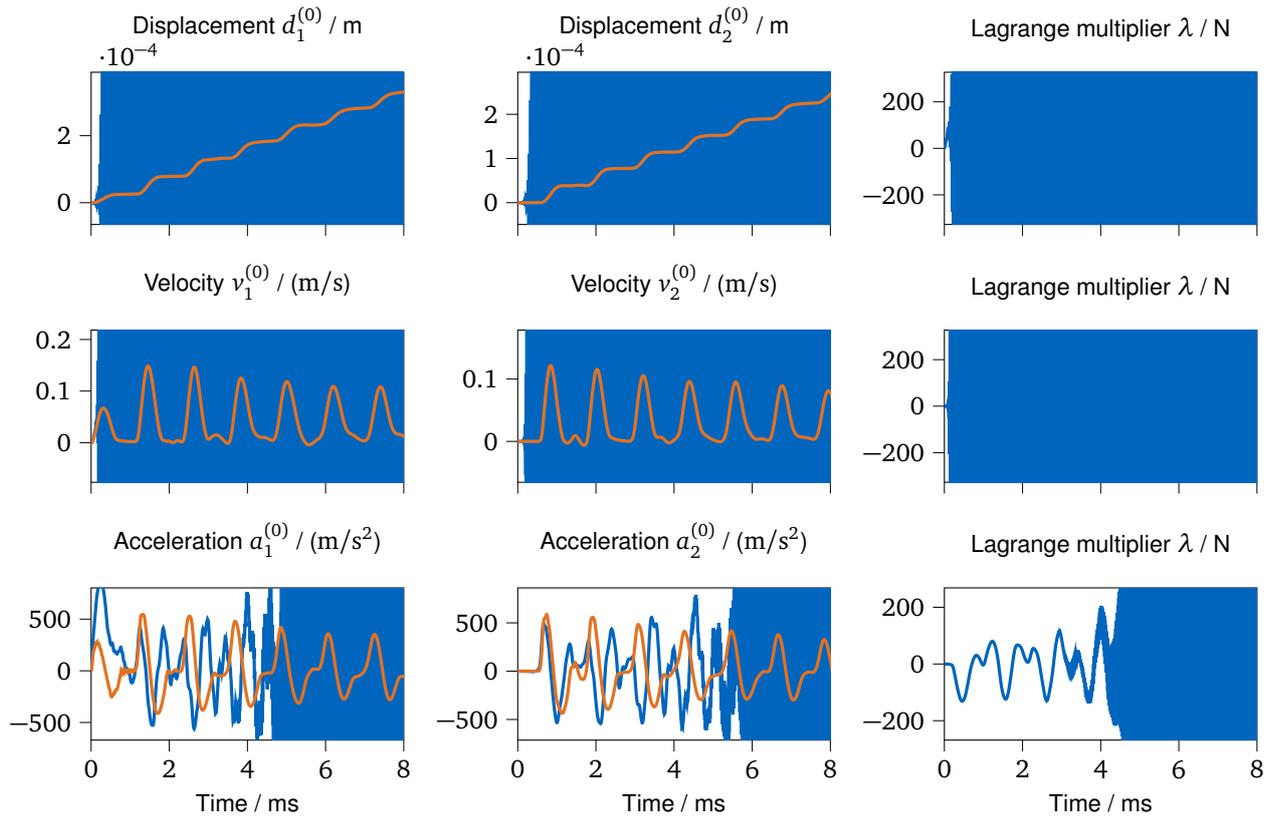
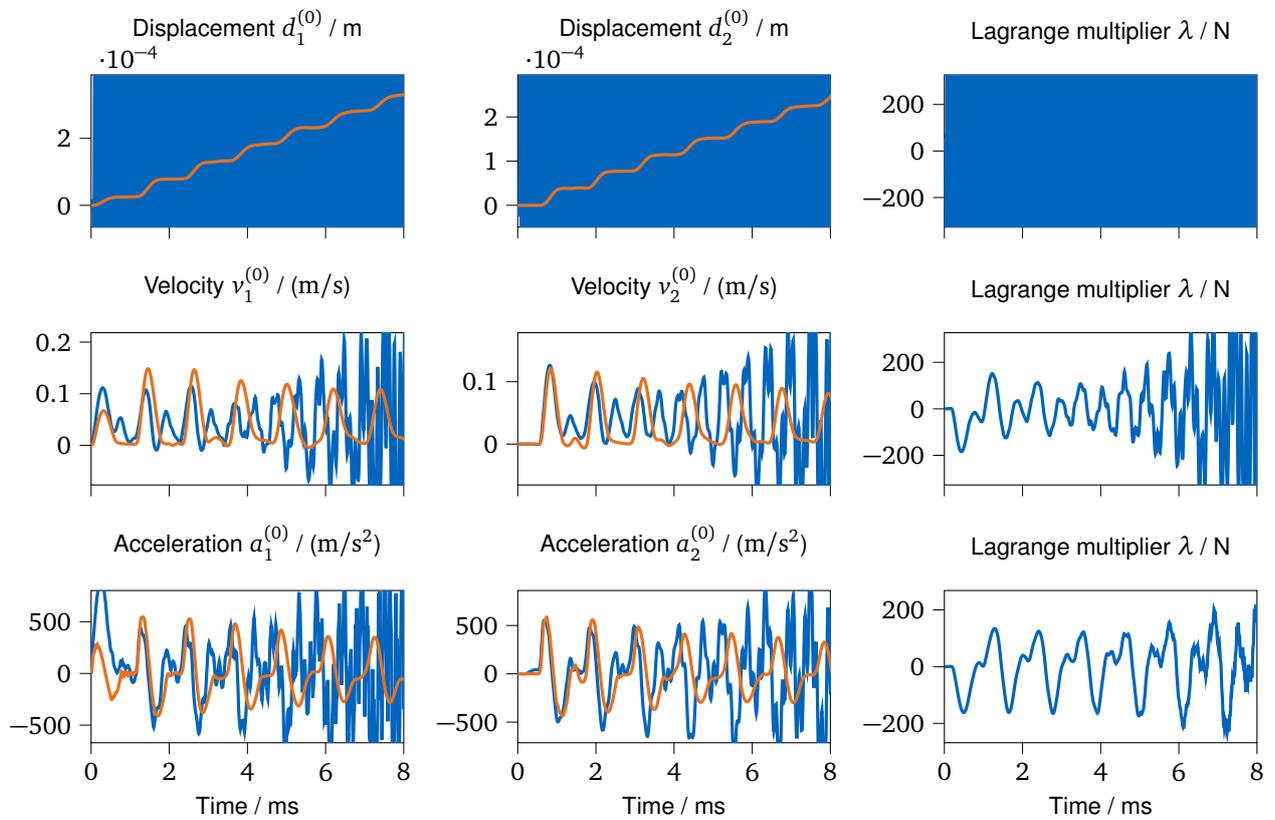
\begin{figure}[ht!]
	\begin{subfigure}{\textwidth}
		\small
		\renewcommand{\IBSlen}{8}
		\renewcommand{\IBSResultCSVdisp}{data/IBS_results_POM_manual_vinyl_d_Frequency.csv}
		\renewcommand{\yAmindisp}{-6.497262187500001e-05}
		\renewcommand{\yAmaxdisp}{0.00038945157890625}
		\renewcommand{\yBmindisp}{-4.9327228125000014e-05}
		\renewcommand{\yBmaxdisp}{0.00029517266953125006}
		\renewcommand{\LMmindisp}{-327.32400556500454}
		\renewcommand{\LMmaxdisp}{327.32400556500454}
		
		\renewcommand{\IBSResultCSVvelo}{data/IBS_results_POM_manual_vinyl_v_Frequency.csv}
		\renewcommand{\yAminvelo}{-0.076767534375}
		\renewcommand{\yAmaxvelo}{0.21826226953125}
		\renewcommand{\yBminvelo}{-0.0656033150390625}
		\renewcommand{\yBmaxvelo}{0.17862257109375}
		\renewcommand{\LMminvelo}{-327.32400556500454}
		\renewcommand{\LMmaxvelo}{327.32400556500454}
		
		\renewcommand{\IBSResultCSVacc}{data/IBS_results_POM_manual_vinyl_a_Frequency.csv}
		\renewcommand{\yAminacc}{-670.4906805419922}
		\renewcommand{\yAmaxacc}{802.0985375976562}
		\renewcommand{\yBminacc}{-713.6285949707031}
		\renewcommand{\yBmaxacc}{861.9254022216796}
		\renewcommand{\LMminacc}{-268.380378112793}
		\renewcommand{\LMmaxacc}{268.380378112793}
		\begin{flushright}
			% This file was created with tikzplotlib v0.10.1.
\begin{tikzpicture}

\definecolor{darkgray176}{RGB}{176,176,176}
\definecolor{gray}{RGB}{128,128,128}
\definecolor{lightgray204}{RGB}{204,204,204}

\begin{groupplot}[group style={group size=3 by 3, horizontal sep=1.5cm, vertical sep=1.4cm}]

% Displacements
\nextgroupplot[
height=3.6cm,
scaled x ticks=manual:{}{\pgfmathparse{#1}},
yticklabel style={
        /pgf/number format/fixed,
        /pgf/number format/precision=3
},
tick align=outside,
tick pos=left,
title={Displacement \(\displaystyle d_1^{(0)}\) / m},
title style={yshift=0.5ex},
width=0.335\textwidth,
x grid style={darkgray176},
xmin=0, xmax=\IBSlen,
xtick style={color=black},
xticklabels={},
y grid style={darkgray176},
ymin=\yAmindisp, ymax=\yAmaxdisp,
ytick style={color=black}
]
\addplot [very thick, TUMBlue]
table[x=t, y=y1IBS, col sep=comma]{\IBSResultCSVdisp};

\addplot [very thick, TUMOrange]
table[x=t, y=y1Meas, col sep=comma]{\IBSResultCSVdisp};

\nextgroupplot[
height=3.6cm,
scaled x ticks=manual:{}{\pgfmathparse{#1}},
yticklabel style={
        /pgf/number format/fixed,
        /pgf/number format/precision=3
},
tick align=outside,
tick pos=left,
title={Displacement \(\displaystyle d_2^{(0)}\) / m},
title style={yshift=0.5ex},
width=0.335\textwidth,
x grid style={darkgray176},
xmin=0, xmax=\IBSlen,
xtick style={color=black},
xticklabels={},
y grid style={darkgray176},
ymin=\yBmindisp, ymax=\yBmaxdisp,
ytick style={color=black}
]
\addplot [very thick, TUMBlue]
table[x=t, y=y2IBS, col sep=comma]{\IBSResultCSVdisp};

\addplot [very thick, TUMOrange]
table[x=t, y=y2Meas, col sep=comma]{\IBSResultCSVdisp};

\nextgroupplot[
height=3.6cm,
tick align=outside,
tick pos=left,
title={Lagrange multiplier \(\displaystyle \lambda\) / N},
title style={yshift=0.5ex},
width=0.335\textwidth,
x grid style={darkgray176},
xmin=0, xmax=\IBSlen,
xtick style={color=black},
xticklabels={},
y grid style={darkgray176},
ymin=\LMmindisp, ymax=\LMmaxdisp,
ytick style={color=black}
]
\addplot [very thick, TUMBlue]
table[x=t, y=lambda, col sep=comma]{\IBSResultCSVdisp};

% Velocities
\nextgroupplot[
height=3.6cm,
scaled x ticks=manual:{}{\pgfmathparse{#1}},
yticklabel style={
        /pgf/number format/fixed,
        /pgf/number format/precision=3
},
tick align=outside,
tick pos=left,
title={Velocity \(\displaystyle v_1^{(0)}\) / (\si[per-mode=symbol]{\meter\per\second})},
title style={yshift=0.5ex},
width=0.335\textwidth,
x grid style={darkgray176},
xmin=0, xmax=\IBSlen,
xtick style={color=black},
xticklabels={},
y grid style={darkgray176},
ymin=\yAminvelo, ymax=\yAmaxvelo,
ytick style={color=black}
]
\addplot [very thick, TUMBlue]
table[x=t, y=y1IBS, col sep=comma]{\IBSResultCSVvelo};

\addplot [very thick, TUMOrange]
table[x=t, y=y1Meas, col sep=comma]{\IBSResultCSVvelo};

\nextgroupplot[
height=3.6cm,
scaled x ticks=manual:{}{\pgfmathparse{#1}},
yticklabel style={
        /pgf/number format/fixed,
        /pgf/number format/precision=3
},
tick align=outside,
tick pos=left,
title={Velocity \(\displaystyle v_2^{(0)}\) / (\si[per-mode=symbol]{\meter\per\second})},
title style={yshift=0.5ex},
width=0.335\textwidth,
x grid style={darkgray176},
xmin=0, xmax=\IBSlen,
xtick style={color=black},
xticklabels={},
y grid style={darkgray176},
ymin=\yBminvelo, ymax=\yBmaxvelo,
ytick style={color=black}
]
\addplot [very thick, TUMBlue]
table[x=t, y=y2IBS, col sep=comma]{\IBSResultCSVvelo};

\addplot [very thick, TUMOrange]
table[x=t, y=y2Meas, col sep=comma]{\IBSResultCSVvelo};

\nextgroupplot[
height=3.6cm,
tick align=outside,
tick pos=left,
title={Lagrange multiplier \(\displaystyle \lambda\) / N},
title style={yshift=0.5ex},
width=0.335\textwidth,
x grid style={darkgray176},
xmin=0, xmax=\IBSlen,
xtick style={color=black},
xticklabels={},
y grid style={darkgray176},
ymin=\LMminvelo, ymax=\LMmaxvelo,
ytick style={color=black}
]
\addplot [very thick, TUMBlue]
table[x=t, y=lambda, col sep=comma]{\IBSResultCSVvelo};

% Accelerations
\nextgroupplot[
height=3.6cm,
scaled x ticks=manual:{}{\pgfmathparse{#1}},
yticklabel style={
        /pgf/number format/fixed,
        /pgf/number format/precision=3
},
tick align=outside,
tick pos=left,
title={Acceleration \(\displaystyle a_1^{(0)}\) / (\si[per-mode=symbol]{\meter\per\second\squared})},
title style={yshift=0.5ex},
width=0.335\textwidth,
x grid style={darkgray176},
xmin=0, xmax=\IBSlen,
xtick style={color=black},
xlabel={Time / ms},
y grid style={darkgray176},
ymin=\yAminacc, ymax=\yAmaxacc,
ytick style={color=black}
]
\addplot [very thick, TUMBlue]
table[x=t, y=y1IBS, col sep=comma]{\IBSResultCSVacc};

\addplot [very thick, TUMOrange]
table[x=t, y=y1Meas, col sep=comma]{\IBSResultCSVacc};

\nextgroupplot[
height=3.6cm,
scaled x ticks=manual:{}{\pgfmathparse{#1}},
yticklabel style={
        /pgf/number format/fixed,
        /pgf/number format/precision=3
},
tick align=outside,
tick pos=left,
title={Acceleration \(\displaystyle a_2^{(0)}\) / (\si[per-mode=symbol]{\meter\per\second\squared})},
title style={yshift=0.5ex},
width=0.335\textwidth,
x grid style={darkgray176},
xmin=0, xmax=\IBSlen,
xtick style={color=black},
xlabel={Time / ms},
y grid style={darkgray176},
ymin=\yBminacc, ymax=\yBmaxacc,
ytick style={color=black}
]
\addplot [very thick, TUMBlue]
table[x=t, y=y2IBS, col sep=comma]{\IBSResultCSVacc};

\addplot [very thick, TUMOrange]
table[x=t, y=y2Meas, col sep=comma]{\IBSResultCSVacc};

\nextgroupplot[
height=3.6cm,
tick align=outside,
tick pos=left,
title={Lagrange multiplier \(\displaystyle \lambda\) / N},
title style={yshift=0.5ex},
width=0.335\textwidth,
x grid style={darkgray176},
xmin=0, xmax=\IBSlen,
xtick style={color=black},
xlabel={Time / ms},
y grid style={darkgray176},
ytick style={color=black},
ymin=\LMminacc, ymax=\LMmaxacc
]
\addplot [very thick, TUMBlue]
table[x=t, y=lambda, col sep=comma]{\IBSResultCSVacc};

\end{groupplot}
\end{tikzpicture}
		\end{flushright}\vspace{-0.4cm}
		\caption{IRFs processed using the \textbf{Frequency Domain} method}\vspace{0.4cm}
		\label{fig:IBS_results_POM_manual_vinylA}
	\end{subfigure}
	\begin{subfigure}{\textwidth}
		\small
		\renewcommand{\IBSlen}{8}
		\renewcommand{\IBSResultCSVdisp}{data/IBS_results_POM_manual_vinyl_d_Time.csv}
		\renewcommand{\yAmindisp}{-6.497262187500001e-05}
		\renewcommand{\yAmaxdisp}{0.00038945157890625}
		\renewcommand{\yBmindisp}{-4.9327228125000014e-05}
		\renewcommand{\yBmaxdisp}{0.00029517266953125006}
		\renewcommand{\LMmindisp}{-327.32400556500454}
		\renewcommand{\LMmaxdisp}{327.32400556500454}
		
		\renewcommand{\IBSResultCSVvelo}{data/IBS_results_POM_manual_vinyl_v_Time.csv}
		\renewcommand{\yAminvelo}{-0.076767534375}
		\renewcommand{\yAmaxvelo}{0.21826226953125}
		\renewcommand{\yBminvelo}{-0.0656033150390625}
		\renewcommand{\yBmaxvelo}{0.17862257109375}
		\renewcommand{\LMminvelo}{-327.32400556500454}
		\renewcommand{\LMmaxvelo}{327.32400556500454}
		
		\renewcommand{\IBSResultCSVacc}{data/IBS_results_POM_manual_vinyl_a_Time.csv}
		\renewcommand{\yAminacc}{-670.4906805419922}
		\renewcommand{\yAmaxacc}{802.0985375976562}
		\renewcommand{\yBminacc}{-713.6285949707031}
		\renewcommand{\yBmaxacc}{861.9254022216796}
		\renewcommand{\LMminacc}{-268.380378112793}
		\renewcommand{\LMmaxacc}{268.380378112793}
		\begin{flushright}
			% This file was created with tikzplotlib v0.10.1.
\begin{tikzpicture}

\definecolor{darkgray176}{RGB}{176,176,176}
\definecolor{gray}{RGB}{128,128,128}
\definecolor{lightgray204}{RGB}{204,204,204}

\begin{groupplot}[group style={group size=3 by 3, horizontal sep=1.5cm, vertical sep=1.4cm}]

% Displacements
\nextgroupplot[
height=3.6cm,
scaled x ticks=manual:{}{\pgfmathparse{#1}},
yticklabel style={
        /pgf/number format/fixed,
        /pgf/number format/precision=3
},
tick align=outside,
tick pos=left,
title={Displacement \(\displaystyle d_1^{(0)}\) / m},
title style={yshift=0.5ex},
width=0.335\textwidth,
x grid style={darkgray176},
xmin=0, xmax=\IBSlen,
xtick style={color=black},
xticklabels={},
y grid style={darkgray176},
ymin=\yAmindisp, ymax=\yAmaxdisp,
ytick style={color=black}
]
\addplot [very thick, TUMBlue]
table[x=t, y=y1IBS, col sep=comma]{\IBSResultCSVdisp};

\addplot [very thick, TUMOrange]
table[x=t, y=y1Meas, col sep=comma]{\IBSResultCSVdisp};

\nextgroupplot[
height=3.6cm,
scaled x ticks=manual:{}{\pgfmathparse{#1}},
yticklabel style={
        /pgf/number format/fixed,
        /pgf/number format/precision=3
},
tick align=outside,
tick pos=left,
title={Displacement \(\displaystyle d_2^{(0)}\) / m},
title style={yshift=0.5ex},
width=0.335\textwidth,
x grid style={darkgray176},
xmin=0, xmax=\IBSlen,
xtick style={color=black},
xticklabels={},
y grid style={darkgray176},
ymin=\yBmindisp, ymax=\yBmaxdisp,
ytick style={color=black}
]
\addplot [very thick, TUMBlue]
table[x=t, y=y2IBS, col sep=comma]{\IBSResultCSVdisp};

\addplot [very thick, TUMOrange]
table[x=t, y=y2Meas, col sep=comma]{\IBSResultCSVdisp};

\nextgroupplot[
height=3.6cm,
tick align=outside,
tick pos=left,
title={Lagrange multiplier \(\displaystyle \lambda\) / N},
title style={yshift=0.5ex},
width=0.335\textwidth,
x grid style={darkgray176},
xmin=0, xmax=\IBSlen,
xtick style={color=black},
xticklabels={},
y grid style={darkgray176},
ymin=\LMmindisp, ymax=\LMmaxdisp,
ytick style={color=black}
]
\addplot [very thick, TUMBlue]
table[x=t, y=lambda, col sep=comma]{\IBSResultCSVdisp};

% Velocities
\nextgroupplot[
height=3.6cm,
scaled x ticks=manual:{}{\pgfmathparse{#1}},
yticklabel style={
        /pgf/number format/fixed,
        /pgf/number format/precision=3
},
tick align=outside,
tick pos=left,
title={Velocity \(\displaystyle v_1^{(0)}\) / (\si[per-mode=symbol]{\meter\per\second})},
title style={yshift=0.5ex},
width=0.335\textwidth,
x grid style={darkgray176},
xmin=0, xmax=\IBSlen,
xtick style={color=black},
xticklabels={},
y grid style={darkgray176},
ymin=\yAminvelo, ymax=\yAmaxvelo,
ytick style={color=black}
]
\addplot [very thick, TUMBlue]
table[x=t, y=y1IBS, col sep=comma]{\IBSResultCSVvelo};

\addplot [very thick, TUMOrange]
table[x=t, y=y1Meas, col sep=comma]{\IBSResultCSVvelo};

\nextgroupplot[
height=3.6cm,
scaled x ticks=manual:{}{\pgfmathparse{#1}},
yticklabel style={
        /pgf/number format/fixed,
        /pgf/number format/precision=3
},
tick align=outside,
tick pos=left,
title={Velocity \(\displaystyle v_2^{(0)}\) / (\si[per-mode=symbol]{\meter\per\second})},
title style={yshift=0.5ex},
width=0.335\textwidth,
x grid style={darkgray176},
xmin=0, xmax=\IBSlen,
xtick style={color=black},
xticklabels={},
y grid style={darkgray176},
ymin=\yBminvelo, ymax=\yBmaxvelo,
ytick style={color=black}
]
\addplot [very thick, TUMBlue]
table[x=t, y=y2IBS, col sep=comma]{\IBSResultCSVvelo};

\addplot [very thick, TUMOrange]
table[x=t, y=y2Meas, col sep=comma]{\IBSResultCSVvelo};

\nextgroupplot[
height=3.6cm,
tick align=outside,
tick pos=left,
title={Lagrange multiplier \(\displaystyle \lambda\) / N},
title style={yshift=0.5ex},
width=0.335\textwidth,
x grid style={darkgray176},
xmin=0, xmax=\IBSlen,
xtick style={color=black},
xticklabels={},
y grid style={darkgray176},
ymin=\LMminvelo, ymax=\LMmaxvelo,
ytick style={color=black}
]
\addplot [very thick, TUMBlue]
table[x=t, y=lambda, col sep=comma]{\IBSResultCSVvelo};

% Accelerations
\nextgroupplot[
height=3.6cm,
scaled x ticks=manual:{}{\pgfmathparse{#1}},
yticklabel style={
        /pgf/number format/fixed,
        /pgf/number format/precision=3
},
tick align=outside,
tick pos=left,
title={Acceleration \(\displaystyle a_1^{(0)}\) / (\si[per-mode=symbol]{\meter\per\second\squared})},
title style={yshift=0.5ex},
width=0.335\textwidth,
x grid style={darkgray176},
xmin=0, xmax=\IBSlen,
xtick style={color=black},
xlabel={Time / ms},
y grid style={darkgray176},
ymin=\yAminacc, ymax=\yAmaxacc,
ytick style={color=black}
]
\addplot [very thick, TUMBlue]
table[x=t, y=y1IBS, col sep=comma]{\IBSResultCSVacc};

\addplot [very thick, TUMOrange]
table[x=t, y=y1Meas, col sep=comma]{\IBSResultCSVacc};

\nextgroupplot[
height=3.6cm,
scaled x ticks=manual:{}{\pgfmathparse{#1}},
yticklabel style={
        /pgf/number format/fixed,
        /pgf/number format/precision=3
},
tick align=outside,
tick pos=left,
title={Acceleration \(\displaystyle a_2^{(0)}\) / (\si[per-mode=symbol]{\meter\per\second\squared})},
title style={yshift=0.5ex},
width=0.335\textwidth,
x grid style={darkgray176},
xmin=0, xmax=\IBSlen,
xtick style={color=black},
xlabel={Time / ms},
y grid style={darkgray176},
ymin=\yBminacc, ymax=\yBmaxacc,
ytick style={color=black}
]
\addplot [very thick, TUMBlue]
table[x=t, y=y2IBS, col sep=comma]{\IBSResultCSVacc};

\addplot [very thick, TUMOrange]
table[x=t, y=y2Meas, col sep=comma]{\IBSResultCSVacc};

\nextgroupplot[
height=3.6cm,
tick align=outside,
tick pos=left,
title={Lagrange multiplier \(\displaystyle \lambda\) / N},
title style={yshift=0.5ex},
width=0.335\textwidth,
x grid style={darkgray176},
xmin=0, xmax=\IBSlen,
xtick style={color=black},
xlabel={Time / ms},
y grid style={darkgray176},
ytick style={color=black},
ymin=\LMminacc, ymax=\LMmaxacc
]
\addplot [very thick, TUMBlue]
table[x=t, y=lambda, col sep=comma]{\IBSResultCSVacc};

\end{groupplot}
\end{tikzpicture}
		\end{flushright}\vspace{-0.4cm}
		\caption{IRFs processed using the \textbf{Time Domain} method}
		\label{fig:IBS_results_POM_manual_vinylB}
	\end{subfigure}	
	\caption{Experimental results of \textbf{POM} rods using manual hammer with \textbf{vinyl tip}, \legendary}
	\label{fig:IBS_results_POM_manual_vinyl}
\end{figure}

In the following, the results of experimentally applying the IBS method are presented, for IRFs calculated from the responses and excitation as measured, i.e.\ with the full original sampling frequency $f_\text{S}=\SI{102.4}{\kilo\hertz}$ and no additional low-pass filter applied. Further, the results are compared based on the domain used for the IRF calculation, where the IBS results using frequency domain IRF identification are shown on the top of the subsequent figures and the ones using the time domain on the bottom, see e.g.\ \cref{fig:IBS_results_POM_manual_vinyl}. Results are shown for IBS based on displacements, velocities and accelerations. The leftmost column shows the response of the assembled system where the IBS force (from reference measurement) is applied and the middle column shows the response at the other side. Depicted in blue is the response determined by IBS, and in orange the reference measurement. In the rightmost column, the Lagrange multiplier applied to enforce interface equilibrium and compatibility is shown.

Note that in order to be able to view the beginning of IBS results that become unstable over time, the limits of the plots are fixed to shortly above the reference measurement magnitudes. The real amplitudes can therefore greatly exceed the shown value ranges. The IBS scheme is evaluated for \SI{8}{\milli\second} for the POM rods and for \SI{2}{\milli\second} for the aluminum rods, showing respectively 6-7 response peaks of the assembled system.

\begin{figure}[ht!]
	\begin{subfigure}{\textwidth}
		\small
		\renewcommand{\IBSlen}{8}
		\renewcommand{\IBSResultCSVdisp}{data/IBS_results_POM_manual_steel_d_Frequency.csv}
		\renewcommand{\yAmindisp}{-4.653330842847936e-05}
		\renewcommand{\yAmaxdisp}{0.0002786072576564038}
		\renewcommand{\yBmindisp}{-5.0873942119324055e-05}
		\renewcommand{\yBmaxdisp}{0.00030443745013690205}
		\renewcommand{\LMmindisp}{-374.7971539306641}
		\renewcommand{\LMmaxdisp}{374.7971539306641}
		
		\renewcommand{\IBSResultCSVvelo}{data/IBS_results_POM_manual_steel_v_Frequency.csv}
		\renewcommand{\yAminvelo}{-0.07895606840029358}
		\renewcommand{\yAmaxvelo}{0.23359584733843802}
		\renewcommand{\yBminvelo}{-0.08551782507449389}
		\renewcommand{\yBmaxvelo}{0.23970942914485932}
		\renewcommand{\LMminvelo}{-374.7971539306641}
		\renewcommand{\LMmaxvelo}{374.7971539306641}
		
		\renewcommand{\IBSResultCSVacc}{data/IBS_results_POM_manual_steel_a_Frequency.csv}
		\renewcommand{\yAminacc}{-1629.6189953613282}
		\renewcommand{\yAmaxacc}{2013.026290283203}
		\renewcommand{\yBminacc}{-1738.612353515625}
		\renewcommand{\yBmaxacc}{2122.8030981445313}
		\renewcommand{\LMminacc}{-403.0341815185547}
		\renewcommand{\LMmaxacc}{403.0341815185547}
		\begin{flushright}
			% This file was created with tikzplotlib v0.10.1.
\begin{tikzpicture}

\definecolor{darkgray176}{RGB}{176,176,176}
\definecolor{gray}{RGB}{128,128,128}
\definecolor{lightgray204}{RGB}{204,204,204}

\begin{groupplot}[group style={group size=3 by 3, horizontal sep=1.5cm, vertical sep=1.4cm}]

% Displacements
\nextgroupplot[
height=3.6cm,
scaled x ticks=manual:{}{\pgfmathparse{#1}},
yticklabel style={
        /pgf/number format/fixed,
        /pgf/number format/precision=3
},
tick align=outside,
tick pos=left,
title={Displacement \(\displaystyle d_1^{(0)}\) / m},
title style={yshift=0.5ex},
width=0.335\textwidth,
x grid style={darkgray176},
xmin=0, xmax=\IBSlen,
xtick style={color=black},
xticklabels={},
y grid style={darkgray176},
ymin=\yAmindisp, ymax=\yAmaxdisp,
ytick style={color=black}
]
\addplot [very thick, TUMBlue]
table[x=t, y=y1IBS, col sep=comma]{\IBSResultCSVdisp};

\addplot [very thick, TUMOrange]
table[x=t, y=y1Meas, col sep=comma]{\IBSResultCSVdisp};

\nextgroupplot[
height=3.6cm,
scaled x ticks=manual:{}{\pgfmathparse{#1}},
yticklabel style={
        /pgf/number format/fixed,
        /pgf/number format/precision=3
},
tick align=outside,
tick pos=left,
title={Displacement \(\displaystyle d_2^{(0)}\) / m},
title style={yshift=0.5ex},
width=0.335\textwidth,
x grid style={darkgray176},
xmin=0, xmax=\IBSlen,
xtick style={color=black},
xticklabels={},
y grid style={darkgray176},
ymin=\yBmindisp, ymax=\yBmaxdisp,
ytick style={color=black}
]
\addplot [very thick, TUMBlue]
table[x=t, y=y2IBS, col sep=comma]{\IBSResultCSVdisp};

\addplot [very thick, TUMOrange]
table[x=t, y=y2Meas, col sep=comma]{\IBSResultCSVdisp};

\nextgroupplot[
height=3.6cm,
tick align=outside,
tick pos=left,
title={Lagrange multiplier \(\displaystyle \lambda\) / N},
title style={yshift=0.5ex},
width=0.335\textwidth,
x grid style={darkgray176},
xmin=0, xmax=\IBSlen,
xtick style={color=black},
xticklabels={},
y grid style={darkgray176},
ymin=\LMmindisp, ymax=\LMmaxdisp,
ytick style={color=black}
]
\addplot [very thick, TUMBlue]
table[x=t, y=lambda, col sep=comma]{\IBSResultCSVdisp};

% Velocities
\nextgroupplot[
height=3.6cm,
scaled x ticks=manual:{}{\pgfmathparse{#1}},
yticklabel style={
        /pgf/number format/fixed,
        /pgf/number format/precision=3
},
tick align=outside,
tick pos=left,
title={Velocity \(\displaystyle v_1^{(0)}\) / (\si[per-mode=symbol]{\meter\per\second})},
title style={yshift=0.5ex},
width=0.335\textwidth,
x grid style={darkgray176},
xmin=0, xmax=\IBSlen,
xtick style={color=black},
xticklabels={},
y grid style={darkgray176},
ymin=\yAminvelo, ymax=\yAmaxvelo,
ytick style={color=black}
]
\addplot [very thick, TUMBlue]
table[x=t, y=y1IBS, col sep=comma]{\IBSResultCSVvelo};

\addplot [very thick, TUMOrange]
table[x=t, y=y1Meas, col sep=comma]{\IBSResultCSVvelo};

\nextgroupplot[
height=3.6cm,
scaled x ticks=manual:{}{\pgfmathparse{#1}},
yticklabel style={
        /pgf/number format/fixed,
        /pgf/number format/precision=3
},
tick align=outside,
tick pos=left,
title={Velocity \(\displaystyle v_2^{(0)}\) / (\si[per-mode=symbol]{\meter\per\second})},
title style={yshift=0.5ex},
width=0.335\textwidth,
x grid style={darkgray176},
xmin=0, xmax=\IBSlen,
xtick style={color=black},
xticklabels={},
y grid style={darkgray176},
ymin=\yBminvelo, ymax=\yBmaxvelo,
ytick style={color=black}
]
\addplot [very thick, TUMBlue]
table[x=t, y=y2IBS, col sep=comma]{\IBSResultCSVvelo};

\addplot [very thick, TUMOrange]
table[x=t, y=y2Meas, col sep=comma]{\IBSResultCSVvelo};

\nextgroupplot[
height=3.6cm,
tick align=outside,
tick pos=left,
title={Lagrange multiplier \(\displaystyle \lambda\) / N},
title style={yshift=0.5ex},
width=0.335\textwidth,
x grid style={darkgray176},
xmin=0, xmax=\IBSlen,
xtick style={color=black},
xticklabels={},
y grid style={darkgray176},
ymin=\LMminvelo, ymax=\LMmaxvelo,
ytick style={color=black}
]
\addplot [very thick, TUMBlue]
table[x=t, y=lambda, col sep=comma]{\IBSResultCSVvelo};

% Accelerations
\nextgroupplot[
height=3.6cm,
scaled x ticks=manual:{}{\pgfmathparse{#1}},
yticklabel style={
        /pgf/number format/fixed,
        /pgf/number format/precision=3
},
tick align=outside,
tick pos=left,
title={Acceleration \(\displaystyle a_1^{(0)}\) / (\si[per-mode=symbol]{\meter\per\second\squared})},
title style={yshift=0.5ex},
width=0.335\textwidth,
x grid style={darkgray176},
xmin=0, xmax=\IBSlen,
xtick style={color=black},
xlabel={Time / ms},
y grid style={darkgray176},
ymin=\yAminacc, ymax=\yAmaxacc,
ytick style={color=black}
]
\addplot [very thick, TUMBlue]
table[x=t, y=y1IBS, col sep=comma]{\IBSResultCSVacc};

\addplot [very thick, TUMOrange]
table[x=t, y=y1Meas, col sep=comma]{\IBSResultCSVacc};

\nextgroupplot[
height=3.6cm,
scaled x ticks=manual:{}{\pgfmathparse{#1}},
yticklabel style={
        /pgf/number format/fixed,
        /pgf/number format/precision=3
},
tick align=outside,
tick pos=left,
title={Acceleration \(\displaystyle a_2^{(0)}\) / (\si[per-mode=symbol]{\meter\per\second\squared})},
title style={yshift=0.5ex},
width=0.335\textwidth,
x grid style={darkgray176},
xmin=0, xmax=\IBSlen,
xtick style={color=black},
xlabel={Time / ms},
y grid style={darkgray176},
ymin=\yBminacc, ymax=\yBmaxacc,
ytick style={color=black}
]
\addplot [very thick, TUMBlue]
table[x=t, y=y2IBS, col sep=comma]{\IBSResultCSVacc};

\addplot [very thick, TUMOrange]
table[x=t, y=y2Meas, col sep=comma]{\IBSResultCSVacc};

\nextgroupplot[
height=3.6cm,
tick align=outside,
tick pos=left,
title={Lagrange multiplier \(\displaystyle \lambda\) / N},
title style={yshift=0.5ex},
width=0.335\textwidth,
x grid style={darkgray176},
xmin=0, xmax=\IBSlen,
xtick style={color=black},
xlabel={Time / ms},
y grid style={darkgray176},
ytick style={color=black},
ymin=\LMminacc, ymax=\LMmaxacc
]
\addplot [very thick, TUMBlue]
table[x=t, y=lambda, col sep=comma]{\IBSResultCSVacc};

\end{groupplot}
\end{tikzpicture}
		\end{flushright}\vspace{-0.4cm}
		\caption{IRFs processed using the \textbf{Frequency Domain} method}\vspace{0.4cm}
		\label{fig:IBS_results_POM_manual_steelA}
	\end{subfigure}
	\begin{subfigure}{\textwidth}
		\small
		\renewcommand{\IBSlen}{8}
		\renewcommand{\IBSResultCSVdisp}{data/IBS_results_POM_manual_steel_d_Time.csv}
		\renewcommand{\yAmindisp}{-4.653330842847936e-05}
		\renewcommand{\yAmaxdisp}{0.0002786072576564038}
		\renewcommand{\yBmindisp}{-5.0873942119324055e-05}
		\renewcommand{\yBmaxdisp}{0.00030443745013690205}
		\renewcommand{\LMmindisp}{-374.7971539306641}
		\renewcommand{\LMmaxdisp}{374.7971539306641}
		
		\renewcommand{\IBSResultCSVvelo}{data/IBS_results_POM_manual_steel_v_Time.csv}
		\renewcommand{\yAminvelo}{-0.07895606840029358}
		\renewcommand{\yAmaxvelo}{0.23359584733843802}
		\renewcommand{\yBminvelo}{-0.08551782507449389}
		\renewcommand{\yBmaxvelo}{0.23970942914485932}
		\renewcommand{\LMminvelo}{-374.7971539306641}
		\renewcommand{\LMmaxvelo}{374.7971539306641}
		
		\renewcommand{\IBSResultCSVacc}{data/IBS_results_POM_manual_steel_a_Time.csv}
		\renewcommand{\yAminacc}{-1629.6189953613282}
		\renewcommand{\yAmaxacc}{2013.026290283203}
		\renewcommand{\yBminacc}{-1738.612353515625}
		\renewcommand{\yBmaxacc}{2122.8030981445313}
		\renewcommand{\LMminacc}{-403.0341815185547}
		\renewcommand{\LMmaxacc}{403.0341815185547}
		\begin{flushright}
			% This file was created with tikzplotlib v0.10.1.
\begin{tikzpicture}

\definecolor{darkgray176}{RGB}{176,176,176}
\definecolor{gray}{RGB}{128,128,128}
\definecolor{lightgray204}{RGB}{204,204,204}

\begin{groupplot}[group style={group size=3 by 3, horizontal sep=1.5cm, vertical sep=1.4cm}]

% Displacements
\nextgroupplot[
height=3.6cm,
scaled x ticks=manual:{}{\pgfmathparse{#1}},
yticklabel style={
        /pgf/number format/fixed,
        /pgf/number format/precision=3
},
tick align=outside,
tick pos=left,
title={Displacement \(\displaystyle d_1^{(0)}\) / m},
title style={yshift=0.5ex},
width=0.335\textwidth,
x grid style={darkgray176},
xmin=0, xmax=\IBSlen,
xtick style={color=black},
xticklabels={},
y grid style={darkgray176},
ymin=\yAmindisp, ymax=\yAmaxdisp,
ytick style={color=black}
]
\addplot [very thick, TUMBlue]
table[x=t, y=y1IBS, col sep=comma]{\IBSResultCSVdisp};

\addplot [very thick, TUMOrange]
table[x=t, y=y1Meas, col sep=comma]{\IBSResultCSVdisp};

\nextgroupplot[
height=3.6cm,
scaled x ticks=manual:{}{\pgfmathparse{#1}},
yticklabel style={
        /pgf/number format/fixed,
        /pgf/number format/precision=3
},
tick align=outside,
tick pos=left,
title={Displacement \(\displaystyle d_2^{(0)}\) / m},
title style={yshift=0.5ex},
width=0.335\textwidth,
x grid style={darkgray176},
xmin=0, xmax=\IBSlen,
xtick style={color=black},
xticklabels={},
y grid style={darkgray176},
ymin=\yBmindisp, ymax=\yBmaxdisp,
ytick style={color=black}
]
\addplot [very thick, TUMBlue]
table[x=t, y=y2IBS, col sep=comma]{\IBSResultCSVdisp};

\addplot [very thick, TUMOrange]
table[x=t, y=y2Meas, col sep=comma]{\IBSResultCSVdisp};

\nextgroupplot[
height=3.6cm,
tick align=outside,
tick pos=left,
title={Lagrange multiplier \(\displaystyle \lambda\) / N},
title style={yshift=0.5ex},
width=0.335\textwidth,
x grid style={darkgray176},
xmin=0, xmax=\IBSlen,
xtick style={color=black},
xticklabels={},
y grid style={darkgray176},
ymin=\LMmindisp, ymax=\LMmaxdisp,
ytick style={color=black}
]
\addplot [very thick, TUMBlue]
table[x=t, y=lambda, col sep=comma]{\IBSResultCSVdisp};

% Velocities
\nextgroupplot[
height=3.6cm,
scaled x ticks=manual:{}{\pgfmathparse{#1}},
yticklabel style={
        /pgf/number format/fixed,
        /pgf/number format/precision=3
},
tick align=outside,
tick pos=left,
title={Velocity \(\displaystyle v_1^{(0)}\) / (\si[per-mode=symbol]{\meter\per\second})},
title style={yshift=0.5ex},
width=0.335\textwidth,
x grid style={darkgray176},
xmin=0, xmax=\IBSlen,
xtick style={color=black},
xticklabels={},
y grid style={darkgray176},
ymin=\yAminvelo, ymax=\yAmaxvelo,
ytick style={color=black}
]
\addplot [very thick, TUMBlue]
table[x=t, y=y1IBS, col sep=comma]{\IBSResultCSVvelo};

\addplot [very thick, TUMOrange]
table[x=t, y=y1Meas, col sep=comma]{\IBSResultCSVvelo};

\nextgroupplot[
height=3.6cm,
scaled x ticks=manual:{}{\pgfmathparse{#1}},
yticklabel style={
        /pgf/number format/fixed,
        /pgf/number format/precision=3
},
tick align=outside,
tick pos=left,
title={Velocity \(\displaystyle v_2^{(0)}\) / (\si[per-mode=symbol]{\meter\per\second})},
title style={yshift=0.5ex},
width=0.335\textwidth,
x grid style={darkgray176},
xmin=0, xmax=\IBSlen,
xtick style={color=black},
xticklabels={},
y grid style={darkgray176},
ymin=\yBminvelo, ymax=\yBmaxvelo,
ytick style={color=black}
]
\addplot [very thick, TUMBlue]
table[x=t, y=y2IBS, col sep=comma]{\IBSResultCSVvelo};

\addplot [very thick, TUMOrange]
table[x=t, y=y2Meas, col sep=comma]{\IBSResultCSVvelo};

\nextgroupplot[
height=3.6cm,
tick align=outside,
tick pos=left,
title={Lagrange multiplier \(\displaystyle \lambda\) / N},
title style={yshift=0.5ex},
width=0.335\textwidth,
x grid style={darkgray176},
xmin=0, xmax=\IBSlen,
xtick style={color=black},
xticklabels={},
y grid style={darkgray176},
ymin=\LMminvelo, ymax=\LMmaxvelo,
ytick style={color=black}
]
\addplot [very thick, TUMBlue]
table[x=t, y=lambda, col sep=comma]{\IBSResultCSVvelo};

% Accelerations
\nextgroupplot[
height=3.6cm,
scaled x ticks=manual:{}{\pgfmathparse{#1}},
yticklabel style={
        /pgf/number format/fixed,
        /pgf/number format/precision=3
},
tick align=outside,
tick pos=left,
title={Acceleration \(\displaystyle a_1^{(0)}\) / (\si[per-mode=symbol]{\meter\per\second\squared})},
title style={yshift=0.5ex},
width=0.335\textwidth,
x grid style={darkgray176},
xmin=0, xmax=\IBSlen,
xtick style={color=black},
xlabel={Time / ms},
y grid style={darkgray176},
ymin=\yAminacc, ymax=\yAmaxacc,
ytick style={color=black}
]
\addplot [very thick, TUMBlue]
table[x=t, y=y1IBS, col sep=comma]{\IBSResultCSVacc};

\addplot [very thick, TUMOrange]
table[x=t, y=y1Meas, col sep=comma]{\IBSResultCSVacc};

\nextgroupplot[
height=3.6cm,
scaled x ticks=manual:{}{\pgfmathparse{#1}},
yticklabel style={
        /pgf/number format/fixed,
        /pgf/number format/precision=3
},
tick align=outside,
tick pos=left,
title={Acceleration \(\displaystyle a_2^{(0)}\) / (\si[per-mode=symbol]{\meter\per\second\squared})},
title style={yshift=0.5ex},
width=0.335\textwidth,
x grid style={darkgray176},
xmin=0, xmax=\IBSlen,
xtick style={color=black},
xlabel={Time / ms},
y grid style={darkgray176},
ymin=\yBminacc, ymax=\yBmaxacc,
ytick style={color=black}
]
\addplot [very thick, TUMBlue]
table[x=t, y=y2IBS, col sep=comma]{\IBSResultCSVacc};

\addplot [very thick, TUMOrange]
table[x=t, y=y2Meas, col sep=comma]{\IBSResultCSVacc};

\nextgroupplot[
height=3.6cm,
tick align=outside,
tick pos=left,
title={Lagrange multiplier \(\displaystyle \lambda\) / N},
title style={yshift=0.5ex},
width=0.335\textwidth,
x grid style={darkgray176},
xmin=0, xmax=\IBSlen,
xtick style={color=black},
xlabel={Time / ms},
y grid style={darkgray176},
ytick style={color=black},
ymin=\LMminacc, ymax=\LMmaxacc
]
\addplot [very thick, TUMBlue]
table[x=t, y=lambda, col sep=comma]{\IBSResultCSVacc};

\end{groupplot}
\end{tikzpicture}
		\end{flushright}\vspace{-0.4cm}
		\caption{IRFs processed using the \textbf{Time Domain} method}
		\label{fig:IBS_results_POM_manual_steelB}
	\end{subfigure}	
	\caption{Experimental results of \textbf{POM} rods using manual hammer with \textbf{steel tip}, \legendary}
	\label{fig:IBS_results_POM_manual_steel}
\end{figure}
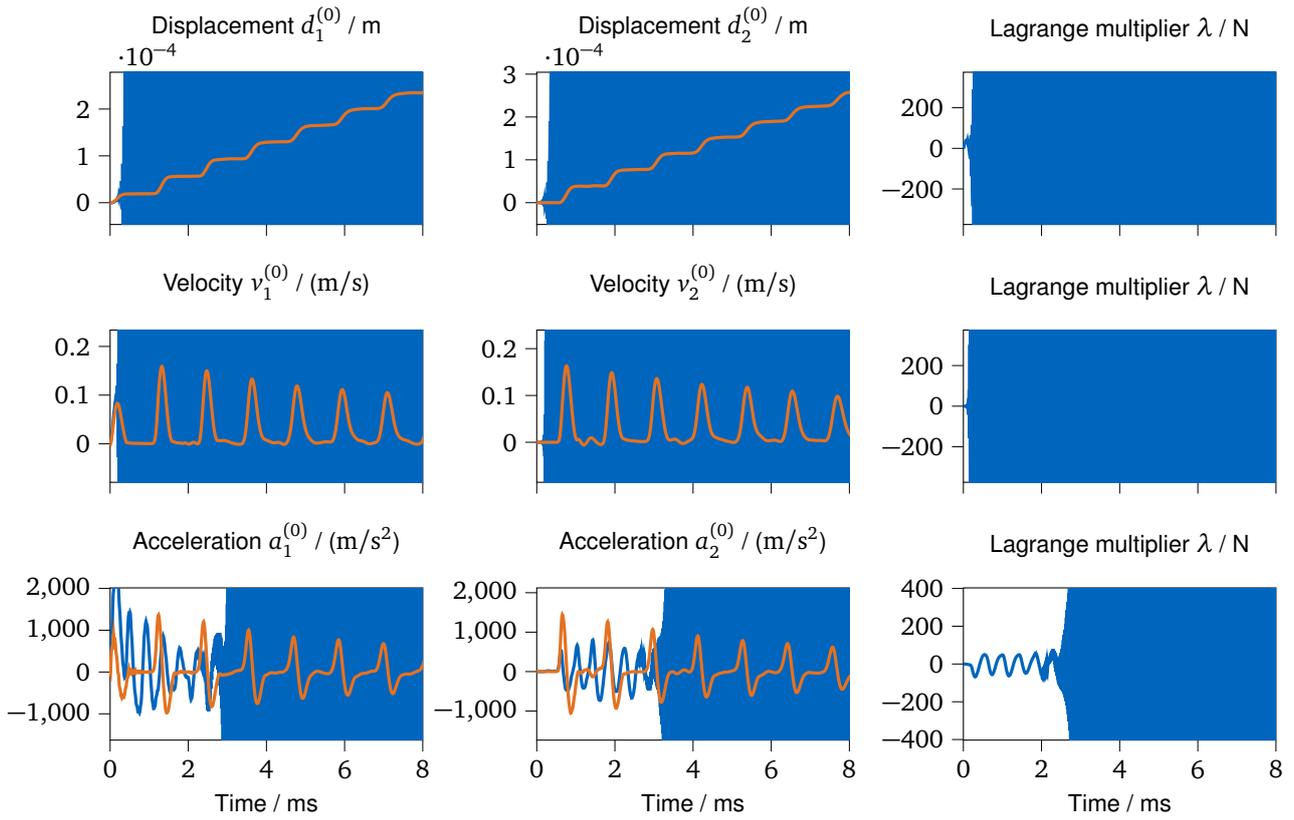
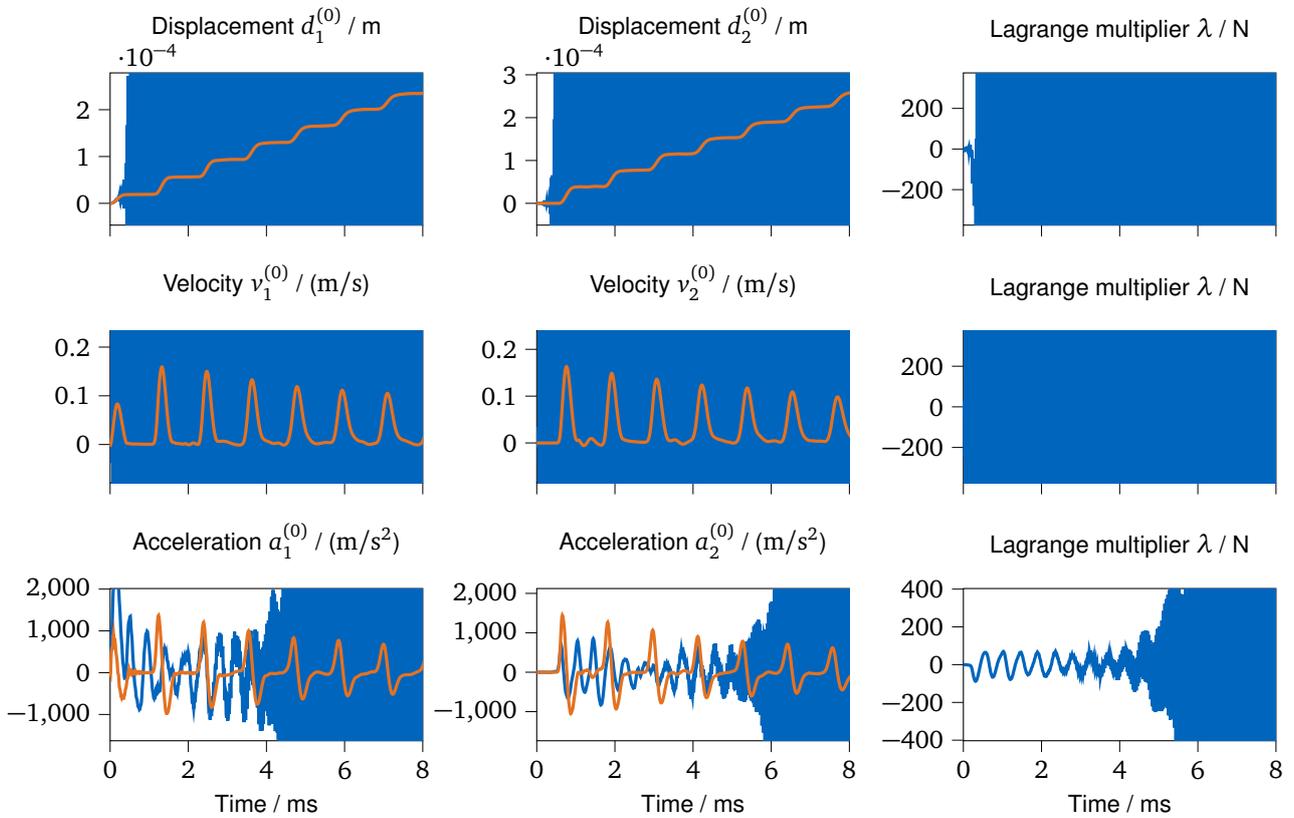

Starting with the results of the POM rods, shown in \cref{fig:IBS_results_POM_manual_vinyl} for the vinyl tip and in \cref{fig:IBS_results_POM_manual_steel} for the steel tip, it can be seen that the displacement-based IBS for both tips and processing domains immediately becomes unstable, i.e.\ the amplitudes of the responses and the Lagrange multiplier grow very large and oscillate.

The same can be seen for velocity responses, except for the result using a vinyl tip and time domain IRFs (\cref{fig:IBS_results_POM_manual_vinylB}). While the result for the velocity response of the assembled system $v_1^{(0)}$ might look reasonable at first, the first two peaks of the driving point response, that line up with the reference measurement, have the same amplitude. This is not plausible considering that the system is excited with an imperfect impulse, i.e.\ energy is added over time and for the first peak not all energy has been added yet. Besides that, the initial response amplitude is significantly wrong with respect to the reference measurement.

Lastly, for acceleration responses, the IBS results generally become unstable at a later point in time. For the measurements using a vinyl tip, the second response peak magnitude is reasonably well estimated with both IRF methods, but using the time domain for the IRF calculation better predicts the following response peaks and is more stable with respect to both, the amplitudes of the responses and the Lagrange multiplier. Nevertheless, the initial response peak is greatly amplified, making the result useless for initial shock amplitude estimation.

The results using a steel tip exhibit coupling issues that manifest in the response $a_1^{(0)}$ showing the dynamics of the unassembled substructure $S_1$ (three times the eigenfrequency), the response $a_2^{(0)}$ being significantly too low and the Lagrange multiplier showing smaller amplitudes than expected based on the applied force.

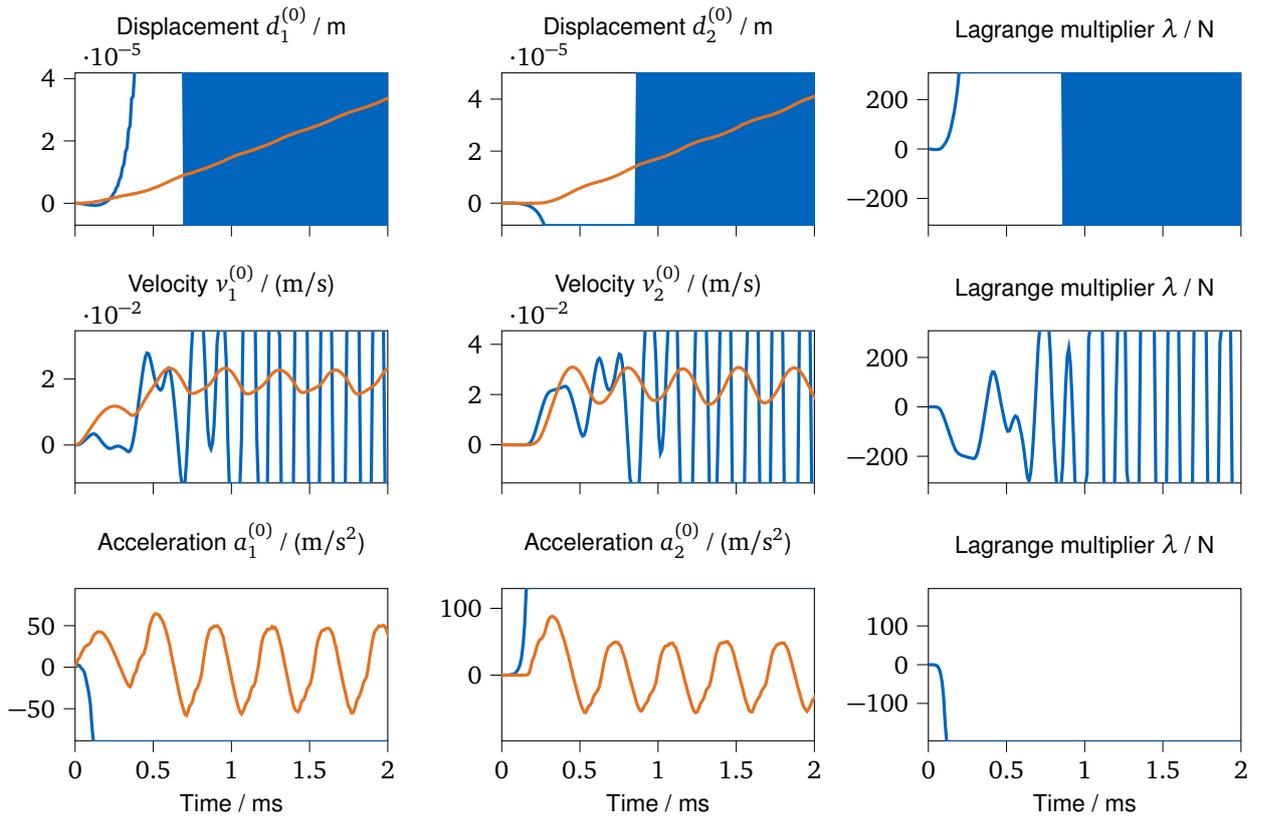
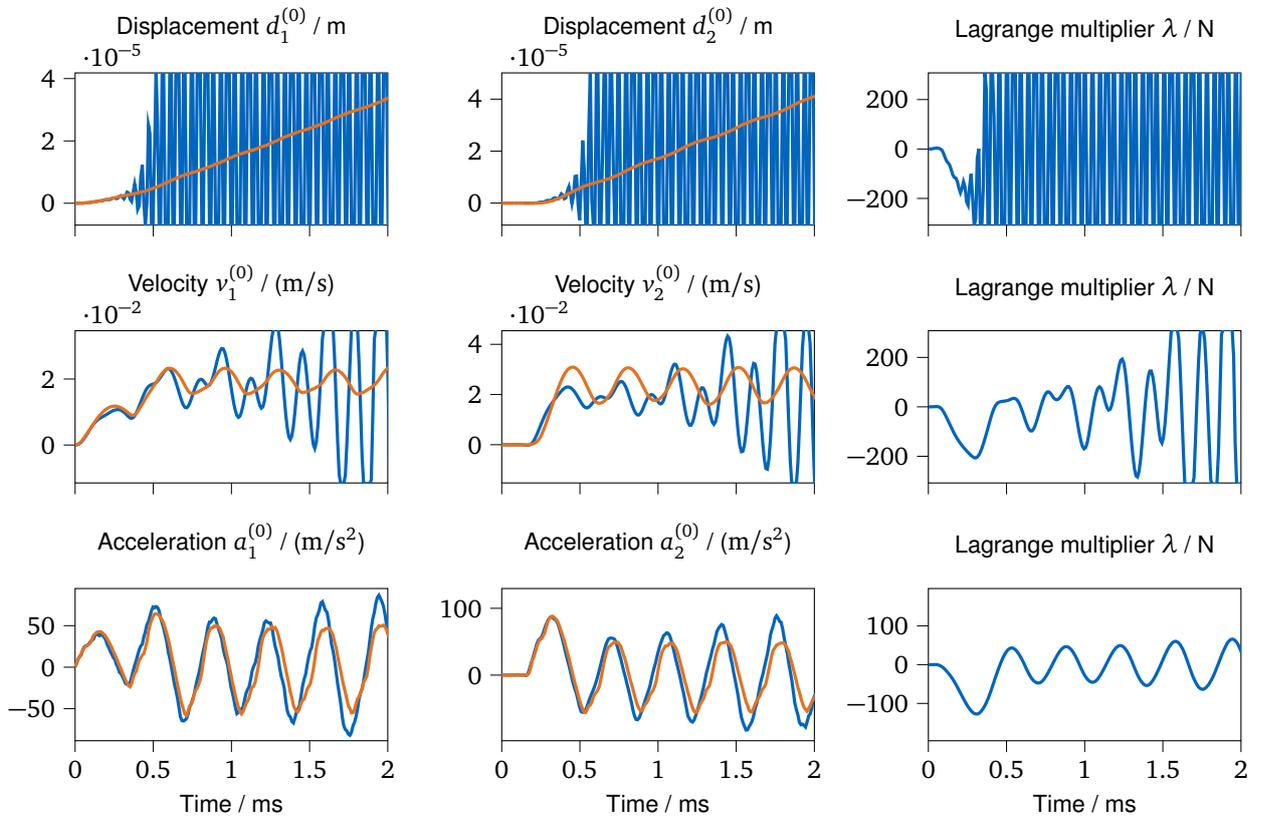
\begin{figure}[ht!]
	\begin{subfigure}{\textwidth}
		\small
		\renewcommand{\IBSlen}{2}
		\renewcommand{\IBSResultCSVdisp}{data/IBS_results_ENAW7075_manual_vinyl_d_Frequency.csv}
		\renewcommand{\yAmindisp}{-7.017584765625001e-06}
		\renewcommand{\yAmaxdisp}{4.182299296875e-05}
		\renewcommand{\yBmindisp}{-8.47856953125e-06}
		\renewcommand{\yBmaxdisp}{5.0089331250000006e-05}
		\renewcommand{\LMmindisp}{-307.9559059135766}
		\renewcommand{\LMmaxdisp}{307.9559059135766}
		
		\renewcommand{\IBSResultCSVvelo}{data/IBS_results_ENAW7075_manual_vinyl_v_Frequency.csv}
		\renewcommand{\yAminvelo}{-0.0114898013671875}
		\renewcommand{\yAmaxvelo}{0.034632754980468757}
		\renewcommand{\yBminvelo}{-0.01522691748046875}
		\renewcommand{\yBmaxvelo}{0.045424248925781244}
		\renewcommand{\LMminvelo}{-307.9559059135766}
		\renewcommand{\LMmaxvelo}{307.9559059135766}
		
		\renewcommand{\IBSResultCSVacc}{data/IBS_results_ENAW7075_manual_vinyl_a_Frequency.csv}
		\renewcommand{\yAminacc}{-88.58306671142579}
		\renewcommand{\yAmaxacc}{94.72701667785644}
		\renewcommand{\yBminacc}{-98.20231491088867}
		\renewcommand{\yBmaxacc}{129.6055009460449}
		\renewcommand{\LMminacc}{-196.43740814208985}
		\renewcommand{\LMmaxacc}{196.43740814208985}
		\begin{flushright}
			% This file was created with tikzplotlib v0.10.1.
\begin{tikzpicture}

\definecolor{darkgray176}{RGB}{176,176,176}
\definecolor{gray}{RGB}{128,128,128}
\definecolor{lightgray204}{RGB}{204,204,204}

\begin{groupplot}[group style={group size=3 by 3, horizontal sep=1.5cm, vertical sep=1.4cm}]

% Displacements
\nextgroupplot[
height=3.6cm,
scaled x ticks=manual:{}{\pgfmathparse{#1}},
yticklabel style={
        /pgf/number format/fixed,
        /pgf/number format/precision=3
},
tick align=outside,
tick pos=left,
title={Displacement \(\displaystyle d_1^{(0)}\) / m},
title style={yshift=0.5ex},
width=0.335\textwidth,
x grid style={darkgray176},
xmin=0, xmax=\IBSlen,
xtick style={color=black},
xticklabels={},
y grid style={darkgray176},
ymin=\yAmindisp, ymax=\yAmaxdisp,
ytick style={color=black}
]
\addplot [very thick, TUMBlue]
table[x=t, y=y1IBS, col sep=comma]{\IBSResultCSVdisp};

\addplot [very thick, TUMOrange]
table[x=t, y=y1Meas, col sep=comma]{\IBSResultCSVdisp};

\nextgroupplot[
height=3.6cm,
scaled x ticks=manual:{}{\pgfmathparse{#1}},
yticklabel style={
        /pgf/number format/fixed,
        /pgf/number format/precision=3
},
tick align=outside,
tick pos=left,
title={Displacement \(\displaystyle d_2^{(0)}\) / m},
title style={yshift=0.5ex},
width=0.335\textwidth,
x grid style={darkgray176},
xmin=0, xmax=\IBSlen,
xtick style={color=black},
xticklabels={},
y grid style={darkgray176},
ymin=\yBmindisp, ymax=\yBmaxdisp,
ytick style={color=black}
]
\addplot [very thick, TUMBlue]
table[x=t, y=y2IBS, col sep=comma]{\IBSResultCSVdisp};

\addplot [very thick, TUMOrange]
table[x=t, y=y2Meas, col sep=comma]{\IBSResultCSVdisp};

\nextgroupplot[
height=3.6cm,
tick align=outside,
tick pos=left,
title={Lagrange multiplier \(\displaystyle \lambda\) / N},
title style={yshift=0.5ex},
width=0.335\textwidth,
x grid style={darkgray176},
xmin=0, xmax=\IBSlen,
xtick style={color=black},
xticklabels={},
y grid style={darkgray176},
ymin=\LMmindisp, ymax=\LMmaxdisp,
ytick style={color=black}
]
\addplot [very thick, TUMBlue]
table[x=t, y=lambda, col sep=comma]{\IBSResultCSVdisp};

% Velocities
\nextgroupplot[
height=3.6cm,
scaled x ticks=manual:{}{\pgfmathparse{#1}},
yticklabel style={
        /pgf/number format/fixed,
        /pgf/number format/precision=3
},
tick align=outside,
tick pos=left,
title={Velocity \(\displaystyle v_1^{(0)}\) / (\si[per-mode=symbol]{\meter\per\second})},
title style={yshift=0.5ex},
width=0.335\textwidth,
x grid style={darkgray176},
xmin=0, xmax=\IBSlen,
xtick style={color=black},
xticklabels={},
y grid style={darkgray176},
ymin=\yAminvelo, ymax=\yAmaxvelo,
ytick style={color=black}
]
\addplot [very thick, TUMBlue]
table[x=t, y=y1IBS, col sep=comma]{\IBSResultCSVvelo};

\addplot [very thick, TUMOrange]
table[x=t, y=y1Meas, col sep=comma]{\IBSResultCSVvelo};

\nextgroupplot[
height=3.6cm,
scaled x ticks=manual:{}{\pgfmathparse{#1}},
yticklabel style={
        /pgf/number format/fixed,
        /pgf/number format/precision=3
},
tick align=outside,
tick pos=left,
title={Velocity \(\displaystyle v_2^{(0)}\) / (\si[per-mode=symbol]{\meter\per\second})},
title style={yshift=0.5ex},
width=0.335\textwidth,
x grid style={darkgray176},
xmin=0, xmax=\IBSlen,
xtick style={color=black},
xticklabels={},
y grid style={darkgray176},
ymin=\yBminvelo, ymax=\yBmaxvelo,
ytick style={color=black}
]
\addplot [very thick, TUMBlue]
table[x=t, y=y2IBS, col sep=comma]{\IBSResultCSVvelo};

\addplot [very thick, TUMOrange]
table[x=t, y=y2Meas, col sep=comma]{\IBSResultCSVvelo};

\nextgroupplot[
height=3.6cm,
tick align=outside,
tick pos=left,
title={Lagrange multiplier \(\displaystyle \lambda\) / N},
title style={yshift=0.5ex},
width=0.335\textwidth,
x grid style={darkgray176},
xmin=0, xmax=\IBSlen,
xtick style={color=black},
xticklabels={},
y grid style={darkgray176},
ymin=\LMminvelo, ymax=\LMmaxvelo,
ytick style={color=black}
]
\addplot [very thick, TUMBlue]
table[x=t, y=lambda, col sep=comma]{\IBSResultCSVvelo};

% Accelerations
\nextgroupplot[
height=3.6cm,
scaled x ticks=manual:{}{\pgfmathparse{#1}},
yticklabel style={
        /pgf/number format/fixed,
        /pgf/number format/precision=3
},
tick align=outside,
tick pos=left,
title={Acceleration \(\displaystyle a_1^{(0)}\) / (\si[per-mode=symbol]{\meter\per\second\squared})},
title style={yshift=0.5ex},
width=0.335\textwidth,
x grid style={darkgray176},
xmin=0, xmax=\IBSlen,
xtick style={color=black},
xlabel={Time / ms},
y grid style={darkgray176},
ymin=\yAminacc, ymax=\yAmaxacc,
ytick style={color=black}
]
\addplot [very thick, TUMBlue]
table[x=t, y=y1IBS, col sep=comma]{\IBSResultCSVacc};

\addplot [very thick, TUMOrange]
table[x=t, y=y1Meas, col sep=comma]{\IBSResultCSVacc};

\nextgroupplot[
height=3.6cm,
scaled x ticks=manual:{}{\pgfmathparse{#1}},
yticklabel style={
        /pgf/number format/fixed,
        /pgf/number format/precision=3
},
tick align=outside,
tick pos=left,
title={Acceleration \(\displaystyle a_2^{(0)}\) / (\si[per-mode=symbol]{\meter\per\second\squared})},
title style={yshift=0.5ex},
width=0.335\textwidth,
x grid style={darkgray176},
xmin=0, xmax=\IBSlen,
xtick style={color=black},
xlabel={Time / ms},
y grid style={darkgray176},
ymin=\yBminacc, ymax=\yBmaxacc,
ytick style={color=black}
]
\addplot [very thick, TUMBlue]
table[x=t, y=y2IBS, col sep=comma]{\IBSResultCSVacc};

\addplot [very thick, TUMOrange]
table[x=t, y=y2Meas, col sep=comma]{\IBSResultCSVacc};

\nextgroupplot[
height=3.6cm,
tick align=outside,
tick pos=left,
title={Lagrange multiplier \(\displaystyle \lambda\) / N},
title style={yshift=0.5ex},
width=0.335\textwidth,
x grid style={darkgray176},
xmin=0, xmax=\IBSlen,
xtick style={color=black},
xlabel={Time / ms},
y grid style={darkgray176},
ytick style={color=black},
ymin=\LMminacc, ymax=\LMmaxacc
]
\addplot [very thick, TUMBlue]
table[x=t, y=lambda, col sep=comma]{\IBSResultCSVacc};

\end{groupplot}
\end{tikzpicture}
		\end{flushright}\vspace{-0.4cm}
		\caption{IRFs processed using the \textbf{Frequency Domain} method}\vspace{0.4cm}
		\label{fig:IBS_results_ENAW7075_manual_vinylA}
	\end{subfigure}
	\begin{subfigure}{\textwidth}
		\small
		\renewcommand{\IBSlen}{2}
		\renewcommand{\IBSResultCSVdisp}{data/IBS_results_ENAW7075_manual_vinyl_d_Time.csv}
		\renewcommand{\yAmindisp}{-7.017584765625001e-06}
		\renewcommand{\yAmaxdisp}{4.182299296875e-05}
		\renewcommand{\yBmindisp}{-8.47856953125e-06}
		\renewcommand{\yBmaxdisp}{5.0089331250000006e-05}
		\renewcommand{\LMmindisp}{-307.9559059135766}
		\renewcommand{\LMmaxdisp}{307.9559059135766}
		
		\renewcommand{\IBSResultCSVvelo}{data/IBS_results_ENAW7075_manual_vinyl_v_Time.csv}
		\renewcommand{\yAminvelo}{-0.0114898013671875}
		\renewcommand{\yAmaxvelo}{0.034632754980468757}
		\renewcommand{\yBminvelo}{-0.01522691748046875}
		\renewcommand{\yBmaxvelo}{0.045424248925781244}
		\renewcommand{\LMminvelo}{-307.9559059135766}
		\renewcommand{\LMmaxvelo}{307.9559059135766}
		
		\renewcommand{\IBSResultCSVacc}{data/IBS_results_ENAW7075_manual_vinyl_a_Time.csv}
		\renewcommand{\yAminacc}{-88.58306671142579}
		\renewcommand{\yAmaxacc}{94.72701667785644}
		\renewcommand{\yBminacc}{-98.20231491088867}
		\renewcommand{\yBmaxacc}{129.6055009460449}
		\renewcommand{\LMminacc}{-196.43740814208985}
		\renewcommand{\LMmaxacc}{196.43740814208985}
		\begin{flushright}
			% This file was created with tikzplotlib v0.10.1.
\begin{tikzpicture}

\definecolor{darkgray176}{RGB}{176,176,176}
\definecolor{gray}{RGB}{128,128,128}
\definecolor{lightgray204}{RGB}{204,204,204}

\begin{groupplot}[group style={group size=3 by 3, horizontal sep=1.5cm, vertical sep=1.4cm}]

% Displacements
\nextgroupplot[
height=3.6cm,
scaled x ticks=manual:{}{\pgfmathparse{#1}},
yticklabel style={
        /pgf/number format/fixed,
        /pgf/number format/precision=3
},
tick align=outside,
tick pos=left,
title={Displacement \(\displaystyle d_1^{(0)}\) / m},
title style={yshift=0.5ex},
width=0.335\textwidth,
x grid style={darkgray176},
xmin=0, xmax=\IBSlen,
xtick style={color=black},
xticklabels={},
y grid style={darkgray176},
ymin=\yAmindisp, ymax=\yAmaxdisp,
ytick style={color=black}
]
\addplot [very thick, TUMBlue]
table[x=t, y=y1IBS, col sep=comma]{\IBSResultCSVdisp};

\addplot [very thick, TUMOrange]
table[x=t, y=y1Meas, col sep=comma]{\IBSResultCSVdisp};

\nextgroupplot[
height=3.6cm,
scaled x ticks=manual:{}{\pgfmathparse{#1}},
yticklabel style={
        /pgf/number format/fixed,
        /pgf/number format/precision=3
},
tick align=outside,
tick pos=left,
title={Displacement \(\displaystyle d_2^{(0)}\) / m},
title style={yshift=0.5ex},
width=0.335\textwidth,
x grid style={darkgray176},
xmin=0, xmax=\IBSlen,
xtick style={color=black},
xticklabels={},
y grid style={darkgray176},
ymin=\yBmindisp, ymax=\yBmaxdisp,
ytick style={color=black}
]
\addplot [very thick, TUMBlue]
table[x=t, y=y2IBS, col sep=comma]{\IBSResultCSVdisp};

\addplot [very thick, TUMOrange]
table[x=t, y=y2Meas, col sep=comma]{\IBSResultCSVdisp};

\nextgroupplot[
height=3.6cm,
tick align=outside,
tick pos=left,
title={Lagrange multiplier \(\displaystyle \lambda\) / N},
title style={yshift=0.5ex},
width=0.335\textwidth,
x grid style={darkgray176},
xmin=0, xmax=\IBSlen,
xtick style={color=black},
xticklabels={},
y grid style={darkgray176},
ymin=\LMmindisp, ymax=\LMmaxdisp,
ytick style={color=black}
]
\addplot [very thick, TUMBlue]
table[x=t, y=lambda, col sep=comma]{\IBSResultCSVdisp};

% Velocities
\nextgroupplot[
height=3.6cm,
scaled x ticks=manual:{}{\pgfmathparse{#1}},
yticklabel style={
        /pgf/number format/fixed,
        /pgf/number format/precision=3
},
tick align=outside,
tick pos=left,
title={Velocity \(\displaystyle v_1^{(0)}\) / (\si[per-mode=symbol]{\meter\per\second})},
title style={yshift=0.5ex},
width=0.335\textwidth,
x grid style={darkgray176},
xmin=0, xmax=\IBSlen,
xtick style={color=black},
xticklabels={},
y grid style={darkgray176},
ymin=\yAminvelo, ymax=\yAmaxvelo,
ytick style={color=black}
]
\addplot [very thick, TUMBlue]
table[x=t, y=y1IBS, col sep=comma]{\IBSResultCSVvelo};

\addplot [very thick, TUMOrange]
table[x=t, y=y1Meas, col sep=comma]{\IBSResultCSVvelo};

\nextgroupplot[
height=3.6cm,
scaled x ticks=manual:{}{\pgfmathparse{#1}},
yticklabel style={
        /pgf/number format/fixed,
        /pgf/number format/precision=3
},
tick align=outside,
tick pos=left,
title={Velocity \(\displaystyle v_2^{(0)}\) / (\si[per-mode=symbol]{\meter\per\second})},
title style={yshift=0.5ex},
width=0.335\textwidth,
x grid style={darkgray176},
xmin=0, xmax=\IBSlen,
xtick style={color=black},
xticklabels={},
y grid style={darkgray176},
ymin=\yBminvelo, ymax=\yBmaxvelo,
ytick style={color=black}
]
\addplot [very thick, TUMBlue]
table[x=t, y=y2IBS, col sep=comma]{\IBSResultCSVvelo};

\addplot [very thick, TUMOrange]
table[x=t, y=y2Meas, col sep=comma]{\IBSResultCSVvelo};

\nextgroupplot[
height=3.6cm,
tick align=outside,
tick pos=left,
title={Lagrange multiplier \(\displaystyle \lambda\) / N},
title style={yshift=0.5ex},
width=0.335\textwidth,
x grid style={darkgray176},
xmin=0, xmax=\IBSlen,
xtick style={color=black},
xticklabels={},
y grid style={darkgray176},
ymin=\LMminvelo, ymax=\LMmaxvelo,
ytick style={color=black}
]
\addplot [very thick, TUMBlue]
table[x=t, y=lambda, col sep=comma]{\IBSResultCSVvelo};

% Accelerations
\nextgroupplot[
height=3.6cm,
scaled x ticks=manual:{}{\pgfmathparse{#1}},
yticklabel style={
        /pgf/number format/fixed,
        /pgf/number format/precision=3
},
tick align=outside,
tick pos=left,
title={Acceleration \(\displaystyle a_1^{(0)}\) / (\si[per-mode=symbol]{\meter\per\second\squared})},
title style={yshift=0.5ex},
width=0.335\textwidth,
x grid style={darkgray176},
xmin=0, xmax=\IBSlen,
xtick style={color=black},
xlabel={Time / ms},
y grid style={darkgray176},
ymin=\yAminacc, ymax=\yAmaxacc,
ytick style={color=black}
]
\addplot [very thick, TUMBlue]
table[x=t, y=y1IBS, col sep=comma]{\IBSResultCSVacc};

\addplot [very thick, TUMOrange]
table[x=t, y=y1Meas, col sep=comma]{\IBSResultCSVacc};

\nextgroupplot[
height=3.6cm,
scaled x ticks=manual:{}{\pgfmathparse{#1}},
yticklabel style={
        /pgf/number format/fixed,
        /pgf/number format/precision=3
},
tick align=outside,
tick pos=left,
title={Acceleration \(\displaystyle a_2^{(0)}\) / (\si[per-mode=symbol]{\meter\per\second\squared})},
title style={yshift=0.5ex},
width=0.335\textwidth,
x grid style={darkgray176},
xmin=0, xmax=\IBSlen,
xtick style={color=black},
xlabel={Time / ms},
y grid style={darkgray176},
ymin=\yBminacc, ymax=\yBmaxacc,
ytick style={color=black}
]
\addplot [very thick, TUMBlue]
table[x=t, y=y2IBS, col sep=comma]{\IBSResultCSVacc};

\addplot [very thick, TUMOrange]
table[x=t, y=y2Meas, col sep=comma]{\IBSResultCSVacc};

\nextgroupplot[
height=3.6cm,
tick align=outside,
tick pos=left,
title={Lagrange multiplier \(\displaystyle \lambda\) / N},
title style={yshift=0.5ex},
width=0.335\textwidth,
x grid style={darkgray176},
xmin=0, xmax=\IBSlen,
xtick style={color=black},
xlabel={Time / ms},
y grid style={darkgray176},
ytick style={color=black},
ymin=\LMminacc, ymax=\LMmaxacc
]
\addplot [very thick, TUMBlue]
table[x=t, y=lambda, col sep=comma]{\IBSResultCSVacc};

\end{groupplot}
\end{tikzpicture}
		\end{flushright}\vspace{-0.4cm}
		\caption{IRFs processed using the \textbf{Time Domain} method}
		\label{fig:IBS_results_ENAW7075_manual_vinylB}
	\end{subfigure}	
	\caption{Experimental results of \textbf{aluminum} rods using manual hammer with \textbf{vinyl tip}, \legendary}
	\label{fig:IBS_results_ENAW7075_manual_vinyl}
\end{figure}

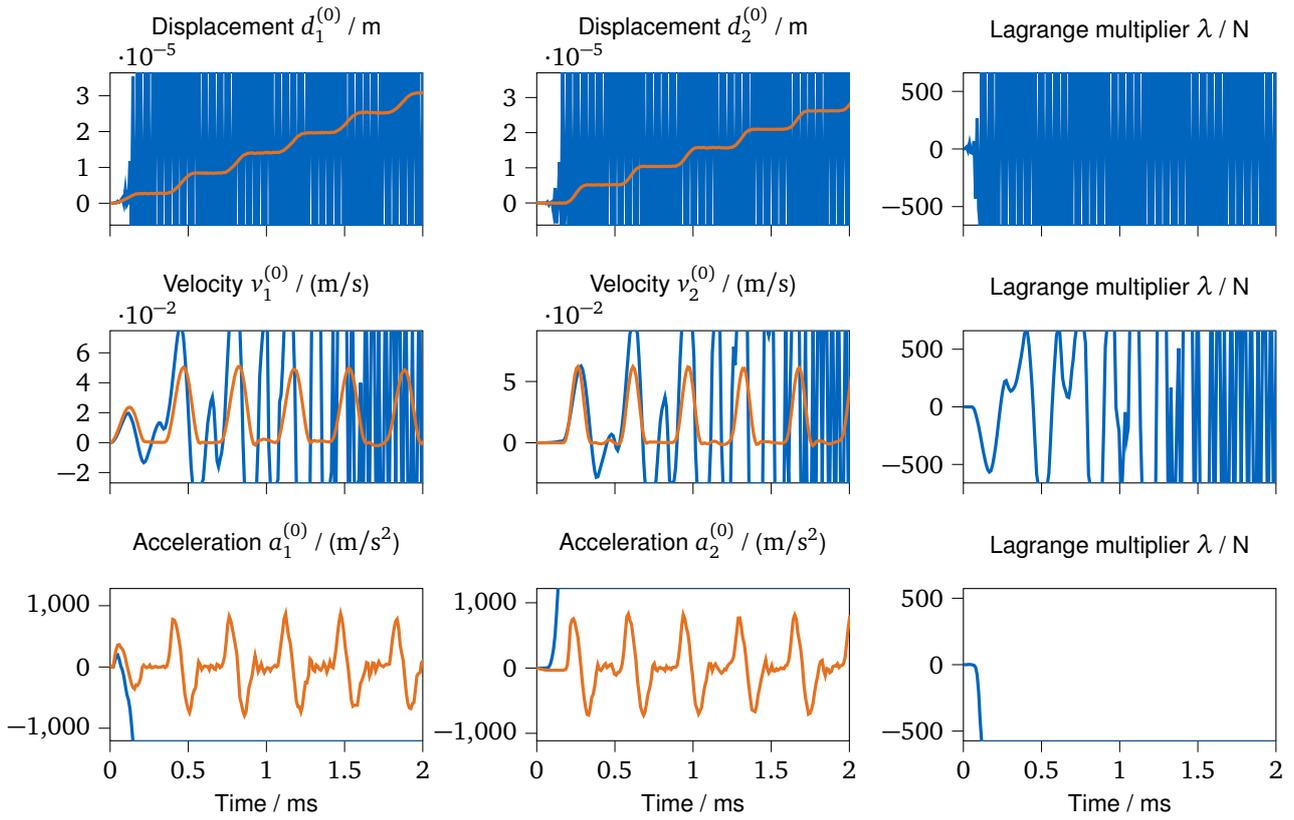
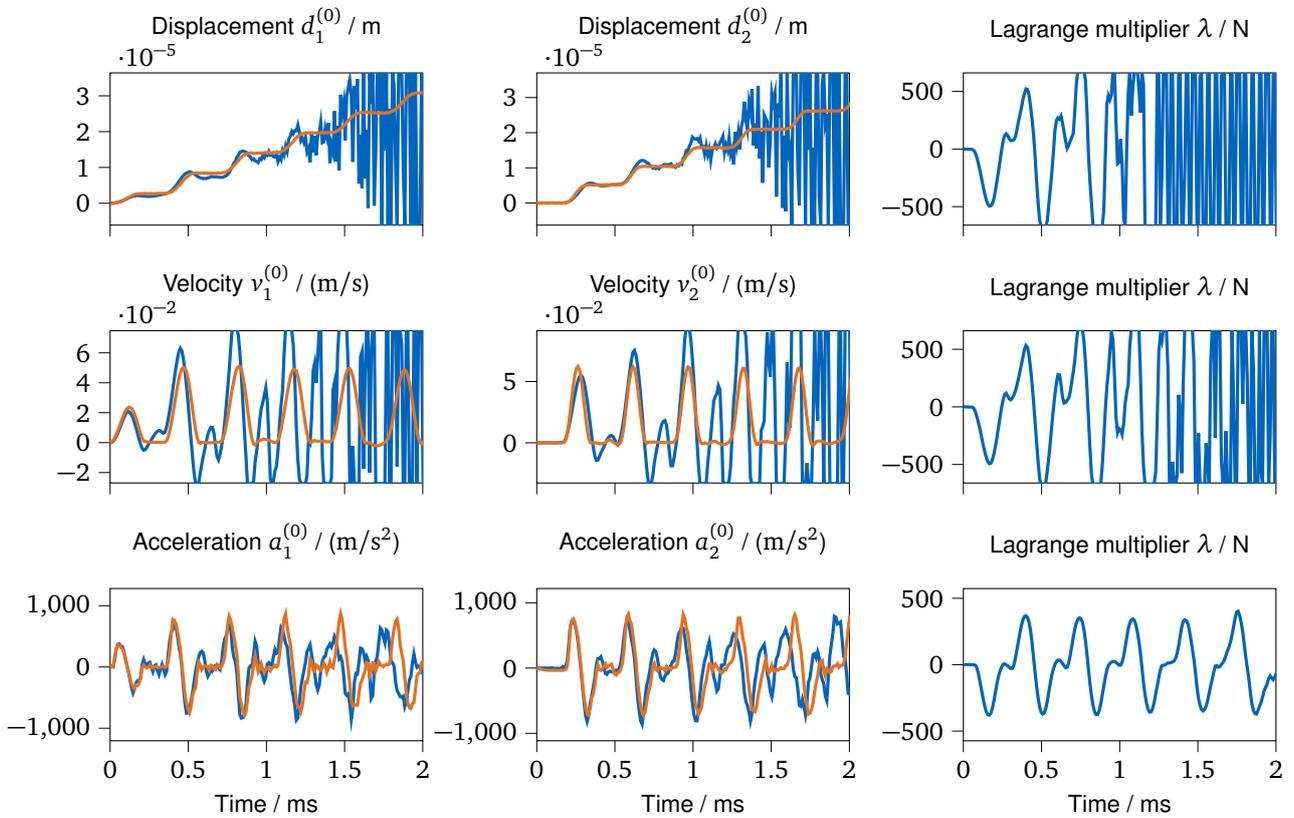
\begin{figure}[ht!]
	\begin{subfigure}{\textwidth}
		\small
		\renewcommand{\IBSlen}{2}
		\renewcommand{\IBSResultCSVdisp}{data/IBS_results_ENAW7075_manual_steel_d_Frequency.csv}
		\renewcommand{\yAmindisp}{-6.104633976065088e-06}
		\renewcommand{\yAmaxdisp}{3.639524549726047e-05}
		\renewcommand{\yBmindisp}{-6.2974566967568535e-06}
		\renewcommand{\yBmaxdisp}{3.701299015324367e-05}
		\renewcommand{\LMmindisp}{-660.6196060180664}
		\renewcommand{\LMmaxdisp}{660.6196060180664}
		
		\renewcommand{\IBSResultCSVvelo}{data/IBS_results_ENAW7075_manual_steel_v_Frequency.csv}
		\renewcommand{\yAminvelo}{-0.026929833833128212}
		\renewcommand{\yAmaxvelo}{0.07476644910871982}
		\renewcommand{\yBminvelo}{-0.03278352747205645}
		\renewcommand{\yBmaxvelo}{0.09145056892186403}
		\renewcommand{\LMminvelo}{-660.6196060180664}
		\renewcommand{\LMmaxvelo}{660.6196060180664}
		
		\renewcommand{\IBSResultCSVacc}{data/IBS_results_ENAW7075_manual_steel_a_Frequency.csv}
		\renewcommand{\yAminacc}{-1204.5668145751954}
		\renewcommand{\yAmaxacc}{1280.624899291992}
		\renewcommand{\yBminacc}{-1114.047120361328}
		\renewcommand{\yBmaxacc}{1218.1015393066407}
		\renewcommand{\LMminacc}{-573.1173266601562}
		\renewcommand{\LMmaxacc}{573.1173266601562}
		\begin{flushright}
			% This file was created with tikzplotlib v0.10.1.
\begin{tikzpicture}

\definecolor{darkgray176}{RGB}{176,176,176}
\definecolor{gray}{RGB}{128,128,128}
\definecolor{lightgray204}{RGB}{204,204,204}

\begin{groupplot}[group style={group size=3 by 3, horizontal sep=1.5cm, vertical sep=1.4cm}]

% Displacements
\nextgroupplot[
height=3.6cm,
scaled x ticks=manual:{}{\pgfmathparse{#1}},
yticklabel style={
        /pgf/number format/fixed,
        /pgf/number format/precision=3
},
tick align=outside,
tick pos=left,
title={Displacement \(\displaystyle d_1^{(0)}\) / m},
title style={yshift=0.5ex},
width=0.335\textwidth,
x grid style={darkgray176},
xmin=0, xmax=\IBSlen,
xtick style={color=black},
xticklabels={},
y grid style={darkgray176},
ymin=\yAmindisp, ymax=\yAmaxdisp,
ytick style={color=black}
]
\addplot [very thick, TUMBlue]
table[x=t, y=y1IBS, col sep=comma]{\IBSResultCSVdisp};

\addplot [very thick, TUMOrange]
table[x=t, y=y1Meas, col sep=comma]{\IBSResultCSVdisp};

\nextgroupplot[
height=3.6cm,
scaled x ticks=manual:{}{\pgfmathparse{#1}},
yticklabel style={
        /pgf/number format/fixed,
        /pgf/number format/precision=3
},
tick align=outside,
tick pos=left,
title={Displacement \(\displaystyle d_2^{(0)}\) / m},
title style={yshift=0.5ex},
width=0.335\textwidth,
x grid style={darkgray176},
xmin=0, xmax=\IBSlen,
xtick style={color=black},
xticklabels={},
y grid style={darkgray176},
ymin=\yBmindisp, ymax=\yBmaxdisp,
ytick style={color=black}
]
\addplot [very thick, TUMBlue]
table[x=t, y=y2IBS, col sep=comma]{\IBSResultCSVdisp};

\addplot [very thick, TUMOrange]
table[x=t, y=y2Meas, col sep=comma]{\IBSResultCSVdisp};

\nextgroupplot[
height=3.6cm,
tick align=outside,
tick pos=left,
title={Lagrange multiplier \(\displaystyle \lambda\) / N},
title style={yshift=0.5ex},
width=0.335\textwidth,
x grid style={darkgray176},
xmin=0, xmax=\IBSlen,
xtick style={color=black},
xticklabels={},
y grid style={darkgray176},
ymin=\LMmindisp, ymax=\LMmaxdisp,
ytick style={color=black}
]
\addplot [very thick, TUMBlue]
table[x=t, y=lambda, col sep=comma]{\IBSResultCSVdisp};

% Velocities
\nextgroupplot[
height=3.6cm,
scaled x ticks=manual:{}{\pgfmathparse{#1}},
yticklabel style={
        /pgf/number format/fixed,
        /pgf/number format/precision=3
},
tick align=outside,
tick pos=left,
title={Velocity \(\displaystyle v_1^{(0)}\) / (\si[per-mode=symbol]{\meter\per\second})},
title style={yshift=0.5ex},
width=0.335\textwidth,
x grid style={darkgray176},
xmin=0, xmax=\IBSlen,
xtick style={color=black},
xticklabels={},
y grid style={darkgray176},
ymin=\yAminvelo, ymax=\yAmaxvelo,
ytick style={color=black}
]
\addplot [very thick, TUMBlue]
table[x=t, y=y1IBS, col sep=comma]{\IBSResultCSVvelo};

\addplot [very thick, TUMOrange]
table[x=t, y=y1Meas, col sep=comma]{\IBSResultCSVvelo};

\nextgroupplot[
height=3.6cm,
scaled x ticks=manual:{}{\pgfmathparse{#1}},
yticklabel style={
        /pgf/number format/fixed,
        /pgf/number format/precision=3
},
tick align=outside,
tick pos=left,
title={Velocity \(\displaystyle v_2^{(0)}\) / (\si[per-mode=symbol]{\meter\per\second})},
title style={yshift=0.5ex},
width=0.335\textwidth,
x grid style={darkgray176},
xmin=0, xmax=\IBSlen,
xtick style={color=black},
xticklabels={},
y grid style={darkgray176},
ymin=\yBminvelo, ymax=\yBmaxvelo,
ytick style={color=black}
]
\addplot [very thick, TUMBlue]
table[x=t, y=y2IBS, col sep=comma]{\IBSResultCSVvelo};

\addplot [very thick, TUMOrange]
table[x=t, y=y2Meas, col sep=comma]{\IBSResultCSVvelo};

\nextgroupplot[
height=3.6cm,
tick align=outside,
tick pos=left,
title={Lagrange multiplier \(\displaystyle \lambda\) / N},
title style={yshift=0.5ex},
width=0.335\textwidth,
x grid style={darkgray176},
xmin=0, xmax=\IBSlen,
xtick style={color=black},
xticklabels={},
y grid style={darkgray176},
ymin=\LMminvelo, ymax=\LMmaxvelo,
ytick style={color=black}
]
\addplot [very thick, TUMBlue]
table[x=t, y=lambda, col sep=comma]{\IBSResultCSVvelo};

% Accelerations
\nextgroupplot[
height=3.6cm,
scaled x ticks=manual:{}{\pgfmathparse{#1}},
yticklabel style={
        /pgf/number format/fixed,
        /pgf/number format/precision=3
},
tick align=outside,
tick pos=left,
title={Acceleration \(\displaystyle a_1^{(0)}\) / (\si[per-mode=symbol]{\meter\per\second\squared})},
title style={yshift=0.5ex},
width=0.335\textwidth,
x grid style={darkgray176},
xmin=0, xmax=\IBSlen,
xtick style={color=black},
xlabel={Time / ms},
y grid style={darkgray176},
ymin=\yAminacc, ymax=\yAmaxacc,
ytick style={color=black}
]
\addplot [very thick, TUMBlue]
table[x=t, y=y1IBS, col sep=comma]{\IBSResultCSVacc};

\addplot [very thick, TUMOrange]
table[x=t, y=y1Meas, col sep=comma]{\IBSResultCSVacc};

\nextgroupplot[
height=3.6cm,
scaled x ticks=manual:{}{\pgfmathparse{#1}},
yticklabel style={
        /pgf/number format/fixed,
        /pgf/number format/precision=3
},
tick align=outside,
tick pos=left,
title={Acceleration \(\displaystyle a_2^{(0)}\) / (\si[per-mode=symbol]{\meter\per\second\squared})},
title style={yshift=0.5ex},
width=0.335\textwidth,
x grid style={darkgray176},
xmin=0, xmax=\IBSlen,
xtick style={color=black},
xlabel={Time / ms},
y grid style={darkgray176},
ymin=\yBminacc, ymax=\yBmaxacc,
ytick style={color=black}
]
\addplot [very thick, TUMBlue]
table[x=t, y=y2IBS, col sep=comma]{\IBSResultCSVacc};

\addplot [very thick, TUMOrange]
table[x=t, y=y2Meas, col sep=comma]{\IBSResultCSVacc};

\nextgroupplot[
height=3.6cm,
tick align=outside,
tick pos=left,
title={Lagrange multiplier \(\displaystyle \lambda\) / N},
title style={yshift=0.5ex},
width=0.335\textwidth,
x grid style={darkgray176},
xmin=0, xmax=\IBSlen,
xtick style={color=black},
xlabel={Time / ms},
y grid style={darkgray176},
ytick style={color=black},
ymin=\LMminacc, ymax=\LMmaxacc
]
\addplot [very thick, TUMBlue]
table[x=t, y=lambda, col sep=comma]{\IBSResultCSVacc};

\end{groupplot}
\end{tikzpicture}
		\end{flushright}\vspace{-0.4cm}
		\caption{IRFs processed using the \textbf{Frequency Domain} method}\vspace{0.4cm}
		\label{fig:IBS_results_ENAW7075_manual_steelA}
	\end{subfigure}
	\begin{subfigure}{\textwidth}
		\small
		\renewcommand{\IBSlen}{2}
		\renewcommand{\IBSResultCSVdisp}{data/IBS_results_ENAW7075_manual_steel_d_Time.csv}
		\renewcommand{\yAmindisp}{-6.104633976065088e-06}
		\renewcommand{\yAmaxdisp}{3.639524549726047e-05}
		\renewcommand{\yBmindisp}{-6.2974566967568535e-06}
		\renewcommand{\yBmaxdisp}{3.701299015324367e-05}
		\renewcommand{\LMmindisp}{-660.6196060180664}
		\renewcommand{\LMmaxdisp}{660.6196060180664}
		
		\renewcommand{\IBSResultCSVvelo}{data/IBS_results_ENAW7075_manual_steel_v_Time.csv}
		\renewcommand{\yAminvelo}{-0.026929833833128212}
		\renewcommand{\yAmaxvelo}{0.07476644910871982}
		\renewcommand{\yBminvelo}{-0.03278352747205645}
		\renewcommand{\yBmaxvelo}{0.09145056892186403}
		\renewcommand{\LMminvelo}{-660.6196060180664}
		\renewcommand{\LMmaxvelo}{660.6196060180664}
		
		\renewcommand{\IBSResultCSVacc}{data/IBS_results_ENAW7075_manual_steel_a_Time.csv}
		\renewcommand{\yAminacc}{-1204.5668145751954}
		\renewcommand{\yAmaxacc}{1280.624899291992}
		\renewcommand{\yBminacc}{-1114.047120361328}
		\renewcommand{\yBmaxacc}{1218.1015393066407}
		\renewcommand{\LMminacc}{-573.1173266601562}
		\renewcommand{\LMmaxacc}{573.1173266601562}
		\begin{flushright}
			% This file was created with tikzplotlib v0.10.1.
\begin{tikzpicture}

\definecolor{darkgray176}{RGB}{176,176,176}
\definecolor{gray}{RGB}{128,128,128}
\definecolor{lightgray204}{RGB}{204,204,204}

\begin{groupplot}[group style={group size=3 by 3, horizontal sep=1.5cm, vertical sep=1.4cm}]

% Displacements
\nextgroupplot[
height=3.6cm,
scaled x ticks=manual:{}{\pgfmathparse{#1}},
yticklabel style={
        /pgf/number format/fixed,
        /pgf/number format/precision=3
},
tick align=outside,
tick pos=left,
title={Displacement \(\displaystyle d_1^{(0)}\) / m},
title style={yshift=0.5ex},
width=0.335\textwidth,
x grid style={darkgray176},
xmin=0, xmax=\IBSlen,
xtick style={color=black},
xticklabels={},
y grid style={darkgray176},
ymin=\yAmindisp, ymax=\yAmaxdisp,
ytick style={color=black}
]
\addplot [very thick, TUMBlue]
table[x=t, y=y1IBS, col sep=comma]{\IBSResultCSVdisp};

\addplot [very thick, TUMOrange]
table[x=t, y=y1Meas, col sep=comma]{\IBSResultCSVdisp};

\nextgroupplot[
height=3.6cm,
scaled x ticks=manual:{}{\pgfmathparse{#1}},
yticklabel style={
        /pgf/number format/fixed,
        /pgf/number format/precision=3
},
tick align=outside,
tick pos=left,
title={Displacement \(\displaystyle d_2^{(0)}\) / m},
title style={yshift=0.5ex},
width=0.335\textwidth,
x grid style={darkgray176},
xmin=0, xmax=\IBSlen,
xtick style={color=black},
xticklabels={},
y grid style={darkgray176},
ymin=\yBmindisp, ymax=\yBmaxdisp,
ytick style={color=black}
]
\addplot [very thick, TUMBlue]
table[x=t, y=y2IBS, col sep=comma]{\IBSResultCSVdisp};

\addplot [very thick, TUMOrange]
table[x=t, y=y2Meas, col sep=comma]{\IBSResultCSVdisp};

\nextgroupplot[
height=3.6cm,
tick align=outside,
tick pos=left,
title={Lagrange multiplier \(\displaystyle \lambda\) / N},
title style={yshift=0.5ex},
width=0.335\textwidth,
x grid style={darkgray176},
xmin=0, xmax=\IBSlen,
xtick style={color=black},
xticklabels={},
y grid style={darkgray176},
ymin=\LMmindisp, ymax=\LMmaxdisp,
ytick style={color=black}
]
\addplot [very thick, TUMBlue]
table[x=t, y=lambda, col sep=comma]{\IBSResultCSVdisp};

% Velocities
\nextgroupplot[
height=3.6cm,
scaled x ticks=manual:{}{\pgfmathparse{#1}},
yticklabel style={
        /pgf/number format/fixed,
        /pgf/number format/precision=3
},
tick align=outside,
tick pos=left,
title={Velocity \(\displaystyle v_1^{(0)}\) / (\si[per-mode=symbol]{\meter\per\second})},
title style={yshift=0.5ex},
width=0.335\textwidth,
x grid style={darkgray176},
xmin=0, xmax=\IBSlen,
xtick style={color=black},
xticklabels={},
y grid style={darkgray176},
ymin=\yAminvelo, ymax=\yAmaxvelo,
ytick style={color=black}
]
\addplot [very thick, TUMBlue]
table[x=t, y=y1IBS, col sep=comma]{\IBSResultCSVvelo};

\addplot [very thick, TUMOrange]
table[x=t, y=y1Meas, col sep=comma]{\IBSResultCSVvelo};

\nextgroupplot[
height=3.6cm,
scaled x ticks=manual:{}{\pgfmathparse{#1}},
yticklabel style={
        /pgf/number format/fixed,
        /pgf/number format/precision=3
},
tick align=outside,
tick pos=left,
title={Velocity \(\displaystyle v_2^{(0)}\) / (\si[per-mode=symbol]{\meter\per\second})},
title style={yshift=0.5ex},
width=0.335\textwidth,
x grid style={darkgray176},
xmin=0, xmax=\IBSlen,
xtick style={color=black},
xticklabels={},
y grid style={darkgray176},
ymin=\yBminvelo, ymax=\yBmaxvelo,
ytick style={color=black}
]
\addplot [very thick, TUMBlue]
table[x=t, y=y2IBS, col sep=comma]{\IBSResultCSVvelo};

\addplot [very thick, TUMOrange]
table[x=t, y=y2Meas, col sep=comma]{\IBSResultCSVvelo};

\nextgroupplot[
height=3.6cm,
tick align=outside,
tick pos=left,
title={Lagrange multiplier \(\displaystyle \lambda\) / N},
title style={yshift=0.5ex},
width=0.335\textwidth,
x grid style={darkgray176},
xmin=0, xmax=\IBSlen,
xtick style={color=black},
xticklabels={},
y grid style={darkgray176},
ymin=\LMminvelo, ymax=\LMmaxvelo,
ytick style={color=black}
]
\addplot [very thick, TUMBlue]
table[x=t, y=lambda, col sep=comma]{\IBSResultCSVvelo};

% Accelerations
\nextgroupplot[
height=3.6cm,
scaled x ticks=manual:{}{\pgfmathparse{#1}},
yticklabel style={
        /pgf/number format/fixed,
        /pgf/number format/precision=3
},
tick align=outside,
tick pos=left,
title={Acceleration \(\displaystyle a_1^{(0)}\) / (\si[per-mode=symbol]{\meter\per\second\squared})},
title style={yshift=0.5ex},
width=0.335\textwidth,
x grid style={darkgray176},
xmin=0, xmax=\IBSlen,
xtick style={color=black},
xlabel={Time / ms},
y grid style={darkgray176},
ymin=\yAminacc, ymax=\yAmaxacc,
ytick style={color=black}
]
\addplot [very thick, TUMBlue]
table[x=t, y=y1IBS, col sep=comma]{\IBSResultCSVacc};

\addplot [very thick, TUMOrange]
table[x=t, y=y1Meas, col sep=comma]{\IBSResultCSVacc};

\nextgroupplot[
height=3.6cm,
scaled x ticks=manual:{}{\pgfmathparse{#1}},
yticklabel style={
        /pgf/number format/fixed,
        /pgf/number format/precision=3
},
tick align=outside,
tick pos=left,
title={Acceleration \(\displaystyle a_2^{(0)}\) / (\si[per-mode=symbol]{\meter\per\second\squared})},
title style={yshift=0.5ex},
width=0.335\textwidth,
x grid style={darkgray176},
xmin=0, xmax=\IBSlen,
xtick style={color=black},
xlabel={Time / ms},
y grid style={darkgray176},
ymin=\yBminacc, ymax=\yBmaxacc,
ytick style={color=black}
]
\addplot [very thick, TUMBlue]
table[x=t, y=y2IBS, col sep=comma]{\IBSResultCSVacc};

\addplot [very thick, TUMOrange]
table[x=t, y=y2Meas, col sep=comma]{\IBSResultCSVacc};

\nextgroupplot[
height=3.6cm,
tick align=outside,
tick pos=left,
title={Lagrange multiplier \(\displaystyle \lambda\) / N},
title style={yshift=0.5ex},
width=0.335\textwidth,
x grid style={darkgray176},
xmin=0, xmax=\IBSlen,
xtick style={color=black},
xlabel={Time / ms},
y grid style={darkgray176},
ytick style={color=black},
ymin=\LMminacc, ymax=\LMmaxacc
]
\addplot [very thick, TUMBlue]
table[x=t, y=lambda, col sep=comma]{\IBSResultCSVacc};

\end{groupplot}
\end{tikzpicture}
		\end{flushright}\vspace{-0.4cm}
		\caption{IRFs processed using the \textbf{Time Domain} method}
		\label{fig:IBS_results_ENAW7075_manual_steelB}
	\end{subfigure}	
	\caption{Experimental results of \textbf{aluminum} rods using manual hammer with \textbf{steel tip}, \legendary}
	\label{fig:IBS_results_ENAW7075_manual_steel}
\end{figure}

For the aluminum rods, \cref{fig:IBS_results_ENAW7075_manual_vinyl,fig:IBS_results_ENAW7075_manual_steel}, the difference in excitation bandwidth can also be seen in the reference responses, where for displacement-based IBS using the vinyl tip mostly the rigid body translation is visible, while using the steel tip also shows the expected stepped response. It might therefore not be surprising that displacement IBS using the vinyl tip is unstable. For the steel tip using frequency domain IRFs also yields results that quickly become unstable, but using time domain IRFs, the magnitudes of the first three response peaks can be reasonably well estimated. After that, the responses and Lagrange multiplier start to oscillate high frequently.

Using velocities, the first two response peaks can be estimated decently using time domain IRFs, while using the frequency domain approach for the IRF calculation leads to worse results that either do not represent the expected system dynamics in the case of the vinyl tip or show worse stability in the case of the steel tip.

Lastly, for acceleration-based IBS the difference between the two IRF calculation domains is the most pronounced. For both tips, the IBS results using frequency domain IRFs immediately diverge to infinity, while using time domain IRFs yields the best amplitude predictions so far, where using both tips the first four response peaks are reasonably close to the reference measurements.

A summary of all experimental IBS results using the full measurement bandwidth can be found in \cref{tab:exp_results}. There, for each combination of material, hammer type, IRF calculation domain and quantity used for IBS, we indicate either how the IBS scheme failed or how many of the expected response peaks could be predicted within a reasonable tolerance with respect to the reference measurement.

\bgroup
\def\arraystretch{1.1}
\begin{table}[ht]
	\centering
	\small
	\begin{tabularx}{\textwidth}{c|Y|Y|Y|Y|Y|Y|Y|Y}
		\multicolumn{9}{c}{\cellcolor{lightgray}\textbf{Results of Experimental IBS Using Full Measurement Bandwidth}} \\\hline
		\textbf{Rod Material} & \multicolumn{4}{c|}{Polyoxymethylene (POM)} & \multicolumn{4}{c}{Aluminum (EN AW 7075)} \\\hline
		
		\textbf{Hammer Type} & \multicolumn{2}{c|}{Manual with Vinyl Tip} & \multicolumn{2}{c|}{Manual with Steel Tip} & \multicolumn{2}{c|}{Manual with Vinyl Tip} & \multicolumn{2}{c}{Manual with Steel Tip} \\\hline
		
		\textbf{IRF Domain} & Frequency & Time & Frequency & Time & Frequency & Time & Frequency & Time \\\hline
		\textbf{Figure} & \Cref{fig:IBS_results_POM_manual_vinylA} & \Cref{fig:IBS_results_POM_manual_vinylB} & \Cref{fig:IBS_results_POM_manual_steelA} & \Cref{fig:IBS_results_POM_manual_steelB} & \Cref{fig:IBS_results_ENAW7075_manual_vinylA} & \Cref{fig:IBS_results_ENAW7075_manual_vinylB} & \Cref{fig:IBS_results_ENAW7075_manual_steelA} & \Cref{fig:IBS_results_ENAW7075_manual_steelB} \\\hline
		Displacement IBS & \trFailUnstable & \trFailUnstable & \trFailUnstable & \trFailUnstable & \trPeakZero & \trPeakOne & \trFailUnstable & \trPeakThree \\\hline
		Velocity IBS & \trFailUnstable & \trPeakZero & \trFailUnstable & \trFailUnstable & \trPeakZero & \trPeakTwo & \trPeakOne & \trPeakOne \\\hline
		Acceleration IBS & \trPeakOne$^1$ & \trPeakTwo$^1$ & \trFailCoupling$^1$ & \trFailCoupling$^1$ & \trFailUnstable & \trPeakFour & \trFailUnstable & \trPeakFour \\\hline
	\end{tabularx} \vspace{0.2cm}
	
	Explanation of symbols: $\Large\boldsymbol{\infty}$: IBS algorithm immediately unstable, $\Large\boldsymbol{\times}$: Coupling issues, \\ $\boldsymbol{\checkmark}$: Number of response peaks within reasonable tolerance of reference measurement \\
	$\boldsymbol{\mathrm{O}}$: No correct response peaks, $^1$: Initial response peak significantly wrong
	\caption{Summary of experimental IBS results of all test cases using the full measurement bandwidth}
	\label{tab:exp_results}
\end{table}
\egroup

As can be seen, especially bad is the performance of IBS with the POM rods, where using displacement or velocity responses led to the IBS scheme immediately becoming unstable. While better results with respect to stability could be found using acceleration responses, in all cases the initial driving point response was greatly inflated, making the result useless for shock analysis. Tremendously better are the results of the aluminum rods. Using accelerations as the IBS quantity and calculating the IRFs in the time domain enabled a decently accurate estimation of the first four driving point response peaks. In almost all cases considered for the aluminum rods, calculating the IRFs in the time domain yielded better IBS results.

Regarding the set goal for the experimental IBS application, namely to correctly predict the first two response peaks of the driving point, this could be achieved for the aluminum rods and all response quantities for at least one of both hammer types. The application of IBS to the POM rods was in no case successful.

\clearpage
% !TeX spellcheck = en_US
\subsection{Analysis of Experimental Results and Issues}

Comparing the two shown IRF calculation methods, it can be seen that in almost all cases the results using IRFs calculated in the time domain are more stable than the ones using the calculation through the frequency domain. While the overlap introduced by the periodic continuation has been compensated for by zero-padding, there still exists cut-off issues, i.e.\ the end of the IRF calculated in the frequency domain has to be continuous with respect to the beginning of the IRF to represent a periodic signal. This might especially be an issue for displacement and velocity IRFs of the aluminum rods as the responses do not fully decay within the measurement window. However, the acceleration responses of the aluminum rods decay fully within the measurement window and yet the results using frequency domain IRFs are significantly worse, becoming unstable immediately.

\begin{figure}[ht]
	\centering
	\small
	\input{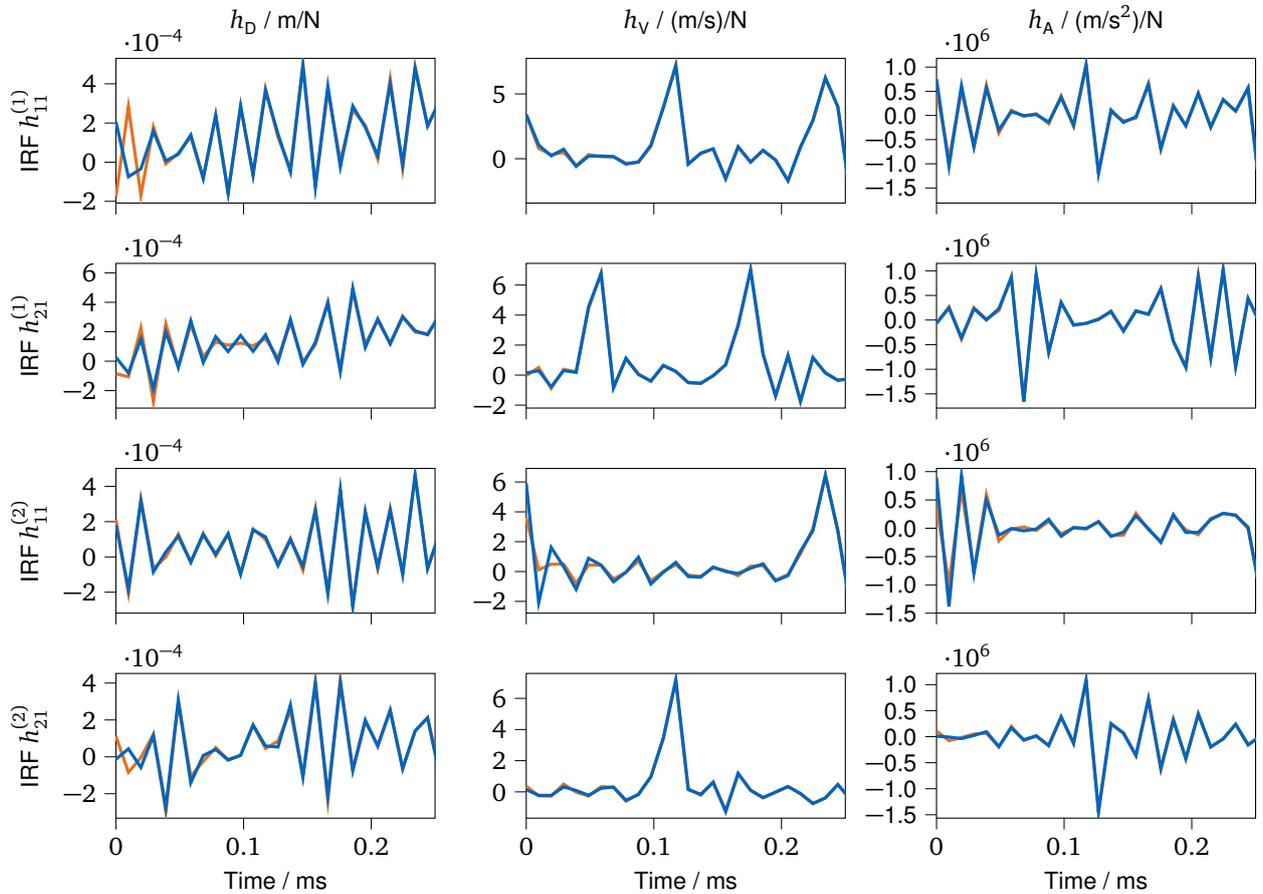}
	\caption{Beginning of substructure IRFs of \textbf{aluminum rods} using a manual hammer with \textbf{steel tip}, comparing both discussed IRF calculation methods, where \legendaryTwo}
	\label{fig:IRF_alu_steel}
\end{figure}

In \Cref{fig:IRF_alu_steel}, each column represents one measured quantity containing all four IRFs required within one IBS evaluation. Looking at the beginning of the IRFs computed for the aluminum rods excited by the manual hammer with steel tip, it can be seen that the time and frequency domain IRFs are different initially. Since the IBS scheme couples the IRFs together, one erroneous IRF might be enough to make it unstable. Also remember that the first IRF values in time $\vH[0]$ determine the Lagrange multipliers in the IBS scheme, see \cref{eq:IBS_LM_from_predictor_D1}.\vspace{-0.15cm}

\vspace{-0.15cm}

This can be seen for example with the displacement IRFs $h_{\text{D}}$, where the initial value of $h_{11}^{(1)}$, the frequency domain driving point IRF of substructure 1, is negative. As mentioned in the theory section, with the algorithm as implemented here this would lead to a shift of all IRFs by one sample, because otherwise, the IBS scheme would be unstable immediately. But then, the first value of $h_{11}^{(2)}$, the driving point response of substructure 2, is negative.

For the acceleration IRFs the culprit might also be the driving point response of substructure 2, where the initial IRF value is about a third of the time domain IRF. Since this first value is used to determine the required Lagrange multiplier, the value being significantly too low results in a Lagrange multiplier that is considerably overestimated. The same can be found for velocities, where also $h_{11}^{(2)}$ is reduced in value compared to the time domain IRF.

The origin of the issue causing the initial IRF values to differ between the two discussed methods is not fully understood at this point in time but is believed to stem from the forced periodization of the frequency domain. Within this paper, therefore, the frequency domain approach is not further developed.

\newpage

Limiting further discussions to the results using time domain IRFs, it can be seen that almost all responses determined using IBS become unstable towards the end of the shown time frame. As simulations using numerically generated responses with identical sample rate and impulse length (not detailed here) showed, this is an issue that only occurs with experimentally obtained IRFs, implying that the instability is caused by errors in the IRFs.

One possibility to determine the error made within the IRF estimation is to perform a procedure that is denoted as a \textit{back-convolution} in this paper. The idea is as follows: An experimental IRF is determined by deconvoluting measured responses and excitation time series. If the IRF estimation is perfect then convoluting the determined IRF with one of the experimentally applied forces should exactly yield the response measured for the respective force. Comparing the result of the back-convolution with the measured response by simply subtracting the values of both time series then enables to estimate the error of the IRF calculation. It is to be noted that this method would also show an error in the IRF estimation when the averaged IRF has less noise than the measured responses or when other measurement errors, like an incorrect impact location or direction, are reduced through the averaging.

\begin{figure}[ht!]
	\small
	\centering
	\renewcommand{\IBSlen}{2}
	\renewcommand{\IBSResultCSVvelo}{data/IBS_results_ENAW7075_manual_steel_v_Time.csv}
	\renewcommand{\yAminvelo}{-0.026929833833128212}
	\renewcommand{\yAmaxvelo}{0.07476644910871982}
	\renewcommand{\yBminvelo}{-0.03278352747205645}
	\renewcommand{\yBmaxvelo}{0.09145056892186403}
	\renewcommand{\LMminvelo}{-660.6196060180664}
	\renewcommand{\LMmaxvelo}{660.6196060180664}
		\begin{subfigure}{\textwidth}
			\centering
			\input{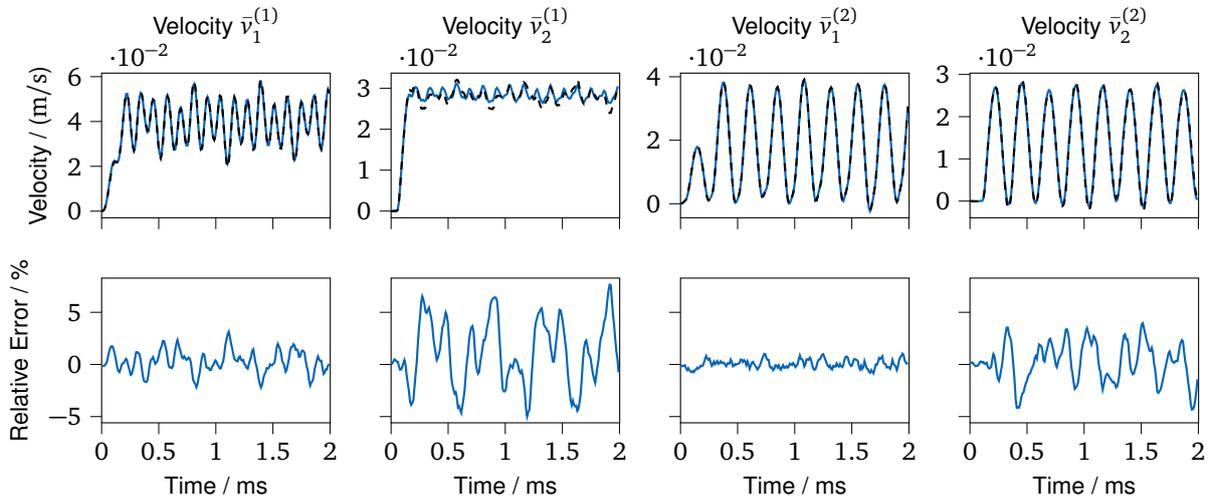}
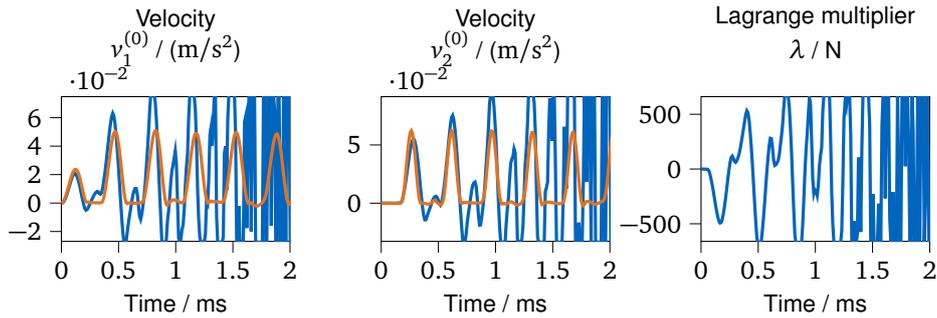
			\caption{Velocity responses $\bar{v}$ calculated through convolution of identified IRFs $h$ with originally applied impact force $f$ shown in blue, compared to measured substructure response in dashed black as well as relative error (normalized to maximum response) between both}
			\label{fig:BC-a}
		\end{subfigure}
		\begin{subfigure}{\textwidth}
			\centering
			% This file was created with tikzplotlib v0.10.1.
\begin{tikzpicture}

\definecolor{darkgray176}{RGB}{176,176,176}
\definecolor{lightgray204}{RGB}{204,204,204}

\begin{groupplot}[group style={group size=4 by 1, horizontal sep=1.2cm, vertical sep=1.2cm}]
\nextgroupplot[
height=3.5cm,
scaled x ticks=manual:{}{\pgfmathparse{#1}},
yticklabel style={
        /pgf/number format/fixed,
        /pgf/number format/precision=3
},
tick align=outside,
tick pos=left,
title={Velocity \\\(\displaystyle v_1^{(0)}\) / (\si[per-mode=symbol]{\meter\per\second\squared})},
title style = {align = center},
title style={yshift=0.5ex},
width=0.27\textwidth,
x grid style={darkgray176},
xmin=0, xmax=\IBSlen,
xlabel={Time / ms},
xtick style={color=black},
y grid style={darkgray176},
ymin=\yAminvelo, ymax=\yAmaxvelo,
ytick style={color=black}
]
\addplot [very thick, TUMBlue]
table[x=t, y=y1IBS, col sep=comma]{\IBSResultCSVvelo};

\addplot [very thick, TUMOrange]
table[x=t, y=y1Meas, col sep=comma]{\IBSResultCSVvelo};

\nextgroupplot[
height=3.5cm,
scaled x ticks=manual:{}{\pgfmathparse{#1}},
yticklabel style={
        /pgf/number format/fixed,
        /pgf/number format/precision=3
},
tick align=outside,
tick pos=left,
title={Velocity \\\(\displaystyle v_2^{(0)}\) / (\si[per-mode=symbol]{\meter\per\second\squared})},
title style = {align = center},
title style={yshift=0.5ex},
width=0.27\textwidth,
x grid style={darkgray176},
xmin=0, xmax=\IBSlen,
xlabel={Time / ms},
xtick style={color=black},
y grid style={darkgray176},
ymin=\yBminvelo, ymax=\yBmaxvelo,
ytick style={color=black}
]
\addplot [very thick, TUMBlue]
table[x=t, y=y2IBS, col sep=comma]{\IBSResultCSVvelo};

\addplot [very thick, TUMOrange]
table[x=t, y=y2Meas, col sep=comma]{\IBSResultCSVvelo};

\nextgroupplot[
height=3.5cm,
tick align=outside,
tick pos=left,
title={Lagrange multiplier \\\(\displaystyle \lambda\vphantom{v_2^{(0)}}\) / N},
title style = {align = center},
title style={yshift=0.5ex},
width=0.27\textwidth,
x grid style={darkgray176},
xmin=0, xmax=\IBSlen,
xlabel={Time / ms},
xtick style={color=black},
y grid style={darkgray176},
ymin=\LMminvelo, ymax=\LMmaxvelo,
ytick style={color=black}
]
\addplot [very thick, TUMBlue]
table[x=t, y=lambda, col sep=comma]{\IBSResultCSVvelo};

\end{groupplot}

\end{tikzpicture}
			\caption{Original velocity-based IBS result using IRFs calculated in the \textbf{time domain}}
			\label{fig:BC-b}
		\end{subfigure}
		\caption{Results of back-convolution for \textbf{aluminum} rods impacted with \textbf{steel tip} and \textbf{velocity} responses}
		\label{fig:back-convolution}\vspace{-0.4cm}
\end{figure}

To get a more meaningful interpretation of the error, a relative error is used, which is normalized to the highest magnitude of the respective response within the viewed period. The results of a back-convolution of all dofs of the aluminum rods excited with a steel tip using velocities are shown in \cref{fig:BC-a}, where the back-convoluted response is depicted in blue and the measured one in dashed black. The previously obtained IBS result is shown in \cref{fig:BC-b}.\vspace{-0.4cm}

As can be seen, some error is present in the determined IRFs that leads to erroneous back-convoluted responses $\bar{v}$, which reaches up to \SI{3}{\percent} for the driving points responses and in the worst case, of the non-driving point of the first substructure, around \SI{7.5}{\percent}. While there exist some errors in the IRFs, the response retrieved using back-convolution in \cref{fig:BC-a} is nevertheless stable and does not exhibit the unstable behavior as observable in the IBS result in \cref{fig:BC-b}, i.e.\ the growing and high frequent oscillations toward the end of the viewed period.

Therefore, it can be argued that the observed instability is most likely introduced by errors of the IRF estimation, but the apparent issue, i.e.\ a growing and high frequent oscillation, is caused by the IBS algorithm itself.

Experimentally, but also in general, any error of the first IRF values $\vH[0]$ will lead to an incorrect prediction of the responses at the interface and with that incorrect Lagrange multipliers. Over time, the inaccurately predicted responses will manifest in additional interface gaps, requiring additional interface forces to close, generating new interface gaps and so on. In the cases where instability can be seen in the experimental IBS results, the Lagrange multipliers also show either instability, i.e.\ growing amplitudes, or a change of the time domain shape, compared to their initial or the expected shape. This could explain the stability issues of experimental IBS applications. 

\begin{figure}[ht!]
	\small
	\renewcommand{\IBSlen}{2}
	\renewcommand{\IBSResultCSVAa}{data/ENAW7075_averaging/IBS_results_exp_ENAW7075_v4_manual_D1_a_E4_n1.csv}
	\renewcommand{\IBSResultCSVBa}{data/ENAW7075_averaging/IBS_results_exp_ENAW7075_v4_manual_steel_soft_D1_a_E4_n1.csv}
	\renewcommand{\IBSResultCSVAb}{data/ENAW7075_averaging/IBS_results_exp_ENAW7075_v4_manual_D1_a_E4_n3.csv}
	\renewcommand{\IBSResultCSVBb}{data/ENAW7075_averaging/IBS_results_exp_ENAW7075_v4_manual_steel_soft_D1_a_E4_n3.csv}
	\renewcommand{\IBSResultCSVAc}{data/ENAW7075_averaging/IBS_results_exp_ENAW7075_v4_manual_D1_a_E4_n5.csv}
	\renewcommand{\IBSResultCSVBc}{data/ENAW7075_averaging/IBS_results_exp_ENAW7075_v4_manual_steel_soft_D1_a_E4_n5.csv}
	\renewcommand{\IBSResultCSVAd}{data/ENAW7075_averaging/IBS_results_exp_ENAW7075_v4_manual_D1_a_E4_n10.csv}
	\renewcommand{\IBSResultCSVBd}{data/ENAW7075_averaging/IBS_results_exp_ENAW7075_v4_manual_steel_soft_D1_a_E4_n10.csv}
	\renewcommand{\IBSResultCSVAe}{data/ENAW7075_averaging/IBS_results_exp_ENAW7075_v4_manual_D1_a_E4_n20.csv}
	\renewcommand{\IBSResultCSVBe}{data/ENAW7075_averaging/IBS_results_exp_ENAW7075_v4_manual_steel_soft_D1_a_E4_n20.csv}
	\renewcommand{\yAmin}{-88.58306671142579}
	\renewcommand{\yAmax}{94.72701667785644}
	\renewcommand{\LMmin}{-196.43740814208985}
	\renewcommand{\LMmax}{196.43740814208985}
	\renewcommand{\yAminB}{-948.4418347167968}
	\renewcommand{\yAmaxB}{1024.4999194335937}
	\renewcommand{\LMminB}{-573.1173266601562}
	\renewcommand{\LMmaxB}{573.1173266601562}
	\begin{minipage}{0.5\textwidth}
		\centering
		\hspace{2.3cm}\textbf{Manual Hammer with Vinyl Tip}
	\end{minipage}
	\begin{minipage}{0.5\textwidth}
		\centering
		\hspace{1.2cm}\textbf{Manual Hammer with Steel Tip}
	\end{minipage}\\\vspace{-0.6cm}
	\begin{flushright}
		\input{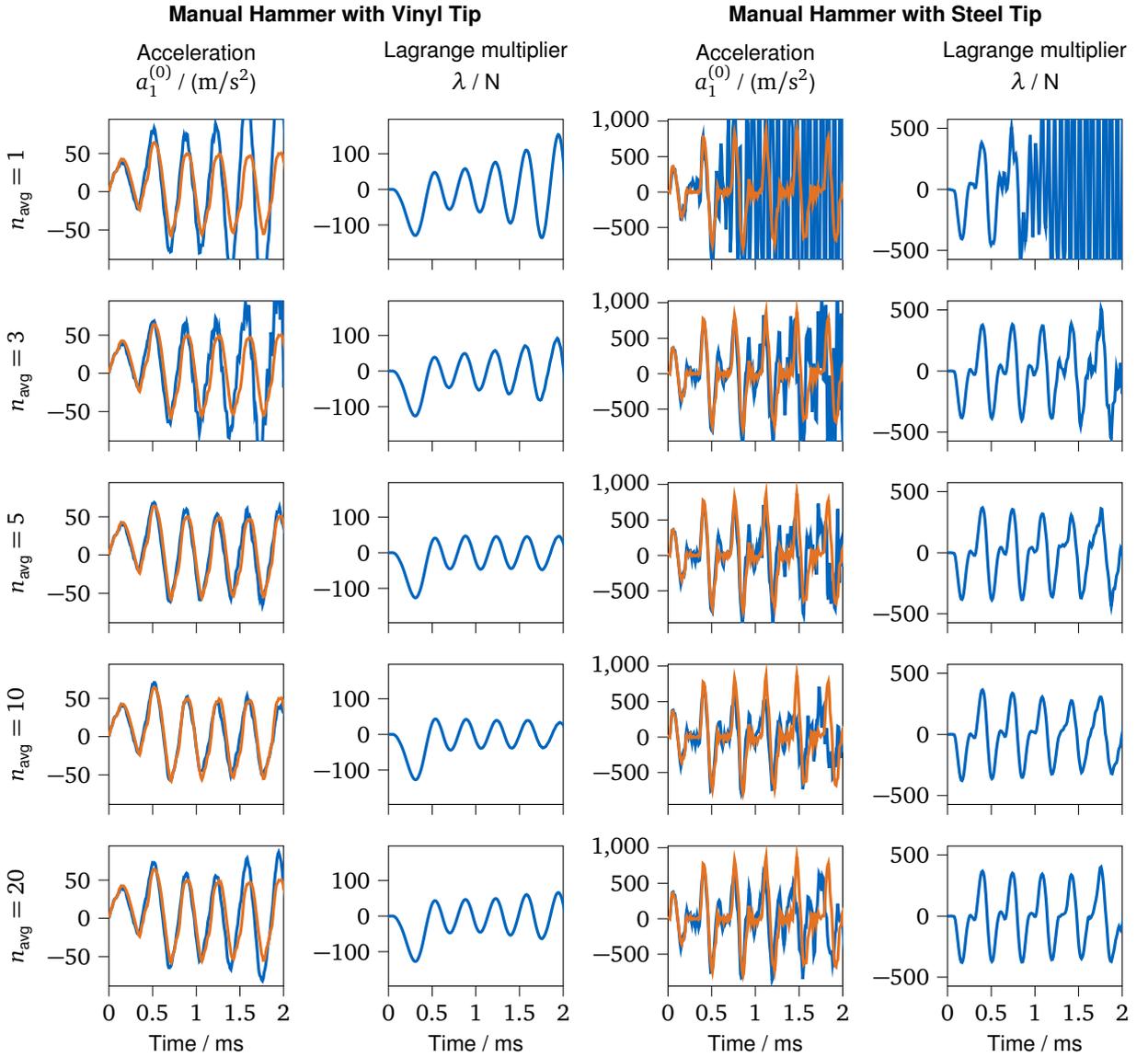}
	\end{flushright}\vspace{-0.4cm}
	\caption{Experimental IBS results of \textbf{aluminum} rods using both hammer types for varying amounts of averages used for the IRF calculation as indicated, \legendary}
	\label{fig:IBS_results_averaging}
\end{figure}

Specifically, the issue is believed to lie in the exact enforcement of the interface compatibility, which in conjunction with the erroneous IRFs leads to the IBS scheme becoming unstable. To remedy this issue, the authors propose two possibilities: Either, the interface compatibility is weakened, e.g.\ by introducing some compliance or damping in the interface, to alleviate the consequences of the erroneous IRFs, or, the IRF estimation itself is improved. Here, the latter approach is further considered, where the first step is to determine the exact cause of the IRF errors. For this, the influence of averaging on the IBS results is investigated. \Cref{fig:IBS_results_averaging} shows the results of the IBS coupling using acceleration responses of the aluminum rods for a varying amount of averages $n_\text{avg}$ used for the IRF calculation, where $n_\text{avg}$ denotes how many responses are taken from the measurement set in the order as acquired.\vspace{-0.1cm}

\vspace{-0.2cm}

For both hammer tips, it can be seen that the IBS result improves, both in terms of stability of the response and the Lagrange multiplier, with an increasing amount of averages up until $n_\text{avg}=10$. Using the full measurement set with $n_\text{avg}=20$ then yields worse results for both tips. For the steel tip results, averaging removes the high frequent oscillations of response and Lagrange multipliers as observable for $n_\text{avg}=1$ or for instance the averaged IBS results using displacements or velocities.

\newpage\restoregeometry

From \cref{fig:IBS_results_averaging} it can be concluded that some averaging of impacts is beneficial for the stability of the IBS results. Unanswered, however, remains the question of which experimental issues are improved when multiple impacts are used. While the influence of noise is reduced by averaging, noise does not appear to be the root cause of the observed instability. As noise is reduced further when more averages are performed, the IBS results should always improve with an increased amount of impacts, when noise is the sole issue. Because this is not the case and further removal of noise using a Hankel truncated SVD filter (not shown here, details can be found for instance in \cite{Fahmy2004}) did not lead to improvements, it can be inferred that, when a sufficient amount of averages are used, noise is not the only cause of the instability of the IBS results.

Since improvements can be seen with averaging, there must exist some variations between the individual impacts. Because the sensor positions are fixed, the greatest variation can be found in the impact locations and the direction, i.e.\ the angle between the tip and the impacted surface.

\begin{figure}[ht]
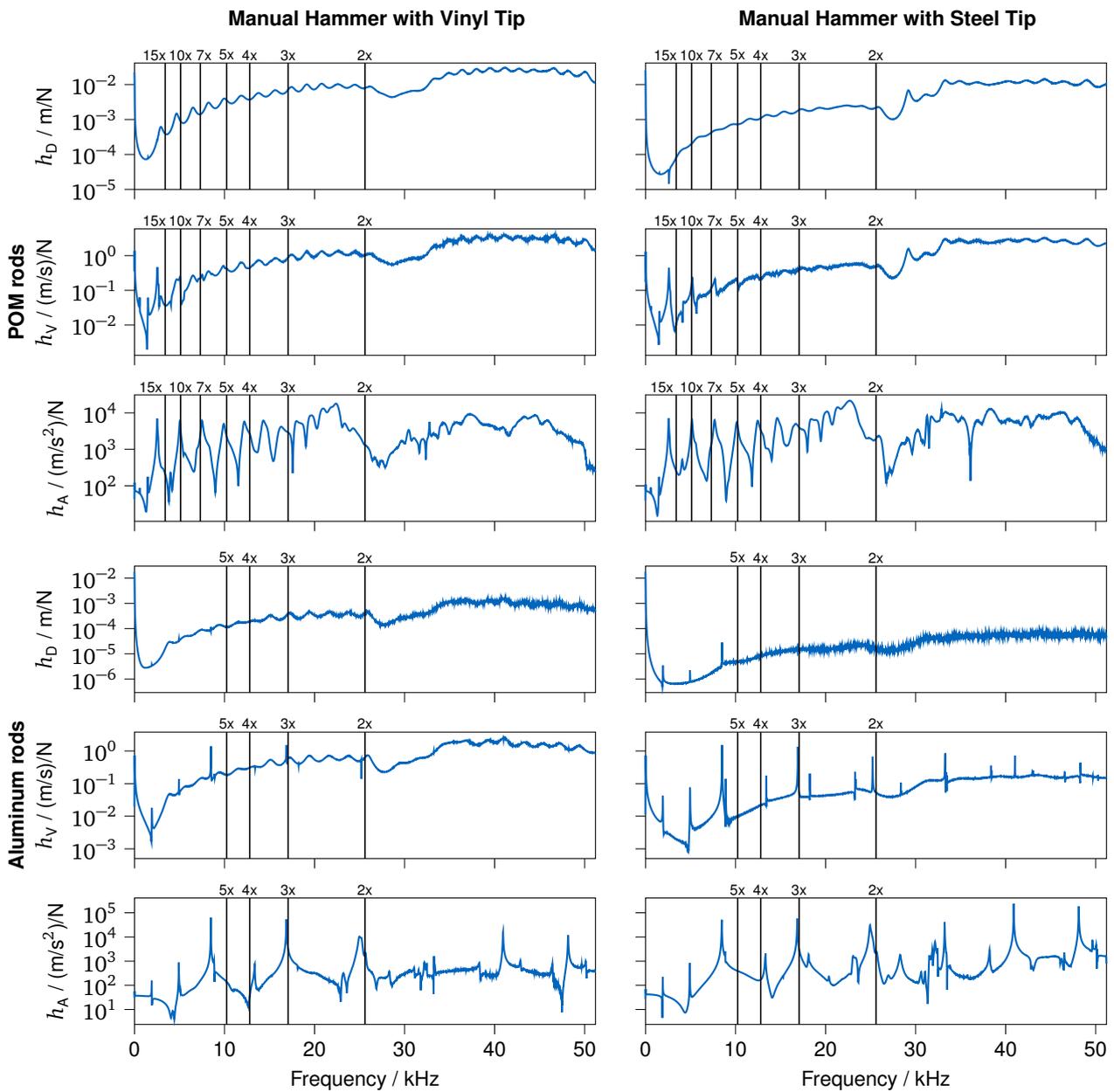

	\small
	\begin{minipage}{0.5\textwidth}
		\centering
		\hspace{2.8cm}\textbf{Manual Hammer with Vinyl Tip}
	\end{minipage}
	\begin{minipage}{0.5\textwidth}
		\centering
		\hspace{1.05cm}\textbf{Manual Hammer with Steel Tip}
	\end{minipage}\vspace{0.2cm}\\
	\begin{subfigure}{0.545\textwidth}
		\input{figures/discussion/DS_estimator_pom_manual_vinyl.tikz}
	\end{subfigure}
	\begin{subfigure}{0.545\textwidth}
		\begin{flushleft}
			\input{figures/discussion/DS_estimator_pom_manual_steel.tikz}
		\end{flushleft}
	\end{subfigure} \\
	\begin{subfigure}{0.545\textwidth}
		\input{figures/discussion/DS_estimator_alu_manual_vinyl.tikz}
	\end{subfigure}
	\begin{subfigure}{0.545\textwidth}
		\begin{flushleft}
			\input{figures/discussion/DS_estimator_alu_manual_steel.tikz}
		\end{flushleft}
	\end{subfigure}
	\caption{Frequency domain view (FFT) of the first substructure's IRFs calculated in the time domain, i.e.\ the FRFs, for both materials and hammer types as labeled. Vertical lines indicate the cut-off point in the frequency domain when a downsampling factor as indicated at the top of the line is used}
	\label{fig:DS_est}
\end{figure}

While such variations between impacts can explain the varying stability of the aluminum rod's IBS results, it does not explain why for the POM rods no satisfactory IBS result was achieved. A reason for this might be found when looking at the frequency content of the IRFs, i.e.\ FRFs, depicted in \cref{fig:DS_est}. Here, the FFT of the first substructure IRFs for all quantities, materials and hammer types is shown. 

\clearpage

Additionally, vertical lines mark the truncation point of the frequency content when downsampling with a factor as indicated on the respective line is applied. As can be seen for the POM rods, for neither combination of IBS quantity and hammer tip, there exists useful signal in the higher frequency areas. Because the acceleration amplitudes scale with the angular frequency squared, the signals contain useful information up to a higher frequency than as they do for velocities (only scaling with the angular frequencies) or displacements (no scaling). 

Since the higher frequency content does not contain any useful information and partially even shows higher amplitudes than the real system responses, e.g.\ for the displacements of the POM rods using the vinyl tip, it seems natural to remove it. This will be further explored in the next section.\vspace{-0.25cm}
% !TeX spellcheck = en_US
\subsection{Results With Removal of Higher Frequency Content}\vspace{-0.15cm}

To remove higher frequency content, two techniques as discussed in the theory section are utilized in this paper: The technique mainly focused on is downsampling, i.e.\ applying a low-pass filter and then removing sample points, because this also yields computational cost reductions. As a comparison, for each shown downsampling result, the results when only the low-pass filter with identical parameters is applied, are shown. The only difference is that no sample points are removed. Due to the specific downsampling implementation, i.e.\ removing sample points without interpolation between samples, only sampling frequencies that can be derived from the original one by dividing with a whole number can be used. Further, the truncation frequency must be high enough in order to keep the first eigenfrequency in the longitudinal direction, as then no relevant system dynamics would remain in the signals, limiting the maximum downsampling factor for the aluminum rods to 5x and for the POM rods to 15x.

Starting with the POM rods, here only the results using a manual hammer with a steel tip are shown, the results using a vinyl tip can be found in the appendix. \Cref{fig:IBS_results_DS_vs_LP_POM_steel_d} depicts the displacement-based results, where in the left column the IBS results using a low-pass filter and downsampling are shown, and the right column shows the results only using low-pass filtering. Here, an increase of the downsampling factor does not strictly lead to an improvement of the stability. The best results using downsampling can be found for a downsampling factor of 5x, which correctly predicts the first two driving point responses. In this case, only applying the low-pass filter is beneficial, where for 5x the first three amplitudes are predicted correctly and for 10x basically the complete considered response is correctly reproduced by IBS. The results with a lower or higher one of the considered downsampling factors are tremendously worse, becoming unstable very quickly.

Different behavior can be seen when using velocities, see \cref{fig:IBS_results_DS_vs_LP_POM_steel_v}, where downsampling by a factor of 4x or higher gives more or less the same results. For a factor of 5x or higher, the first two driving points are decently captured, a great improvement over the original results, which become unstable immediately. Only applying a low-pass filter gives worse results for 10x and 15x, compared to downsampling.

While for the acceleration of the POM rods, the IBS coupling seems to fail, which is noticeable in the response of the assembled system showing the eigenfrequency of the first unassembled substructure and the Lagrange multiplier being too low, downsampling by a factor of 3x enables good estimation of the first four response peaks. Further downsampling leads to an increased instability of the Lagrange multipliers.

Interestingly, when a downsampling factor of 15x is used, after approximately \SI{2}{\milli\second} the Lagrange multiplier oscillates between the exact same value, once positive, once negative, while the result with only the low-pass filter applied does not show this behavior. It is believed that at this point the time that passes between the calculation of two Lagrange multipliers is too long, causing this issue.

For the vinyl tip results (see \cref{fig:IBS_results_DS_vs_LP_POM_vinyl_d,fig:IBS_results_DS_vs_LP_POM_vinyl_v,fig:IBS_results_DS_vs_LP_POM_vinyl_a} in the appendix) basically the same observations can be made.

The achievable improvements for the aluminum rod's IBS results are smaller, most likely because the excitation bandwidth is better and therefore the signal is usable up to a higher frequency than with the POM rods.

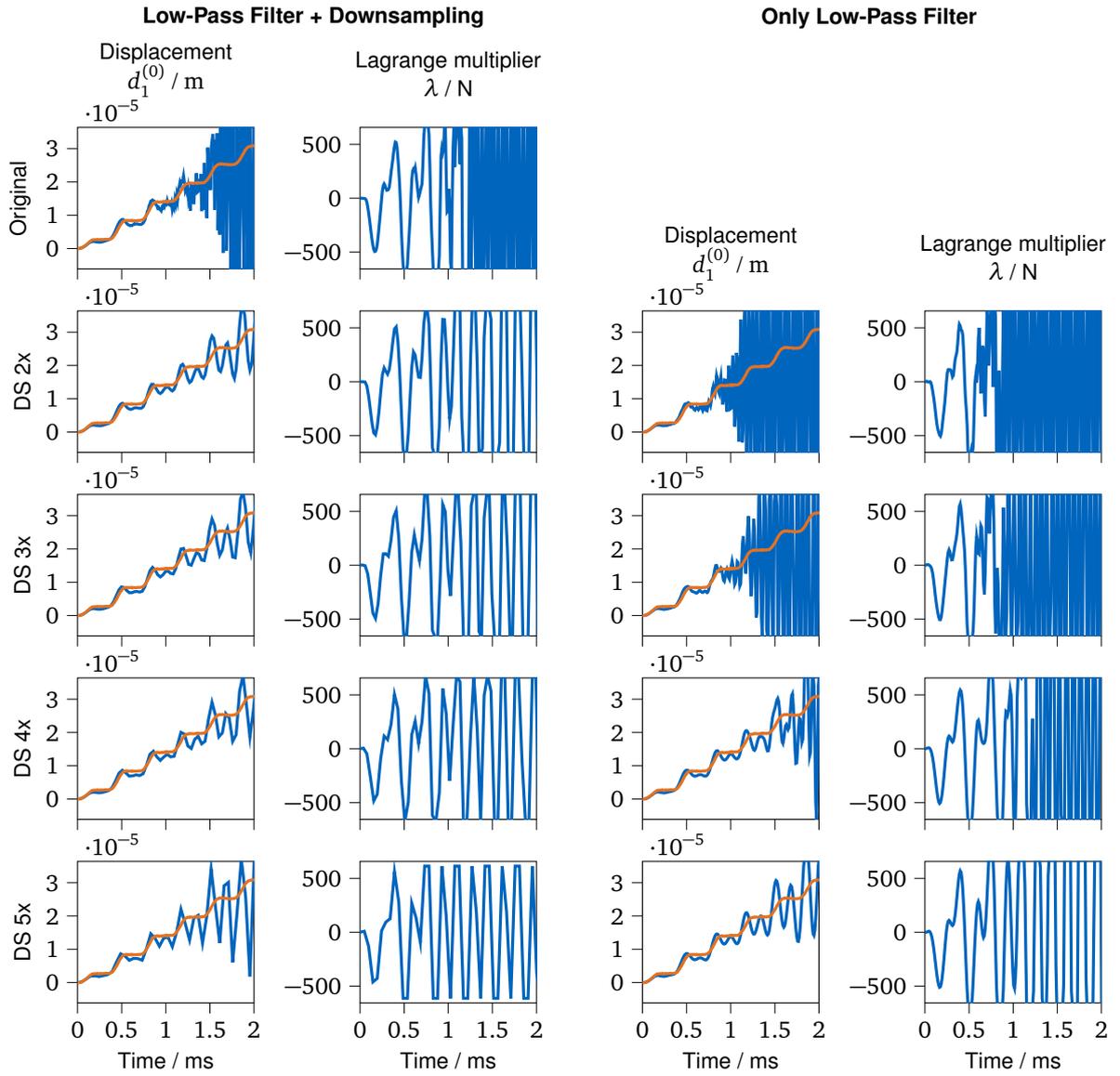
\begin{figure}[ht!]
	\small
	\renewcommand{\IBSlen}{2}
	\renewcommand{\IBSResultCSVdisp}{data/IBS_results_ENAW7075_manual_steel_d_Time.csv}
	\renewcommand{\IBSResultCSVAa}{data/ENAW7075_manual_steel_DS/IBS_results_ENAW7075_manual_steel_d_DS_2.csv}
	\renewcommand{\IBSResultCSVBa}{data/ENAW7075_manual_steel_LPDS/IBS_results_ENAW7075_manual_steel_d_LPDS_2.csv}
	\renewcommand{\IBSResultCSVAb}{data/ENAW7075_manual_steel_DS/IBS_results_ENAW7075_manual_steel_d_DS_3.csv}
	\renewcommand{\IBSResultCSVBb}{data/ENAW7075_manual_steel_LPDS/IBS_results_ENAW7075_manual_steel_d_LPDS_3.csv}
	\renewcommand{\IBSResultCSVAc}{data/ENAW7075_manual_steel_DS/IBS_results_ENAW7075_manual_steel_d_DS_4.csv}
	\renewcommand{\IBSResultCSVBc}{data/ENAW7075_manual_steel_LPDS/IBS_results_ENAW7075_manual_steel_d_LPDS_4.csv}
	\renewcommand{\IBSResultCSVAd}{data/ENAW7075_manual_steel_DS/IBS_results_ENAW7075_manual_steel_d_DS_5.csv}
	\renewcommand{\IBSResultCSVBd}{data/ENAW7075_manual_steel_LPDS/IBS_results_ENAW7075_manual_steel_d_LPDS_5.csv}
	\renewcommand{\yAmin}{-6.104633976065088e-06}
	\renewcommand{\yAmax}{3.639524549726047e-05}
	\renewcommand{\LMmin}{-656.911739006243}
	\renewcommand{\LMmax}{656.911739006243}
	\begin{minipage}{0.5\textwidth}
		\centering
		\hspace{2.8cm}\textbf{Low-Pass Filter + Downsampling}
	\end{minipage}
	\begin{minipage}{0.5\textwidth}
		\centering
		\hspace{1.4cm}\textbf{Only Low-Pass Filter}
	\end{minipage}\\\vspace{-0.6cm}
	\begin{flushright}
		% This file was created with tikzplotlib v0.10.1.
\begin{tikzpicture}

\definecolor{darkgray176}{RGB}{176,176,176}
\definecolor{gray}{RGB}{128,128,128}
\definecolor{lightgray204}{RGB}{204,204,204}

\begin{groupplot}[group style={group size=4 by 5, horizontal sep=1.5cm, vertical sep=0.6cm}]

% DS 1 - A
\nextgroupplot[
height=3.6cm,
scaled x ticks=manual:{}{\pgfmathparse{#1}},
yticklabel style={
        /pgf/number format/fixed,
        /pgf/number format/precision=3
},
tick align=outside,
tick pos=left,
title={Displacement \\\(\displaystyle d_1^{(0)}\) / \si[per-mode=symbol]{\meter}},
title style = {align = center},
title style={yshift=0.7ex},
width=0.24\textwidth,
x grid style={darkgray176},
xmin=0, xmax=\IBSlen,
ylabel={Original},
xtick style={color=black},
xticklabels={},
y grid style={darkgray176},
ymin=\yAmin, ymax=\yAmax,
ytick style={color=black}
]
\addplot [very thick, TUMBlue]
table[x=t, y=y1IBS, col sep=comma]{\IBSResultCSVdisp};

\addplot [very thick, TUMOrange]
table[x=t, y=y1Meas, col sep=comma]{\IBSResultCSVdisp};

\nextgroupplot[
height=3.6cm,
tick align=outside,
tick pos=left,
title={Lagrange multiplier \\\(\displaystyle \lambda\) / N},
title style = {align = center},
title style={yshift=0.7ex},
width=0.24\textwidth,
x grid style={darkgray176},
xmin=0, xmax=\IBSlen,
xtick style={color=black},
xticklabels={},
y grid style={darkgray176},
ymin=\LMmin, ymax=\LMmax,
ytick style={color=black}
]
\addplot [very thick, TUMBlue]
table[x=t, y=lambda, col sep=comma]{\IBSResultCSVdisp};

% DS 1 - B
\nextgroupplot[
height=3.6cm,
width=0.24\textwidth,
x grid style=none,
xticklabels={},
yticklabels={},
xtick style={draw=none},
ytick style={draw=none},
hide x axis,
hide y axis
]

\nextgroupplot[
height=3.6cm,
width=0.24\textwidth,
x grid style=none,
xticklabels={},
yticklabels={},
xtick style={draw=none},
ytick style={draw=none},
hide x axis,
hide y axis
]

% DS 2 - A
\nextgroupplot[
height=3.6cm,
scaled x ticks=manual:{}{\pgfmathparse{#1}},
yticklabel style={
        /pgf/number format/fixed,
        /pgf/number format/precision=3
},
tick align=outside,
tick pos=left,
width=0.24\textwidth,
x grid style={darkgray176},
xmin=0, xmax=\IBSlen,
ylabel={DS 2x},
xtick style={color=black},
xticklabels={},
y grid style={darkgray176},
ymin=\yAmin, ymax=\yAmax,
ytick style={color=black}
]
\addplot [very thick, TUMBlue]
table[x=t, y=y1IBS, col sep=comma]{\IBSResultCSVAa};

\addplot [very thick, TUMOrange]
table[x=t, y=y1Meas, col sep=comma]{\IBSResultCSVAa};

\nextgroupplot[
height=3.6cm,
tick align=outside,
tick pos=left,
width=0.24\textwidth,
x grid style={darkgray176},
xmin=0, xmax=\IBSlen,
xtick style={color=black},
xticklabels={},
y grid style={darkgray176},
ymin=\LMmin, ymax=\LMmax,
ytick style={color=black}
]
\addplot [very thick, TUMBlue]
table[x=t, y=lambda, col sep=comma]{\IBSResultCSVAa};

% DS 2 - B
\nextgroupplot[
height=3.6cm,
scaled x ticks=manual:{}{\pgfmathparse{#1}},
yticklabel style={
        /pgf/number format/fixed,
        /pgf/number format/precision=3
},
tick align=outside,
tick pos=left,
title={Displacement \\\(\displaystyle d_1^{(0)}\) / \si[per-mode=symbol]{\meter}},
title style = {align = center},
title style={yshift=0.7ex},
width=0.24\textwidth,
x grid style={darkgray176},
xmin=0, xmax=\IBSlen,
xtick style={color=black},
xticklabels={},
y grid style={darkgray176},
ymin=\yAmin, ymax=\yAmax,
ytick style={color=black}
]
\addplot [very thick, TUMBlue]
table[x=t, y=y1IBS, col sep=comma]{\IBSResultCSVBa};

\addplot [very thick, TUMOrange]
table[x=t, y=y1Meas, col sep=comma]{\IBSResultCSVBa};

\nextgroupplot[
height=3.6cm,
tick align=outside,
tick pos=left,
title={Lagrange multiplier \\\(\displaystyle \lambda\) / N},
title style = {align = center},
title style={yshift=0.7ex},
width=0.24\textwidth,
x grid style={darkgray176},
xmin=0, xmax=\IBSlen,
xtick style={color=black},
xticklabels={},
y grid style={darkgray176},
ymin=\LMmin, ymax=\LMmax,
ytick style={color=black}
]
\addplot [very thick, TUMBlue]
table[x=t, y=lambda, col sep=comma]{\IBSResultCSVBa};

% DS 3 - A
\nextgroupplot[
height=3.6cm,
scaled x ticks=manual:{}{\pgfmathparse{#1}},
yticklabel style={
        /pgf/number format/fixed,
        /pgf/number format/precision=3
},
tick align=outside,
tick pos=left,
width=0.24\textwidth,
x grid style={darkgray176},
xmin=0, xmax=\IBSlen,
ylabel={DS 3x},
xtick style={color=black},
xticklabels={},
y grid style={darkgray176},
ymin=\yAmin, ymax=\yAmax,
ytick style={color=black}
]
\addplot [very thick, TUMBlue]
table[x=t, y=y1IBS, col sep=comma]{\IBSResultCSVAb};

\addplot [very thick, TUMOrange]
table[x=t, y=y1Meas, col sep=comma]{\IBSResultCSVAb};

\nextgroupplot[
height=3.6cm,
tick align=outside,
tick pos=left,
width=0.24\textwidth,
x grid style={darkgray176},
xmin=0, xmax=\IBSlen,
xtick style={color=black},
xticklabels={},
y grid style={darkgray176},
ymin=\LMmin, ymax=\LMmax,
ytick style={color=black}
]
\addplot [very thick, TUMBlue]
table[x=t, y=lambda, col sep=comma]{\IBSResultCSVAb};

% DS 3 - B
\nextgroupplot[
height=3.6cm,
scaled x ticks=manual:{}{\pgfmathparse{#1}},
yticklabel style={
        /pgf/number format/fixed,
        /pgf/number format/precision=3
},
tick align=outside,
tick pos=left,
width=0.24\textwidth,
x grid style={darkgray176},
xmin=0, xmax=\IBSlen,
xtick style={color=black},
xticklabels={},
y grid style={darkgray176},
ymin=\yAmin, ymax=\yAmax,
ytick style={color=black}
]
\addplot [very thick, TUMBlue]
table[x=t, y=y1IBS, col sep=comma]{\IBSResultCSVBb};

\addplot [very thick, TUMOrange]
table[x=t, y=y1Meas, col sep=comma]{\IBSResultCSVBb};

\nextgroupplot[
height=3.6cm,
tick align=outside,
tick pos=left,
width=0.24\textwidth,
x grid style={darkgray176},
xmin=0, xmax=\IBSlen,
xtick style={color=black},
xticklabels={},
y grid style={darkgray176},
ymin=\LMmin, ymax=\LMmax,
ytick style={color=black}
]
\addplot [very thick, TUMBlue]
table[x=t, y=lambda, col sep=comma]{\IBSResultCSVBb};

% DS 4 - A
\nextgroupplot[
height=3.6cm,
scaled x ticks=manual:{}{\pgfmathparse{#1}},
yticklabel style={
        /pgf/number format/fixed,
        /pgf/number format/precision=3
},
tick align=outside,
tick pos=left,
width=0.24\textwidth,
x grid style={darkgray176},
xmin=0, xmax=\IBSlen,
ylabel={DS 4x},
xtick style={color=black},
xticklabels={},
y grid style={darkgray176},
ymin=\yAmin, ymax=\yAmax,
ytick style={color=black}
]
\addplot [very thick, TUMBlue]
table[x=t, y=y1IBS, col sep=comma]{\IBSResultCSVAc};

\addplot [very thick, TUMOrange]
table[x=t, y=y1Meas, col sep=comma]{\IBSResultCSVAc};

\nextgroupplot[
height=3.6cm,
tick align=outside,
tick pos=left,
width=0.24\textwidth,
x grid style={darkgray176},
xmin=0, xmax=\IBSlen,
xtick style={color=black},
xticklabels={},
y grid style={darkgray176},
ymin=\LMmin, ymax=\LMmax,
ytick style={color=black}
]
\addplot [very thick, TUMBlue]
table[x=t, y=lambda, col sep=comma]{\IBSResultCSVAc};

% DS 4 - B
\nextgroupplot[
height=3.6cm,
scaled x ticks=manual:{}{\pgfmathparse{#1}},
yticklabel style={
        /pgf/number format/fixed,
        /pgf/number format/precision=3
},
tick align=outside,
tick pos=left,
width=0.24\textwidth,
x grid style={darkgray176},
xmin=0, xmax=\IBSlen,
xtick style={color=black},
xticklabels={},
y grid style={darkgray176},
ymin=\yAmin, ymax=\yAmax,
ytick style={color=black}
]
\addplot [very thick, TUMBlue]
table[x=t, y=y1IBS, col sep=comma]{\IBSResultCSVBc};

\addplot [very thick, TUMOrange]
table[x=t, y=y1Meas, col sep=comma]{\IBSResultCSVBc};

\nextgroupplot[
height=3.6cm,
tick align=outside,
tick pos=left,
width=0.24\textwidth,
x grid style={darkgray176},
xmin=0, xmax=\IBSlen,
xtick style={color=black},
xticklabels={},
y grid style={darkgray176},
ymin=\LMmin, ymax=\LMmax,
ytick style={color=black}
]
\addplot [very thick, TUMBlue]
table[x=t, y=lambda, col sep=comma]{\IBSResultCSVBc};

% DS 5 - A
\nextgroupplot[
height=3.6cm,
scaled x ticks=manual:{}{\pgfmathparse{#1}},
yticklabel style={
        /pgf/number format/fixed,
        /pgf/number format/precision=3
},
tick align=outside,
tick pos=left,
width=0.24\textwidth,
x grid style={darkgray176},
xmin=0, xmax=\IBSlen,
ylabel={DS 5x},
xtick style={color=black},
xlabel={Time / ms},
y grid style={darkgray176},
ymin=\yAmin, ymax=\yAmax,
ytick style={color=black}
]
\addplot [very thick, TUMBlue]
table[x=t, y=y1IBS, col sep=comma]{\IBSResultCSVAd};

\addplot [very thick, TUMOrange]
table[x=t, y=y1Meas, col sep=comma]{\IBSResultCSVAd};

\nextgroupplot[
height=3.6cm,
tick align=outside,
tick pos=left,
width=0.24\textwidth,
x grid style={darkgray176},
xmin=0, xmax=\IBSlen,
xtick style={color=black},
xlabel={Time / ms},
y grid style={darkgray176},
ymin=\LMmin, ymax=\LMmax,
ytick style={color=black}
]
\addplot [very thick, TUMBlue]
table[x=t, y=lambda, col sep=comma]{\IBSResultCSVAd};

% DS 5 - B
\nextgroupplot[
height=3.6cm,
scaled x ticks=manual:{}{\pgfmathparse{#1}},
yticklabel style={
        /pgf/number format/fixed,
        /pgf/number format/precision=3
},
tick align=outside,
tick pos=left,
width=0.24\textwidth,
x grid style={darkgray176},
xmin=0, xmax=\IBSlen,
xtick style={color=black},
xlabel={Time / ms},
y grid style={darkgray176},
ymin=\yAmin, ymax=\yAmax,
ytick style={color=black}
]
\addplot [very thick, TUMBlue]
table[x=t, y=y1IBS, col sep=comma]{\IBSResultCSVBd};

\addplot [very thick, TUMOrange]
table[x=t, y=y1Meas, col sep=comma]{\IBSResultCSVBd};

\nextgroupplot[
height=3.6cm,
tick align=outside,
tick pos=left,
width=0.24\textwidth,
x grid style={darkgray176},
xmin=0, xmax=\IBSlen,
xtick style={color=black},
xlabel={Time / ms},
y grid style={darkgray176},
ymin=\LMmin, ymax=\LMmax,
ytick style={color=black}
]
\addplot [very thick, TUMBlue]
table[x=t, y=lambda, col sep=comma]{\IBSResultCSVBd};

\end{groupplot}
\end{tikzpicture}
	\end{flushright}\vspace{-0.4cm}
	\caption{Comparison of experimental results using a low-pass filter with downsampling versus only low-pass filtering, applied to \textbf{displacements} of \textbf{aluminum} rods using a manual hammer with \textbf{steel tip}, \legendary}
	\label{fig:IBS_results_DS_vs_LP_ENAW7075_steel_d}
\end{figure}

For both hammer types, the displacement results can be improved, where for the steel tip (\cref{fig:IBS_results_DS_vs_LP_ENAW7075_steel_d}) mainly the high frequent oscillations are removed using downsampling. In all cases, the response and the Lagrange multiplier become unstable towards the end of the viewed response. For downsampling by a factor of 5x, the same oscillatory behavior of the Lagrange multiplier as observed with the accelerations of the POM rods and 15x can be seen.

No meaningful improvement could be achieved for the velocity-based results, therefore not shown below.

Lastly, for the accelerations using a vinyl tip (\cref{fig:IBS_results_DS_vs_LP_ENAW7075_vinyl_a}), an increase of downsampling factor leads to a reduction of the growing amplitudes of both, the driving point response and the Lagrange multiplier. Using only a low-pass filter does not yield the same stability improvements and the results of the various factors are less distinct. For the results using a steel tip, a reduction of the considered frequency bandwidth either gave very similar or worse results in terms of stability. Applying downsampling by a factor of 2x already removes significant dynamics observable in the reference measurement, leading to an underestimation of the second response peak.

\begin{figure}[ht!]
	\small
	\renewcommand{\IBSlen}{8}
	\renewcommand{\IBSResultCSVdisp}{data/IBS_results_POM_manual_steel_d_Time.csv}
	\renewcommand{\IBSResultCSVAa}{data/POM_manual_steel_DS/IBS_results_POM_manual_steel_d_DS_2.csv}
	\renewcommand{\IBSResultCSVBa}{data/POM_manual_steel_LPDS/IBS_results_POM_manual_steel_d_LPDS_2.csv}
	\renewcommand{\IBSResultCSVAb}{data/POM_manual_steel_DS/IBS_results_POM_manual_steel_d_DS_3.csv}
	\renewcommand{\IBSResultCSVBb}{data/POM_manual_steel_LPDS/IBS_results_POM_manual_steel_d_LPDS_3.csv}
	\renewcommand{\IBSResultCSVAc}{data/POM_manual_steel_DS/IBS_results_POM_manual_steel_d_DS_4.csv}
	\renewcommand{\IBSResultCSVBc}{data/POM_manual_steel_LPDS/IBS_results_POM_manual_steel_d_LPDS_4.csv}
	\renewcommand{\IBSResultCSVAd}{data/POM_manual_steel_DS/IBS_results_POM_manual_steel_d_DS_5.csv}
	\renewcommand{\IBSResultCSVBd}{data/POM_manual_steel_LPDS/IBS_results_POM_manual_steel_d_LPDS_5.csv}
	\renewcommand{\IBSResultCSVAe}{data/POM_manual_steel_DS/IBS_results_POM_manual_steel_d_DS_7.csv}
	\renewcommand{\IBSResultCSVBe}{data/POM_manual_steel_LPDS/IBS_results_POM_manual_steel_d_LPDS_7.csv}
	\renewcommand{\IBSResultCSVAf}{data/POM_manual_steel_DS/IBS_results_POM_manual_steel_d_DS_10.csv}
	\renewcommand{\IBSResultCSVBf}{data/POM_manual_steel_LPDS/IBS_results_POM_manual_steel_d_LPDS_10.csv}
	\renewcommand{\IBSResultCSVAg}{data/POM_manual_steel_DS/IBS_results_POM_manual_steel_d_DS_15.csv}
	\renewcommand{\IBSResultCSVBg}{data/POM_manual_steel_LPDS/IBS_results_POM_manual_steel_d_LPDS_15.csv}
	\renewcommand{\yAmin}{-4.653330842847936e-05}
	\renewcommand{\yAmax}{0.0002786072576564038}
	\renewcommand{\LMmin}{-371.51661247466654}
	\renewcommand{\LMmax}{371.51661247466654}
	\begin{minipage}{0.5\textwidth}
		\centering
		\hspace{2.8cm}\textbf{Low-Pass Filter + Downsampling}
	\end{minipage}
	\begin{minipage}{0.5\textwidth}
		\centering
		\hspace{1.4cm}\textbf{Only Low-Pass Filter}
	\end{minipage}\\\vspace{-0.6cm}
	\begin{flushright}
		\input{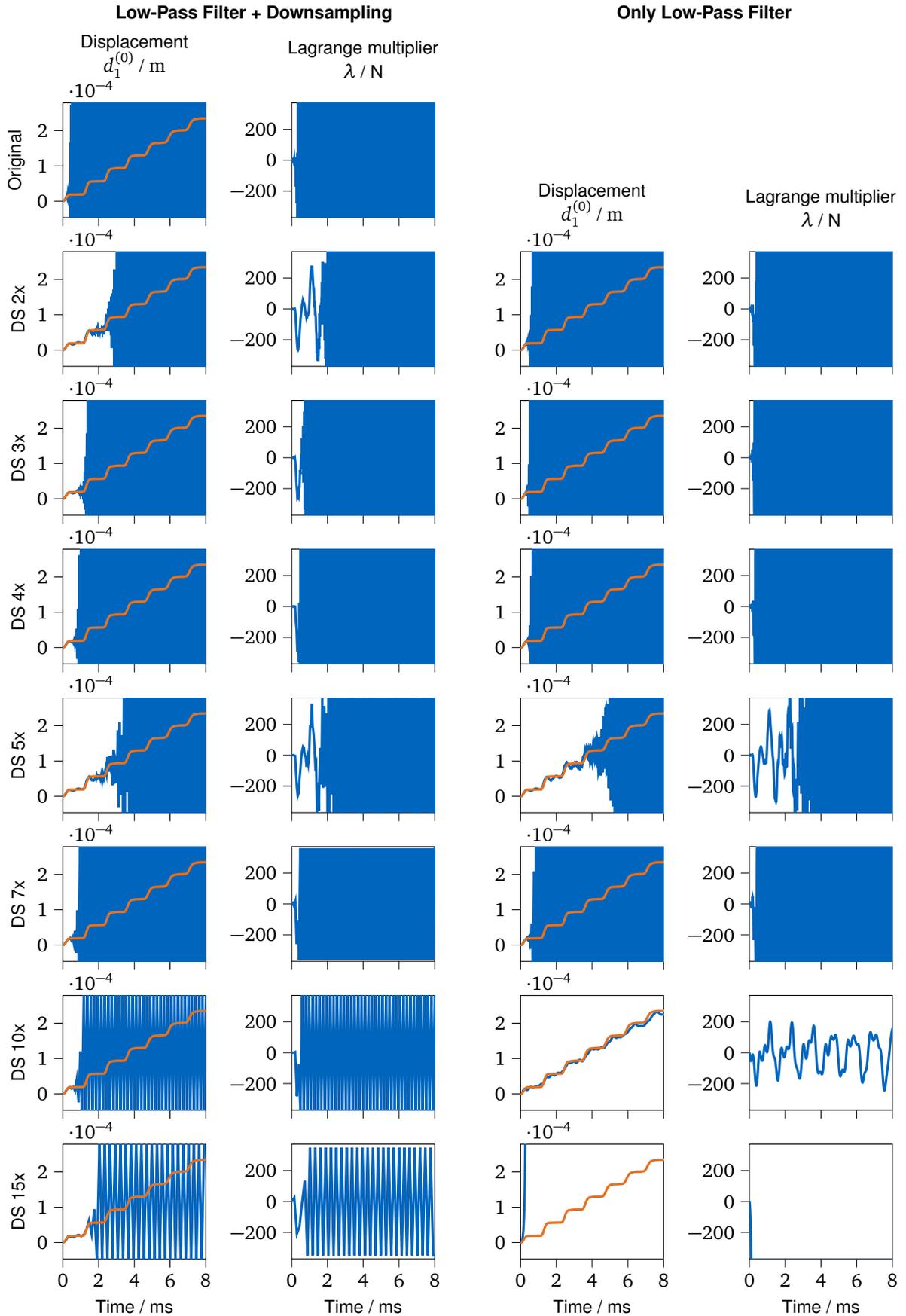}
	\end{flushright}\vspace{-0.4cm}
	\caption{Comparison of experimental results using a low-pass filter with downsampling versus only low-pass filtering, applied to \textbf{displacements} of \textbf{POM} rods using a manual hammer with \textbf{steel tip}, \legendary}
	\label{fig:IBS_results_DS_vs_LP_POM_steel_d}
\end{figure}

\begin{figure}[ht!]
	\small
	\renewcommand{\IBSlen}{8}
	\renewcommand{\IBSResultCSVvelo}{data/IBS_results_POM_manual_steel_v_Time.csv}
	\renewcommand{\IBSResultCSVAa}{data/POM_manual_steel_DS/IBS_results_POM_manual_steel_v_DS_2.csv}
	\renewcommand{\IBSResultCSVBa}{data/POM_manual_steel_LPDS/IBS_results_POM_manual_steel_v_LPDS_2.csv}
	\renewcommand{\IBSResultCSVAb}{data/POM_manual_steel_DS/IBS_results_POM_manual_steel_v_DS_3.csv}
	\renewcommand{\IBSResultCSVBb}{data/POM_manual_steel_LPDS/IBS_results_POM_manual_steel_v_LPDS_3.csv}
	\renewcommand{\IBSResultCSVAc}{data/POM_manual_steel_DS/IBS_results_POM_manual_steel_v_DS_4.csv}
	\renewcommand{\IBSResultCSVBc}{data/POM_manual_steel_LPDS/IBS_results_POM_manual_steel_v_LPDS_4.csv}
	\renewcommand{\IBSResultCSVAd}{data/POM_manual_steel_DS/IBS_results_POM_manual_steel_v_DS_5.csv}
	\renewcommand{\IBSResultCSVBd}{data/POM_manual_steel_LPDS/IBS_results_POM_manual_steel_v_LPDS_5.csv}
	\renewcommand{\IBSResultCSVAe}{data/POM_manual_steel_DS/IBS_results_POM_manual_steel_v_DS_7.csv}
	\renewcommand{\IBSResultCSVBe}{data/POM_manual_steel_LPDS/IBS_results_POM_manual_steel_v_LPDS_7.csv}
	\renewcommand{\IBSResultCSVAf}{data/POM_manual_steel_DS/IBS_results_POM_manual_steel_v_DS_10.csv}
	\renewcommand{\IBSResultCSVBf}{data/POM_manual_steel_LPDS/IBS_results_POM_manual_steel_v_LPDS_10.csv}
	\renewcommand{\IBSResultCSVAg}{data/POM_manual_steel_DS/IBS_results_POM_manual_steel_v_DS_15.csv}
	\renewcommand{\IBSResultCSVBg}{data/POM_manual_steel_LPDS/IBS_results_POM_manual_steel_v_LPDS_15.csv}
	\renewcommand{\yAmin}{-0.07895606840029358}
	\renewcommand{\yAmax}{0.23359584733843802}
	\renewcommand{\LMmin}{-371.51661247466654}
	\renewcommand{\LMmax}{371.51661247466654}
	\begin{minipage}{0.5\textwidth}
		\centering
		\hspace{2.8cm}\textbf{Low-Pass Filter + Downsampling}
	\end{minipage}
	\begin{minipage}{0.5\textwidth}
		\centering
		\hspace{1.4cm}\textbf{Only Low-Pass Filter}
	\end{minipage}\\\vspace{-0.6cm}
	\begin{flushright}
		\input{figures/experimental/IBS_results_DS_vs_LP_POM_velo.tikz}
	\end{flushright}\vspace{-0.4cm}
	\caption{Comparison of experimental results using a low-pass filter with downsampling versus only low-pass filtering, applied to \textbf{velocities} of \textbf{POM} rods using a manual hammer with \textbf{steel tip}, \legendary}
	\label{fig:IBS_results_DS_vs_LP_POM_steel_v}
\end{figure}

\begin{figure}[ht!]
	\small
	\renewcommand{\IBSlen}{8}
	\renewcommand{\IBSResultCSVacc}{data/IBS_results_POM_manual_steel_a_Time.csv}
	\renewcommand{\IBSResultCSVAa}{data/POM_manual_steel_DS/IBS_results_POM_manual_steel_a_DS_2.csv}
	\renewcommand{\IBSResultCSVBa}{data/POM_manual_steel_LPDS/IBS_results_POM_manual_steel_a_LPDS_2.csv}
	\renewcommand{\IBSResultCSVAb}{data/POM_manual_steel_DS/IBS_results_POM_manual_steel_a_DS_3.csv}
	\renewcommand{\IBSResultCSVBb}{data/POM_manual_steel_LPDS/IBS_results_POM_manual_steel_a_LPDS_3.csv}
	\renewcommand{\IBSResultCSVAc}{data/POM_manual_steel_DS/IBS_results_POM_manual_steel_a_DS_4.csv}
	\renewcommand{\IBSResultCSVBc}{data/POM_manual_steel_LPDS/IBS_results_POM_manual_steel_a_LPDS_4.csv}
	\renewcommand{\IBSResultCSVAd}{data/POM_manual_steel_DS/IBS_results_POM_manual_steel_a_DS_5.csv}
	\renewcommand{\IBSResultCSVBd}{data/POM_manual_steel_LPDS/IBS_results_POM_manual_steel_a_LPDS_5.csv}
	\renewcommand{\IBSResultCSVAe}{data/POM_manual_steel_DS/IBS_results_POM_manual_steel_a_DS_7.csv}
	\renewcommand{\IBSResultCSVBe}{data/POM_manual_steel_LPDS/IBS_results_POM_manual_steel_a_LPDS_7.csv}
	\renewcommand{\IBSResultCSVAf}{data/POM_manual_steel_DS/IBS_results_POM_manual_steel_a_DS_10.csv}
	\renewcommand{\IBSResultCSVBf}{data/POM_manual_steel_LPDS/IBS_results_POM_manual_steel_a_LPDS_10.csv}
	\renewcommand{\IBSResultCSVAg}{data/POM_manual_steel_DS/IBS_results_POM_manual_steel_a_DS_15.csv}
	\renewcommand{\IBSResultCSVBg}{data/POM_manual_steel_LPDS/IBS_results_POM_manual_steel_a_LPDS_15.csv}
	\renewcommand{\yAmin}{-1629.6189953613282}
	\renewcommand{\yAmax}{2013.026290283203}
	\renewcommand{\LMmin}{-403.0341815185547}
	\renewcommand{\LMmax}{403.0341815185547}
	\begin{minipage}{0.5\textwidth}
		\centering
		\hspace{2.8cm}\textbf{Low-Pass Filter + Downsampling}
	\end{minipage}
	\begin{minipage}{0.5\textwidth}
		\centering
		\hspace{1.4cm}\textbf{Only Low-Pass Filter}
	\end{minipage}\\\vspace{-0.6cm}
	\begin{flushright}
		\input{figures/experimental/IBS_results_DS_vs_LP_POM_acc.tikz}
	\end{flushright}\vspace{-0.4cm}
	\caption{Comparison of experimental results using a low-pass filter with downsampling versus only low-pass filtering, applied to \textbf{accelerations} of \textbf{POM} rods using a manual hammer with \textbf{steel tip}, \legendary}
	\label{fig:IBS_results_DS_vs_LP_POM_steel_a}
\end{figure}

\begin{figure}[ht!]
	\small
	\renewcommand{\IBSlen}{2}
	\renewcommand{\IBSResultCSVacc}{data/IBS_results_ENAW7075_manual_vinyl_a_Time.csv}
	\renewcommand{\IBSResultCSVAa}{data/ENAW7075_manual_vinyl_DS/IBS_results_ENAW7075_manual_vinyl_a_DS_2.csv}
	\renewcommand{\IBSResultCSVBa}{data/ENAW7075_manual_vinyl_LPDS/IBS_results_ENAW7075_manual_vinyl_a_LPDS_2.csv}
	\renewcommand{\IBSResultCSVAb}{data/ENAW7075_manual_vinyl_DS/IBS_results_ENAW7075_manual_vinyl_a_DS_3.csv}
	\renewcommand{\IBSResultCSVBb}{data/ENAW7075_manual_vinyl_LPDS/IBS_results_ENAW7075_manual_vinyl_a_LPDS_3.csv}
	\renewcommand{\IBSResultCSVAc}{data/ENAW7075_manual_vinyl_DS/IBS_results_ENAW7075_manual_vinyl_a_DS_4.csv}
	\renewcommand{\IBSResultCSVBc}{data/ENAW7075_manual_vinyl_LPDS/IBS_results_ENAW7075_manual_vinyl_a_LPDS_4.csv}
	\renewcommand{\IBSResultCSVAd}{data/ENAW7075_manual_vinyl_DS/IBS_results_ENAW7075_manual_vinyl_a_DS_5.csv}
	\renewcommand{\IBSResultCSVBd}{data/ENAW7075_manual_vinyl_LPDS/IBS_results_ENAW7075_manual_vinyl_a_LPDS_5.csv}
	\renewcommand{\yAmin}{-88.58306671142579}
	\renewcommand{\yAmax}{94.72701667785644}
	\renewcommand{\LMmin}{-194.3283048905131}
	\renewcommand{\LMmax}{194.3283048905131}
	\begin{minipage}{0.5\textwidth}
		\centering
		\hspace{2.8cm}\textbf{Low-Pass Filter + Downsampling}
	\end{minipage}
	\begin{minipage}{0.5\textwidth}
		\centering
		\hspace{1.4cm}\textbf{Only Low-Pass Filter}
	\end{minipage}\\\vspace{-0.6cm}
	\begin{flushright}
		% This file was created with tikzplotlib v0.10.1.
\begin{tikzpicture}

\definecolor{darkgray176}{RGB}{176,176,176}
\definecolor{gray}{RGB}{128,128,128}
\definecolor{lightgray204}{RGB}{204,204,204}

\begin{groupplot}[group style={group size=4 by 5, horizontal sep=1.5cm, vertical sep=0.6cm}]

% DS 1 - A
\nextgroupplot[
height=3.6cm,
scaled x ticks=manual:{}{\pgfmathparse{#1}},
yticklabel style={
        /pgf/number format/fixed,
        /pgf/number format/precision=3
},
tick align=outside,
tick pos=left,
title={Acceleration \\\(\displaystyle a_1^{(0)}\) / (\si[per-mode=symbol]{\meter\per\second\squared})},
title style = {align = center},
title style={yshift=0.7ex},
width=0.24\textwidth,
x grid style={darkgray176},
xmin=0, xmax=\IBSlen,
ylabel={Original},
xtick style={color=black},
xticklabels={},
y grid style={darkgray176},
ymin=\yAmin, ymax=\yAmax,
ytick style={color=black}
]
\addplot [very thick, TUMBlue]
table[x=t, y=y1IBS, col sep=comma]{\IBSResultCSVacc};

\addplot [very thick, TUMOrange]
table[x=t, y=y1Meas, col sep=comma]{\IBSResultCSVacc};

\nextgroupplot[
height=3.6cm,
tick align=outside,
tick pos=left,
title={Lagrange multiplier \\\(\displaystyle \lambda\) / N},
title style = {align = center},
title style={yshift=0.7ex},
width=0.24\textwidth,
x grid style={darkgray176},
xmin=0, xmax=\IBSlen,
xtick style={color=black},
xticklabels={},
y grid style={darkgray176},
ymin=\LMmin, ymax=\LMmax,
ytick style={color=black}
]
\addplot [very thick, TUMBlue]
table[x=t, y=lambda, col sep=comma]{\IBSResultCSVacc};

% DS 1 - B
\nextgroupplot[
height=3.6cm,
width=0.24\textwidth,
x grid style=none,
xticklabels={},
yticklabels={},
xtick style={draw=none},
ytick style={draw=none},
hide x axis,
hide y axis
]

\nextgroupplot[
height=3.6cm,
width=0.24\textwidth,
x grid style=none,
xticklabels={},
yticklabels={},
xtick style={draw=none},
ytick style={draw=none},
hide x axis,
hide y axis
]

% DS 2 - A
\nextgroupplot[
height=3.6cm,
scaled x ticks=manual:{}{\pgfmathparse{#1}},
yticklabel style={
        /pgf/number format/fixed,
        /pgf/number format/precision=3
},
tick align=outside,
tick pos=left,
width=0.24\textwidth,
x grid style={darkgray176},
xmin=0, xmax=\IBSlen,
ylabel={DS 2x},
xtick style={color=black},
xticklabels={},
y grid style={darkgray176},
ymin=\yAmin, ymax=\yAmax,
ytick style={color=black}
]
\addplot [very thick, TUMBlue]
table[x=t, y=y1IBS, col sep=comma]{\IBSResultCSVAa};

\addplot [very thick, TUMOrange]
table[x=t, y=y1Meas, col sep=comma]{\IBSResultCSVAa};

\nextgroupplot[
height=3.6cm,
tick align=outside,
tick pos=left,
width=0.24\textwidth,
x grid style={darkgray176},
xmin=0, xmax=\IBSlen,
xtick style={color=black},
xticklabels={},
y grid style={darkgray176},
ymin=\LMmin, ymax=\LMmax,
ytick style={color=black}
]
\addplot [very thick, TUMBlue]
table[x=t, y=lambda, col sep=comma]{\IBSResultCSVAa};

% DS 2 - B
\nextgroupplot[
height=3.6cm,
scaled x ticks=manual:{}{\pgfmathparse{#1}},
yticklabel style={
        /pgf/number format/fixed,
        /pgf/number format/precision=3
},
tick align=outside,
tick pos=left,
title={Acceleration \\\(\displaystyle a_1^{(0)}\) / (\si[per-mode=symbol]{\meter\per\second\squared})},
title style = {align = center},
title style={yshift=0.7ex},
width=0.24\textwidth,
x grid style={darkgray176},
xmin=0, xmax=\IBSlen,
xtick style={color=black},
xticklabels={},
y grid style={darkgray176},
ymin=\yAmin, ymax=\yAmax,
ytick style={color=black}
]
\addplot [very thick, TUMBlue]
table[x=t, y=y1IBS, col sep=comma]{\IBSResultCSVBa};

\addplot [very thick, TUMOrange]
table[x=t, y=y1Meas, col sep=comma]{\IBSResultCSVBa};

\nextgroupplot[
height=3.6cm,
tick align=outside,
tick pos=left,
title={Lagrange multiplier \\\(\displaystyle \lambda\) / N},
title style = {align = center},
title style={yshift=0.7ex},
width=0.24\textwidth,
x grid style={darkgray176},
xmin=0, xmax=\IBSlen,
xtick style={color=black},
xticklabels={},
y grid style={darkgray176},
ymin=\LMmin, ymax=\LMmax,
ytick style={color=black}
]
\addplot [very thick, TUMBlue]
table[x=t, y=lambda, col sep=comma]{\IBSResultCSVBa};

% DS 3 - A
\nextgroupplot[
height=3.6cm,
scaled x ticks=manual:{}{\pgfmathparse{#1}},
yticklabel style={
        /pgf/number format/fixed,
        /pgf/number format/precision=3
},
tick align=outside,
tick pos=left,
width=0.24\textwidth,
x grid style={darkgray176},
xmin=0, xmax=\IBSlen,
ylabel={DS 3x},
xtick style={color=black},
xticklabels={},
y grid style={darkgray176},
ymin=\yAmin, ymax=\yAmax,
ytick style={color=black}
]
\addplot [very thick, TUMBlue]
table[x=t, y=y1IBS, col sep=comma]{\IBSResultCSVAb};

\addplot [very thick, TUMOrange]
table[x=t, y=y1Meas, col sep=comma]{\IBSResultCSVAb};

\nextgroupplot[
height=3.6cm,
tick align=outside,
tick pos=left,
width=0.24\textwidth,
x grid style={darkgray176},
xmin=0, xmax=\IBSlen,
xtick style={color=black},
xticklabels={},
y grid style={darkgray176},
ymin=\LMmin, ymax=\LMmax,
ytick style={color=black}
]
\addplot [very thick, TUMBlue]
table[x=t, y=lambda, col sep=comma]{\IBSResultCSVAb};

% DS 3 - B
\nextgroupplot[
height=3.6cm,
scaled x ticks=manual:{}{\pgfmathparse{#1}},
yticklabel style={
        /pgf/number format/fixed,
        /pgf/number format/precision=3
},
tick align=outside,
tick pos=left,
width=0.24\textwidth,
x grid style={darkgray176},
xmin=0, xmax=\IBSlen,
xtick style={color=black},
xticklabels={},
y grid style={darkgray176},
ymin=\yAmin, ymax=\yAmax,
ytick style={color=black}
]
\addplot [very thick, TUMBlue]
table[x=t, y=y1IBS, col sep=comma]{\IBSResultCSVBb};

\addplot [very thick, TUMOrange]
table[x=t, y=y1Meas, col sep=comma]{\IBSResultCSVBb};

\nextgroupplot[
height=3.6cm,
tick align=outside,
tick pos=left,
width=0.24\textwidth,
x grid style={darkgray176},
xmin=0, xmax=\IBSlen,
xtick style={color=black},
xticklabels={},
y grid style={darkgray176},
ymin=\LMmin, ymax=\LMmax,
ytick style={color=black}
]
\addplot [very thick, TUMBlue]
table[x=t, y=lambda, col sep=comma]{\IBSResultCSVBb};

% DS 4 - A
\nextgroupplot[
height=3.6cm,
scaled x ticks=manual:{}{\pgfmathparse{#1}},
yticklabel style={
        /pgf/number format/fixed,
        /pgf/number format/precision=3
},
tick align=outside,
tick pos=left,
width=0.24\textwidth,
x grid style={darkgray176},
xmin=0, xmax=\IBSlen,
ylabel={DS 4x},
xtick style={color=black},
xticklabels={},
y grid style={darkgray176},
ymin=\yAmin, ymax=\yAmax,
ytick style={color=black}
]
\addplot [very thick, TUMBlue]
table[x=t, y=y1IBS, col sep=comma]{\IBSResultCSVAc};

\addplot [very thick, TUMOrange]
table[x=t, y=y1Meas, col sep=comma]{\IBSResultCSVAc};

\nextgroupplot[
height=3.6cm,
tick align=outside,
tick pos=left,
width=0.24\textwidth,
x grid style={darkgray176},
xmin=0, xmax=\IBSlen,
xtick style={color=black},
xticklabels={},
y grid style={darkgray176},
ymin=\LMmin, ymax=\LMmax,
ytick style={color=black}
]
\addplot [very thick, TUMBlue]
table[x=t, y=lambda, col sep=comma]{\IBSResultCSVAc};

% DS 4 - B
\nextgroupplot[
height=3.6cm,
scaled x ticks=manual:{}{\pgfmathparse{#1}},
yticklabel style={
        /pgf/number format/fixed,
        /pgf/number format/precision=3
},
tick align=outside,
tick pos=left,
width=0.24\textwidth,
x grid style={darkgray176},
xmin=0, xmax=\IBSlen,
xtick style={color=black},
xticklabels={},
y grid style={darkgray176},
ymin=\yAmin, ymax=\yAmax,
ytick style={color=black}
]
\addplot [very thick, TUMBlue]
table[x=t, y=y1IBS, col sep=comma]{\IBSResultCSVBc};

\addplot [very thick, TUMOrange]
table[x=t, y=y1Meas, col sep=comma]{\IBSResultCSVBc};

\nextgroupplot[
height=3.6cm,
tick align=outside,
tick pos=left,
width=0.24\textwidth,
x grid style={darkgray176},
xmin=0, xmax=\IBSlen,
xtick style={color=black},
xticklabels={},
y grid style={darkgray176},
ymin=\LMmin, ymax=\LMmax,
ytick style={color=black}
]
\addplot [very thick, TUMBlue]
table[x=t, y=lambda, col sep=comma]{\IBSResultCSVBc};

% DS 5 - A
\nextgroupplot[
height=3.6cm,
scaled x ticks=manual:{}{\pgfmathparse{#1}},
yticklabel style={
        /pgf/number format/fixed,
        /pgf/number format/precision=3
},
tick align=outside,
tick pos=left,
width=0.24\textwidth,
x grid style={darkgray176},
xmin=0, xmax=\IBSlen,
ylabel={DS 5x},
xtick style={color=black},
xlabel={Time / ms},
y grid style={darkgray176},
ymin=\yAmin, ymax=\yAmax,
ytick style={color=black}
]
\addplot [very thick, TUMBlue]
table[x=t, y=y1IBS, col sep=comma]{\IBSResultCSVAd};

\addplot [very thick, TUMOrange]
table[x=t, y=y1Meas, col sep=comma]{\IBSResultCSVAd};

\nextgroupplot[
height=3.6cm,
tick align=outside,
tick pos=left,
width=0.24\textwidth,
x grid style={darkgray176},
xmin=0, xmax=\IBSlen,
xtick style={color=black},
xlabel={Time / ms},
y grid style={darkgray176},
ymin=\LMmin, ymax=\LMmax,
ytick style={color=black}
]
\addplot [very thick, TUMBlue]
table[x=t, y=lambda, col sep=comma]{\IBSResultCSVAd};

% DS 5 - B
\nextgroupplot[
height=3.6cm,
scaled x ticks=manual:{}{\pgfmathparse{#1}},
yticklabel style={
        /pgf/number format/fixed,
        /pgf/number format/precision=3
},
tick align=outside,
tick pos=left,
width=0.24\textwidth,
x grid style={darkgray176},
xmin=0, xmax=\IBSlen,
xtick style={color=black},
xlabel={Time / ms},
y grid style={darkgray176},
ymin=\yAmin, ymax=\yAmax,
ytick style={color=black}
]
\addplot [very thick, TUMBlue]
table[x=t, y=y1IBS, col sep=comma]{\IBSResultCSVBd};

\addplot [very thick, TUMOrange]
table[x=t, y=y1Meas, col sep=comma]{\IBSResultCSVBd};

\nextgroupplot[
height=3.6cm,
tick align=outside,
tick pos=left,
width=0.24\textwidth,
x grid style={darkgray176},
xmin=0, xmax=\IBSlen,
xtick style={color=black},
xlabel={Time / ms},
y grid style={darkgray176},
ymin=\LMmin, ymax=\LMmax,
ytick style={color=black}
]
\addplot [very thick, TUMBlue]
table[x=t, y=lambda, col sep=comma]{\IBSResultCSVBd};

\end{groupplot}
\end{tikzpicture}
	\end{flushright}\vspace{-0.4cm}
	\caption{Comparison of experimental results using a low-pass filter with downsampling versus only low-pass filtering, applied to \textbf{accelerations} of \textbf{aluminum} rods using a manual hammer with \textbf{vinyl tip}, \legendary}
	\label{fig:IBS_results_DS_vs_LP_ENAW7075_vinyl_a}
\end{figure}
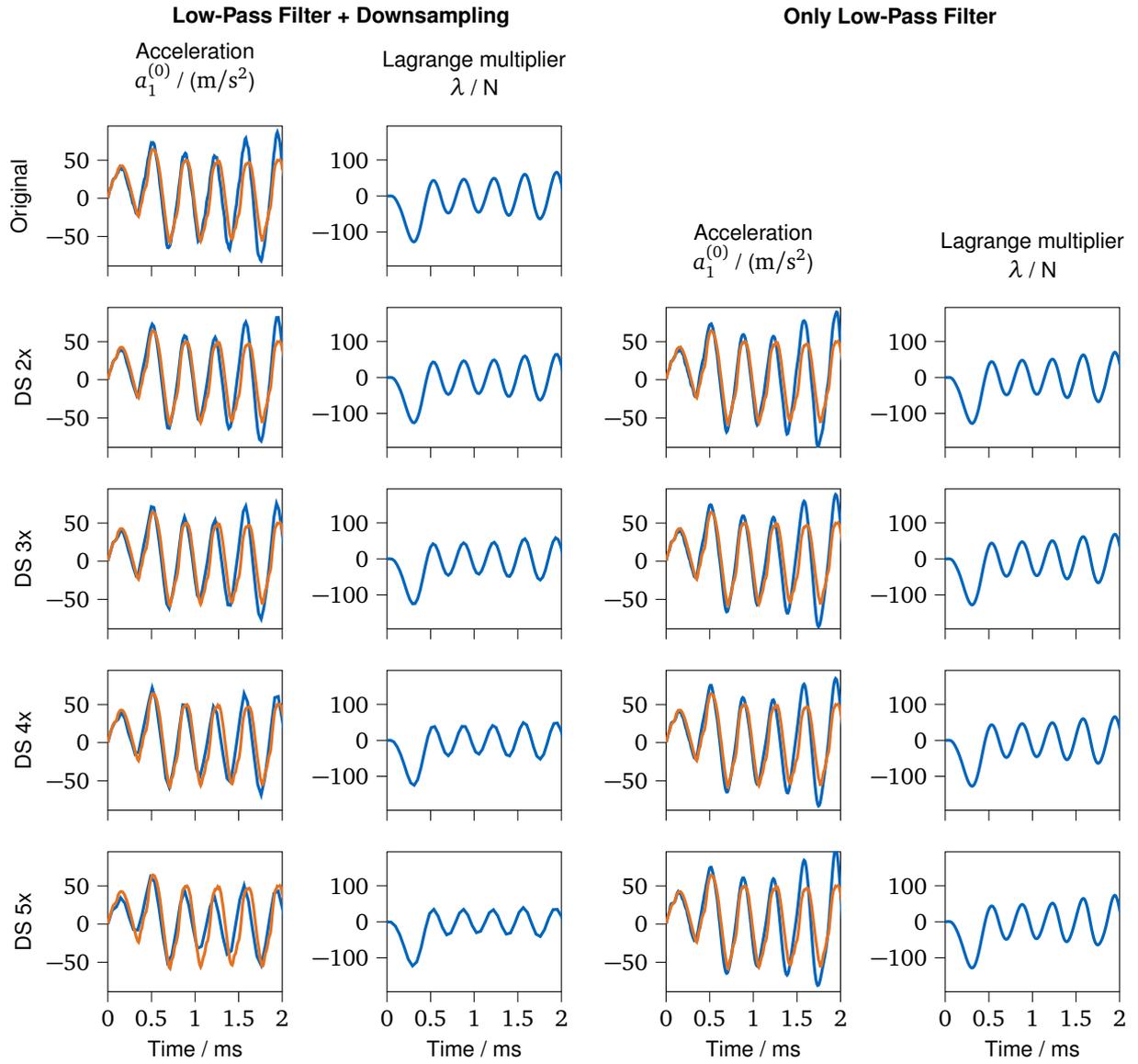

\begin{figure}[ht!]
	\small
	\renewcommand{\IBSlen}{2}
	\renewcommand{\IBSResultCSVacc}{data/IBS_results_ENAW7075_manual_steel_a_Time.csv}
	\renewcommand{\IBSResultCSVAa}{data/ENAW7075_manual_steel_DS/IBS_results_ENAW7075_manual_steel_a_DS_2.csv}
	\renewcommand{\IBSResultCSVBa}{data/ENAW7075_manual_steel_LPDS/IBS_results_ENAW7075_manual_steel_a_LPDS_2.csv}
	\renewcommand{\IBSResultCSVAb}{data/ENAW7075_manual_steel_DS/IBS_results_ENAW7075_manual_steel_a_DS_3.csv}
	\renewcommand{\IBSResultCSVBb}{data/ENAW7075_manual_steel_LPDS/IBS_results_ENAW7075_manual_steel_a_LPDS_3.csv}
	\renewcommand{\IBSResultCSVAc}{data/ENAW7075_manual_steel_DS/IBS_results_ENAW7075_manual_steel_a_DS_4.csv}
	\renewcommand{\IBSResultCSVBc}{data/ENAW7075_manual_steel_LPDS/IBS_results_ENAW7075_manual_steel_a_LPDS_4.csv}
	\renewcommand{\IBSResultCSVAd}{data/ENAW7075_manual_steel_DS/IBS_results_ENAW7075_manual_steel_a_DS_5.csv}
	\renewcommand{\IBSResultCSVBd}{data/ENAW7075_manual_steel_LPDS/IBS_results_ENAW7075_manual_steel_a_LPDS_5.csv}
	\renewcommand{\yAmin}{-1204.5668145751954}
	\renewcommand{\yAmax}{1280.624899291992}
	\renewcommand{\LMmin}{-556.4107557846804}
	\renewcommand{\LMmax}{556.4107557846804}
	\begin{minipage}{0.5\textwidth}
		\centering
		\hspace{2.8cm}\textbf{Low-Pass Filter + Downsampling}
	\end{minipage}
	\begin{minipage}{0.5\textwidth}
		\centering
		\hspace{1.4cm}\textbf{Only Low-Pass Filter}
	\end{minipage}\\\vspace{-0.6cm}
	\begin{flushright}
		% This file was created with tikzplotlib v0.10.1.
\begin{tikzpicture}

\definecolor{darkgray176}{RGB}{176,176,176}
\definecolor{gray}{RGB}{128,128,128}
\definecolor{lightgray204}{RGB}{204,204,204}

\begin{groupplot}[group style={group size=4 by 5, horizontal sep=1.5cm, vertical sep=0.6cm}]

% DS 1 - A
\nextgroupplot[
height=3.6cm,
scaled x ticks=manual:{}{\pgfmathparse{#1}},
yticklabel style={
        /pgf/number format/fixed,
        /pgf/number format/precision=3
},
tick align=outside,
tick pos=left,
title={Acceleration \\\(\displaystyle a_1^{(0)}\) / (\si[per-mode=symbol]{\meter\per\second\squared})},
title style = {align = center},
title style={yshift=0.7ex},
width=0.24\textwidth,
x grid style={darkgray176},
xmin=0, xmax=\IBSlen,
ylabel={Original},
xtick style={color=black},
xticklabels={},
y grid style={darkgray176},
ymin=\yAmin, ymax=\yAmax,
ytick style={color=black}
]
\addplot [very thick, TUMBlue]
table[x=t, y=y1IBS, col sep=comma]{\IBSResultCSVacc};

\addplot [very thick, TUMOrange]
table[x=t, y=y1Meas, col sep=comma]{\IBSResultCSVacc};

\nextgroupplot[
height=3.6cm,
tick align=outside,
tick pos=left,
title={Lagrange multiplier \\\(\displaystyle \lambda\) / N},
title style = {align = center},
title style={yshift=0.7ex},
width=0.24\textwidth,
x grid style={darkgray176},
xmin=0, xmax=\IBSlen,
xtick style={color=black},
xticklabels={},
y grid style={darkgray176},
ymin=\LMmin, ymax=\LMmax,
ytick style={color=black}
]
\addplot [very thick, TUMBlue]
table[x=t, y=lambda, col sep=comma]{\IBSResultCSVacc};

% DS 1 - B
\nextgroupplot[
height=3.6cm,
width=0.24\textwidth,
x grid style=none,
xticklabels={},
yticklabels={},
xtick style={draw=none},
ytick style={draw=none},
hide x axis,
hide y axis
]

\nextgroupplot[
height=3.6cm,
width=0.24\textwidth,
x grid style=none,
xticklabels={},
yticklabels={},
xtick style={draw=none},
ytick style={draw=none},
hide x axis,
hide y axis
]

% DS 2 - A
\nextgroupplot[
height=3.6cm,
scaled x ticks=manual:{}{\pgfmathparse{#1}},
yticklabel style={
        /pgf/number format/fixed,
        /pgf/number format/precision=3
},
tick align=outside,
tick pos=left,
width=0.24\textwidth,
x grid style={darkgray176},
xmin=0, xmax=\IBSlen,
ylabel={DS 2x},
xtick style={color=black},
xticklabels={},
y grid style={darkgray176},
ymin=\yAmin, ymax=\yAmax,
ytick style={color=black}
]
\addplot [very thick, TUMBlue]
table[x=t, y=y1IBS, col sep=comma]{\IBSResultCSVAa};

\addplot [very thick, TUMOrange]
table[x=t, y=y1Meas, col sep=comma]{\IBSResultCSVAa};

\nextgroupplot[
height=3.6cm,
tick align=outside,
tick pos=left,
width=0.24\textwidth,
x grid style={darkgray176},
xmin=0, xmax=\IBSlen,
xtick style={color=black},
xticklabels={},
y grid style={darkgray176},
ymin=\LMmin, ymax=\LMmax,
ytick style={color=black}
]
\addplot [very thick, TUMBlue]
table[x=t, y=lambda, col sep=comma]{\IBSResultCSVAa};

% DS 2 - B
\nextgroupplot[
height=3.6cm,
scaled x ticks=manual:{}{\pgfmathparse{#1}},
yticklabel style={
        /pgf/number format/fixed,
        /pgf/number format/precision=3
},
tick align=outside,
tick pos=left,
title={Acceleration \\\(\displaystyle a_1^{(0)}\) / (\si[per-mode=symbol]{\meter\per\second\squared})},
title style = {align = center},
title style={yshift=0.7ex},
width=0.24\textwidth,
x grid style={darkgray176},
xmin=0, xmax=\IBSlen,
xtick style={color=black},
xticklabels={},
y grid style={darkgray176},
ymin=\yAmin, ymax=\yAmax,
ytick style={color=black}
]
\addplot [very thick, TUMBlue]
table[x=t, y=y1IBS, col sep=comma]{\IBSResultCSVBa};

\addplot [very thick, TUMOrange]
table[x=t, y=y1Meas, col sep=comma]{\IBSResultCSVBa};

\nextgroupplot[
height=3.6cm,
tick align=outside,
tick pos=left,
title={Lagrange multiplier \\\(\displaystyle \lambda\) / N},
title style = {align = center},
title style={yshift=0.7ex},
width=0.24\textwidth,
x grid style={darkgray176},
xmin=0, xmax=\IBSlen,
xtick style={color=black},
xticklabels={},
y grid style={darkgray176},
ymin=\LMmin, ymax=\LMmax,
ytick style={color=black}
]
\addplot [very thick, TUMBlue]
table[x=t, y=lambda, col sep=comma]{\IBSResultCSVBa};

% DS 3 - A
\nextgroupplot[
height=3.6cm,
scaled x ticks=manual:{}{\pgfmathparse{#1}},
yticklabel style={
        /pgf/number format/fixed,
        /pgf/number format/precision=3
},
tick align=outside,
tick pos=left,
width=0.24\textwidth,
x grid style={darkgray176},
xmin=0, xmax=\IBSlen,
ylabel={DS 3x},
xtick style={color=black},
xticklabels={},
y grid style={darkgray176},
ymin=\yAmin, ymax=\yAmax,
ytick style={color=black}
]
\addplot [very thick, TUMBlue]
table[x=t, y=y1IBS, col sep=comma]{\IBSResultCSVAb};

\addplot [very thick, TUMOrange]
table[x=t, y=y1Meas, col sep=comma]{\IBSResultCSVAb};

\nextgroupplot[
height=3.6cm,
tick align=outside,
tick pos=left,
width=0.24\textwidth,
x grid style={darkgray176},
xmin=0, xmax=\IBSlen,
xtick style={color=black},
xticklabels={},
y grid style={darkgray176},
ymin=\LMmin, ymax=\LMmax,
ytick style={color=black}
]
\addplot [very thick, TUMBlue]
table[x=t, y=lambda, col sep=comma]{\IBSResultCSVAb};

% DS 3 - B
\nextgroupplot[
height=3.6cm,
scaled x ticks=manual:{}{\pgfmathparse{#1}},
yticklabel style={
        /pgf/number format/fixed,
        /pgf/number format/precision=3
},
tick align=outside,
tick pos=left,
width=0.24\textwidth,
x grid style={darkgray176},
xmin=0, xmax=\IBSlen,
xtick style={color=black},
xticklabels={},
y grid style={darkgray176},
ymin=\yAmin, ymax=\yAmax,
ytick style={color=black}
]
\addplot [very thick, TUMBlue]
table[x=t, y=y1IBS, col sep=comma]{\IBSResultCSVBb};

\addplot [very thick, TUMOrange]
table[x=t, y=y1Meas, col sep=comma]{\IBSResultCSVBb};

\nextgroupplot[
height=3.6cm,
tick align=outside,
tick pos=left,
width=0.24\textwidth,
x grid style={darkgray176},
xmin=0, xmax=\IBSlen,
xtick style={color=black},
xticklabels={},
y grid style={darkgray176},
ymin=\LMmin, ymax=\LMmax,
ytick style={color=black}
]
\addplot [very thick, TUMBlue]
table[x=t, y=lambda, col sep=comma]{\IBSResultCSVBb};

% DS 4 - A
\nextgroupplot[
height=3.6cm,
scaled x ticks=manual:{}{\pgfmathparse{#1}},
yticklabel style={
        /pgf/number format/fixed,
        /pgf/number format/precision=3
},
tick align=outside,
tick pos=left,
width=0.24\textwidth,
x grid style={darkgray176},
xmin=0, xmax=\IBSlen,
ylabel={DS 4x},
xtick style={color=black},
xticklabels={},
y grid style={darkgray176},
ymin=\yAmin, ymax=\yAmax,
ytick style={color=black}
]
\addplot [very thick, TUMBlue]
table[x=t, y=y1IBS, col sep=comma]{\IBSResultCSVAc};

\addplot [very thick, TUMOrange]
table[x=t, y=y1Meas, col sep=comma]{\IBSResultCSVAc};

\nextgroupplot[
height=3.6cm,
tick align=outside,
tick pos=left,
width=0.24\textwidth,
x grid style={darkgray176},
xmin=0, xmax=\IBSlen,
xtick style={color=black},
xticklabels={},
y grid style={darkgray176},
ymin=\LMmin, ymax=\LMmax,
ytick style={color=black}
]
\addplot [very thick, TUMBlue]
table[x=t, y=lambda, col sep=comma]{\IBSResultCSVAc};

% DS 4 - B
\nextgroupplot[
height=3.6cm,
scaled x ticks=manual:{}{\pgfmathparse{#1}},
yticklabel style={
        /pgf/number format/fixed,
        /pgf/number format/precision=3
},
tick align=outside,
tick pos=left,
width=0.24\textwidth,
x grid style={darkgray176},
xmin=0, xmax=\IBSlen,
xtick style={color=black},
xticklabels={},
y grid style={darkgray176},
ymin=\yAmin, ymax=\yAmax,
ytick style={color=black}
]
\addplot [very thick, TUMBlue]
table[x=t, y=y1IBS, col sep=comma]{\IBSResultCSVBc};

\addplot [very thick, TUMOrange]
table[x=t, y=y1Meas, col sep=comma]{\IBSResultCSVBc};

\nextgroupplot[
height=3.6cm,
tick align=outside,
tick pos=left,
width=0.24\textwidth,
x grid style={darkgray176},
xmin=0, xmax=\IBSlen,
xtick style={color=black},
xticklabels={},
y grid style={darkgray176},
ymin=\LMmin, ymax=\LMmax,
ytick style={color=black}
]
\addplot [very thick, TUMBlue]
table[x=t, y=lambda, col sep=comma]{\IBSResultCSVBc};

% DS 5 - A
\nextgroupplot[
height=3.6cm,
scaled x ticks=manual:{}{\pgfmathparse{#1}},
yticklabel style={
        /pgf/number format/fixed,
        /pgf/number format/precision=3
},
tick align=outside,
tick pos=left,
width=0.24\textwidth,
x grid style={darkgray176},
xmin=0, xmax=\IBSlen,
ylabel={DS 5x},
xtick style={color=black},
xlabel={Time / ms},
y grid style={darkgray176},
ymin=\yAmin, ymax=\yAmax,
ytick style={color=black}
]
\addplot [very thick, TUMBlue]
table[x=t, y=y1IBS, col sep=comma]{\IBSResultCSVAd};

\addplot [very thick, TUMOrange]
table[x=t, y=y1Meas, col sep=comma]{\IBSResultCSVAd};

\nextgroupplot[
height=3.6cm,
tick align=outside,
tick pos=left,
width=0.24\textwidth,
x grid style={darkgray176},
xmin=0, xmax=\IBSlen,
xtick style={color=black},
xlabel={Time / ms},
y grid style={darkgray176},
ymin=\LMmin, ymax=\LMmax,
ytick style={color=black}
]
\addplot [very thick, TUMBlue]
table[x=t, y=lambda, col sep=comma]{\IBSResultCSVAd};

% DS 5 - B
\nextgroupplot[
height=3.6cm,
scaled x ticks=manual:{}{\pgfmathparse{#1}},
yticklabel style={
        /pgf/number format/fixed,
        /pgf/number format/precision=3
},
tick align=outside,
tick pos=left,
width=0.24\textwidth,
x grid style={darkgray176},
xmin=0, xmax=\IBSlen,
xtick style={color=black},
xlabel={Time / ms},
y grid style={darkgray176},
ymin=\yAmin, ymax=\yAmax,
ytick style={color=black}
]
\addplot [very thick, TUMBlue]
table[x=t, y=y1IBS, col sep=comma]{\IBSResultCSVBd};

\addplot [very thick, TUMOrange]
table[x=t, y=y1Meas, col sep=comma]{\IBSResultCSVBd};

\nextgroupplot[
height=3.6cm,
tick align=outside,
tick pos=left,
width=0.24\textwidth,
x grid style={darkgray176},
xmin=0, xmax=\IBSlen,
xtick style={color=black},
xlabel={Time / ms},
y grid style={darkgray176},
ymin=\LMmin, ymax=\LMmax,
ytick style={color=black}
]
\addplot [very thick, TUMBlue]
table[x=t, y=lambda, col sep=comma]{\IBSResultCSVBd};

\end{groupplot}
\end{tikzpicture}
	\end{flushright}\vspace{-0.4cm}
	\caption{Comparison of experimental results using a low-pass filter with downsampling versus only low-pass filtering, applied to \textbf{accelerations} of \textbf{aluminum} rods using a manual hammer with \textbf{steel tip}, \legendary}
	\label{fig:IBS_results_DS_vs_LP_ENAW7075_steel_a}
\end{figure}
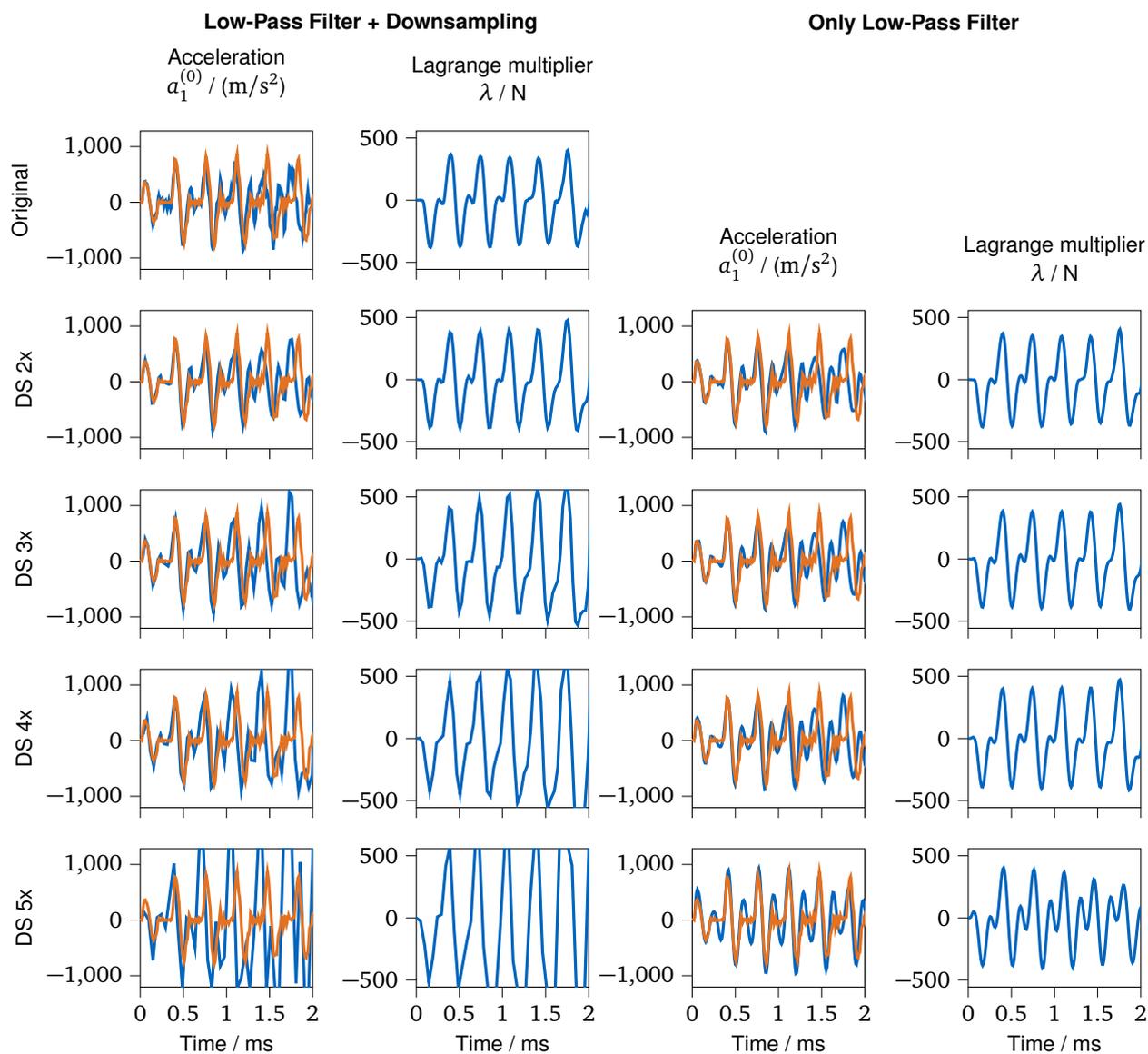

\clearpage
% !TeX spellcheck = en_US
\section{Discussion}

The experimental IBS results of the POM and aluminum rods for both utilized hammer tips as well as for the full and reduced measurement bandwidth are summarized in \cref{tab:exp_results_MKII}. For the reduced bandwidth, the best result achieved, out of the considered downsampling factors, is shown.

\bgroup
\def\arraystretch{1.1}
\begin{table}[ht]
	\centering
	\small
	\begin{tabularx}{\textwidth}{c|Y|Y|Y|Y|Y|Y|Y|Y}
		\multicolumn{9}{c}{\cellcolor{lightgray}\textbf{Results of Experimental IBS Using Full versus Reduced Measurement Bandwidth}} \\\hline
		\textbf{Rod Material} & \multicolumn{4}{c|}{Polyoxymethylene (POM)} & \multicolumn{4}{c}{Aluminum (EN AW 7075)} \\\hline
		
		\textbf{Hammer Type} & \multicolumn{2}{c|}{Manual with Vinyl Tip} & \multicolumn{2}{c|}{Manual with Steel Tip} & \multicolumn{2}{c|}{Manual with Vinyl Tip} & \multicolumn{2}{c}{Manual with Steel Tip} \\\hline
		
		\textbf{Bandwidth} & Original & Reduced & Original & Reduced & Original & Reduced & Original & Reduced \\\hline
		
		Displacement IBS & \trFailUnstable & \trPeakFive & \trFailUnstable & \trPeakFive & \trPeakZero & \trPeakThree & \trPeakThree & \trPeakFour \\\hline
		Velocity IBS & \trPeakZero & \trPeakTwo & \trFailUnstable & \trPeakTwo & \trPeakTwo & \trPeakTwo & \trPeakOne & \trPeakOne \\\hline
		Acceleration IBS & \trPeakTwo$^1$ & \trPeakFive & \trFailCoupling & \trPeakFour & \trPeakFour & \trPeakFour & \trPeakFour & \trPeakThree \\\hline
	\end{tabularx} \vspace{0.2cm}
	
	Explanation of symbols: $\Large\boldsymbol{\infty}$: IBS algorithm immediately unstable, $\Large\boldsymbol{\times}$: Coupling issues, \\ $\boldsymbol{\checkmark}$: Number of response peaks within reasonable tolerance of reference measurement \\
	$\boldsymbol{\mathrm{O}}$: No correct response peaks, $^1$: Initial response peak significantly wrong
	\caption{Summary of experimental IBS results of POM and aluminum rods using displacement, velocity and acceleration responses, excited by a manual hammer with either a vinyl or steel tip. Results using original bandwidth represent the IBS results with no filtering or downsampling applied to the responses, reduced bandwidth represents the best result achieved}
	\label{tab:exp_results_MKII}
\end{table}
\egroup

Regarding the set goal for a successful application for shock response estimation, i.e.\ correctly predicting at least the first two driving point response peaks, this could be achieved for every material and response quantity for at least one excitation type. For the POM rods, this was only possible by reducing the considered frequency bandwidth, either by downsampling with a low-pass filter or only low-pass filtering. Out of the two discussed methods for limiting the frequency content, none was clearly advantageous over the other. In most cases, where both methods performed similarly, downsampling is preferable over only low-pass filtering, because this yields computational advantages, see also \cref{fig:computational_performance}, depicting the computational cost for the IRF calculation of the aluminum rods using a manual hammer with a steel tip, depended on the number of averages and the used downsampling factor.

\begin{figure}[ht]
	\small
	\centering
	% This file was created with tikzplotlib v0.10.1.
\begin{tikzpicture}

\definecolor{darkgray176}{RGB}{176,176,176}
\definecolor{lightgray204}{RGB}{204,204,204}
% Color cycle
\definecolor{steelblue31119180}{RGB}{0,51,89} % Pantone 540
\definecolor{darkorange25512714}{RGB}{0,82,147} % Pantone 301
\definecolor{forestgreen4416044}{RGB}{0,101,189} % Pantone 300 C
\definecolor{crimson2143940}{RGB}{100,160,200} % Pantone 542
\definecolor{mediumpurple148103189}{RGB}{152,198,234} % Pantone 283

\begin{groupplot}[group style={group size=1 by 1, horizontal sep=1.2cm}]
\nextgroupplot[
height=5.5cm,
legend cell align={left},
legend style={fill opacity=0.8, draw opacity=1, text opacity=1},
legend pos=outer north east,
log basis y={10},
tick align=outside,
tick pos=left,
width=0.5\textwidth,
x grid style={darkgray176},
xlabel={Downsampling Factor},
xmin=1, xmax=5,
xtick style={color=black},
y grid style={darkgray176},
ymin=0.0055972, ymax=1.2801,
ymode=log,
ytick style={color=black},
ytick={0.0001,0.001,0.01,0.1,1,10,100},
yticklabels={
  \(\displaystyle {10^{-4}}\),
  \(\displaystyle {10^{-3}}\),
  \(\displaystyle {10^{-2}}\),
  \(\displaystyle {10^{-1}}\),
  \(\displaystyle {10^{0}}\),
  \(\displaystyle {10^{1}}\),
  \(\displaystyle {10^{2}}\)
}
]
\addplot [very thick, steelblue31119180]
table {%
1 1
2 0.39765
3 0.24569
4 0.17629
5 0.13497
};
\addlegendentry{$n_\text{avg}=20$}
\addplot [very thick, darkorange25512714]
table {%
1 0.49929
2 0.20603
3 0.11639
4 0.085062
5 0.067816
};
\addlegendentry{$n_\text{avg}=10$}
\addplot [very thick, forestgreen4416044]
table {%
1 0.2541
2 0.10872
3 0.059735
4 0.044322
5 0.034741
};
\addlegendentry{$n_\text{avg}=5$}
\addplot [very thick, crimson2143940]
table {%
1 0.16588
2 0.0664
3 0.036907
4 0.029409
5 0.021245
};
\addlegendentry{$n_\text{avg}=3$}
\addplot [very thick, mediumpurple148103189]
table {%
1 0.054486
2 0.021328
3 0.012414
4 0.010081
5 0.0071649
};
\addlegendentry{$n_\text{avg}=1$}
\addplot [very thick, steelblue31119180, dashed, forget plot]
table {%
1 1
2 0.5
3 0.3333
4 0.25
5 0.2
};
\addplot [very thick, darkorange25512714, dashed, forget plot]
table {%
1 0.49929
2 0.24965
3 0.16641
4 0.12482
5 0.099858
};
\addplot [very thick, forestgreen4416044, dashed, forget plot]
table {%
1 0.2541
2 0.12705
3 0.084693
4 0.063526
5 0.050821
};
\addplot [very thick, crimson2143940, dashed, forget plot]
table {%
1 0.16588
2 0.082938
3 0.055286
4 0.041469
5 0.033175
};
\addplot [very thick, mediumpurple148103189, dashed, forget plot]
table {%
1 0.054486
2 0.027243
3 0.01816
4 0.013622
5 0.010897
};
\end{groupplot}

\end{tikzpicture}\vspace{-0.2cm}
	\caption{Illustration of single IRF computation cost depended on the number of averages and utilized downsampling factor. Individual computation time of all four IRFs required for IBS application averaged and normalized to longest computation duration ($n_\text{avg}=20$, no downsampling). Dashed lines represent theoretical linear scaling with the number of time points, i.e.\ downsampling by 2x yields \SI{50}{\percent} computation time}
	\label{fig:computational_performance}
\end{figure}
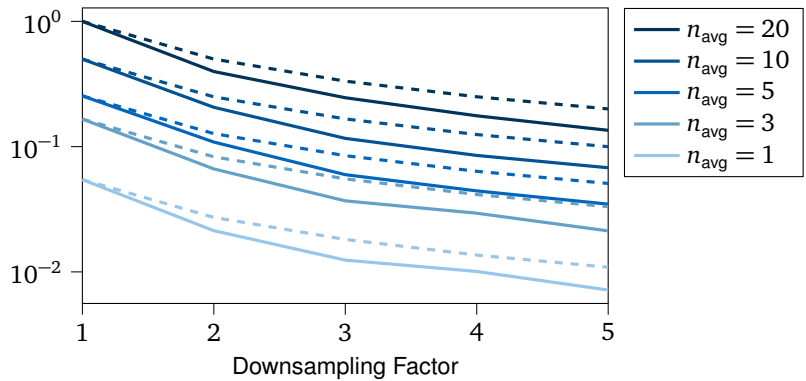\vspace{-0.15cm}

As can be seen, downsampling by a factor of 2 reduces the computational costs by \SI{60}{\percent}, compared to not limiting the frequency content or only applying a low-pass filter. Unclear at this point is how an appropriate downsampling factor can be determined based on the measured signals, instead of deciding based on the retrieved IBS results. A look at the FRFs, as shown in \cref{fig:DS_est}, can give an indication of a practicable frequency cut-off point. In some cases, however, other downsampling factors proved to be better.

It is to be noted that downsampling is necessary because the acquired frequency bandwidth is greater than what is reasonably excited. Assuming similar performance of the low-pass filter applied for the downsampling and the ones utilized within the measurement system, downsampling after the measurement is equivalent to acquiring the responses at a lower sample rate. Nevertheless, the optimal frequency bandwidth might not be known a priori, meaning that a limitation of the frequency content might still be necessary after the measurement.

For unknown reasons, using velocity responses leads to worse IBS results for the considered cases. Because the displacement responses are acquired from the velocity responses through a time integration in the Laservibrometer controller, it seems unlikely that erroneous measurements are the cause for the worse performance.
% !TeX spellcheck = en_US
\section{Conclusion and Outlook}

The most robust IBS results were found for the aluminum rods using acceleration responses, which required no limitation of the frequency content, most likely due to the accelerations scaling with the angular frequency squared, i.e.\ more signal at high frequencies, and the better excitation bandwidth due to the harder material pairing. Considering the results with a reduced bandwidth, good results could also be achieved using displacements. 

In this paper, it was shown that the experimental application of the impulse-based substructuring method for the prediction of the shock responses of rods considered one-dimensional is possible using displacement, velocity or acceleration responses. While this might not be useful on its own, the basis for experimental IBS application has been laid by highlighting a viable impulse response function estimation method as well as first improvement possibilities and, first of all, showing that an experimental IBS application is in fact possible.

Within future works, besides trying to seek out criteria for selecting a suitable downsampling factor based on the measured responses and excitation, the performance of different low-pass filters should be evaluated. Further improvements of the experimental IBS performance can potentially be achieved in the future by improving the IRF estimation. Since the cause of the instability is believed to originate from the interaction of erroneous IRFs, another viable improvement approach might be to weaken the interface compatibility condition.

% Statements
% !TeX spellcheck = en_US

\paragraph{Data availability}
The measured, unedited time series of excitation and response for both materials and hammer types are available at: \url{https://doi.org/10.14459/2023mp1729648}, \cite{IBS1DexpData}.

\paragraph{CRediT authorship contribution statement:} 
\textbf{O.~M.~Zobel:} Conceptualization; Data curation; Formal analysis; Investigation; Methodology; Software; Validation; Visualization; Writing - original draft; Writing - review \& editing. 
\textbf{F.~Trainotti:} Conceptualization, Methodology, Investigation, Writing - review \& editing. 
\textbf{D.~J.~Rixen:} Conceptualization, Resources, Writing - review \& editing, Supervision.

\paragraph{Declaration of Competing Interest:} 
The authors declare that they have no known competing financial interests or personal relationships that could have appeared to influence the work reported in this paper.

% Bibliography
\printbibliography

% Appendix
\appendix
% !TeX spellcheck = en_US
\newpage
\section{Additional Results With Removal of Higher Frequency Content}

\begin{figure}[ht!]
	\small
	\renewcommand{\IBSlen}{8}
	\renewcommand{\IBSResultCSVdisp}{data/IBS_results_POM_manual_vinyl_d_Time.csv}
	\renewcommand{\IBSResultCSVAa}{data/POM_manual_vinyl_DS/IBS_results_POM_manual_vinyl_d_DS_2.csv}
	\renewcommand{\IBSResultCSVBa}{data/POM_manual_vinyl_LPDS/IBS_results_POM_manual_vinyl_d_LPDS_2.csv}
	\renewcommand{\IBSResultCSVAb}{data/POM_manual_vinyl_DS/IBS_results_POM_manual_vinyl_d_DS_3.csv}
	\renewcommand{\IBSResultCSVBb}{data/POM_manual_vinyl_LPDS/IBS_results_POM_manual_vinyl_d_LPDS_3.csv}
	\renewcommand{\IBSResultCSVAc}{data/POM_manual_vinyl_DS/IBS_results_POM_manual_vinyl_d_DS_4.csv}
	\renewcommand{\IBSResultCSVBc}{data/POM_manual_vinyl_LPDS/IBS_results_POM_manual_vinyl_d_LPDS_4.csv}
	\renewcommand{\IBSResultCSVAd}{data/POM_manual_vinyl_DS/IBS_results_POM_manual_vinyl_d_DS_5.csv}
	\renewcommand{\IBSResultCSVBd}{data/POM_manual_vinyl_LPDS/IBS_results_POM_manual_vinyl_d_LPDS_5.csv}
	\renewcommand{\IBSResultCSVAe}{data/POM_manual_vinyl_DS/IBS_results_POM_manual_vinyl_d_DS_7.csv}
	\renewcommand{\IBSResultCSVBe}{data/POM_manual_vinyl_LPDS/IBS_results_POM_manual_vinyl_d_LPDS_7.csv}
	\renewcommand{\IBSResultCSVAf}{data/POM_manual_vinyl_DS/IBS_results_POM_manual_vinyl_d_DS_10.csv}
	\renewcommand{\IBSResultCSVBf}{data/POM_manual_vinyl_LPDS/IBS_results_POM_manual_vinyl_d_LPDS_10.csv}
	\renewcommand{\IBSResultCSVAg}{data/POM_manual_vinyl_DS/IBS_results_POM_manual_vinyl_d_DS_15.csv}
	\renewcommand{\IBSResultCSVBg}{data/POM_manual_vinyl_LPDS/IBS_results_POM_manual_vinyl_d_LPDS_15.csv}
	\renewcommand{\yAmin}{-6.497262187500001e-05}
	\renewcommand{\yAmax}{0.00038945157890625}
	\renewcommand{\LMmin}{-324.1821672220685}
	\renewcommand{\LMmax}{324.1821672220685}
	\begin{minipage}{0.5\textwidth}
		\centering
		\hspace{2.8cm}\textbf{Low-Pass Filter + Downsampling}
	\end{minipage}
	\begin{minipage}{0.5\textwidth}
		\centering
		\hspace{1.4cm}\textbf{Only Low-Pass Filter}
	\end{minipage}\\\vspace{-0.6cm}
	\begin{flushright}
		\input{figures/experimental/IBS_results_DS_vs_LP_POM_disp.tikz}
	\end{flushright}\vspace{-0.4cm}
	\caption{Comparison of experimental results using a low-pass filter with downsampling versus only low-pass filtering, applied to \textbf{displacements} of \textbf{POM} rods using manual hammer with \textbf{vinyl tip}, \legendary}
	\label{fig:IBS_results_DS_vs_LP_POM_vinyl_d}
\end{figure}

\begin{figure}[ht!]
	\small
	\renewcommand{\IBSlen}{8}
	\renewcommand{\IBSResultCSVvelo}{data/IBS_results_POM_manual_vinyl_v_Time.csv}
	\renewcommand{\IBSResultCSVAa}{data/POM_manual_vinyl_DS/IBS_results_POM_manual_vinyl_v_DS_2.csv}
	\renewcommand{\IBSResultCSVBa}{data/POM_manual_vinyl_LPDS/IBS_results_POM_manual_vinyl_v_LPDS_2.csv}
	\renewcommand{\IBSResultCSVAb}{data/POM_manual_vinyl_DS/IBS_results_POM_manual_vinyl_v_DS_3.csv}
	\renewcommand{\IBSResultCSVBb}{data/POM_manual_vinyl_LPDS/IBS_results_POM_manual_vinyl_v_LPDS_3.csv}
	\renewcommand{\IBSResultCSVAc}{data/POM_manual_vinyl_DS/IBS_results_POM_manual_vinyl_v_DS_4.csv}
	\renewcommand{\IBSResultCSVBc}{data/POM_manual_vinyl_LPDS/IBS_results_POM_manual_vinyl_v_LPDS_4.csv}
	\renewcommand{\IBSResultCSVAd}{data/POM_manual_vinyl_DS/IBS_results_POM_manual_vinyl_v_DS_5.csv}
	\renewcommand{\IBSResultCSVBd}{data/POM_manual_vinyl_LPDS/IBS_results_POM_manual_vinyl_v_LPDS_5.csv}
	\renewcommand{\IBSResultCSVAe}{data/POM_manual_vinyl_DS/IBS_results_POM_manual_vinyl_v_DS_7.csv}
	\renewcommand{\IBSResultCSVBe}{data/POM_manual_vinyl_LPDS/IBS_results_POM_manual_vinyl_v_LPDS_7.csv}
	\renewcommand{\IBSResultCSVAf}{data/POM_manual_vinyl_DS/IBS_results_POM_manual_vinyl_v_DS_10.csv}
	\renewcommand{\IBSResultCSVBf}{data/POM_manual_vinyl_LPDS/IBS_results_POM_manual_vinyl_v_LPDS_10.csv}
	\renewcommand{\IBSResultCSVAg}{data/POM_manual_vinyl_DS/IBS_results_POM_manual_vinyl_v_DS_15.csv}
	\renewcommand{\IBSResultCSVBg}{data/POM_manual_vinyl_LPDS/IBS_results_POM_manual_vinyl_v_LPDS_15.csv}
	\renewcommand{\yAmin}{-0.076767534375}
	\renewcommand{\yAmax}{0.21826226953125}
	\renewcommand{\LMmin}{-324.1821672220685}
	\renewcommand{\LMmax}{324.1821672220685}
	\begin{minipage}{0.5\textwidth}
		\centering
		\hspace{2.8cm}\textbf{Low-Pass Filter + Downsampling}
	\end{minipage}
	\begin{minipage}{0.5\textwidth}
		\centering
		\hspace{1.4cm}\textbf{Only Low-Pass Filter}
	\end{minipage}\\\vspace{-0.6cm}
	\begin{flushright}
		\input{figures/experimental/IBS_results_DS_vs_LP_POM_velo.tikz}
	\end{flushright}\vspace{-0.4cm}
	\caption{Comparison of experimental results using a low-pass filter with downsampling versus only low-pass filtering, applied to \textbf{velocities} of \textbf{POM} rods using manual hammer with \textbf{vinyl tip}, \legendary}
	\label{fig:IBS_results_DS_vs_LP_POM_vinyl_v}
\end{figure}

\begin{figure}[ht!]
	\small
	\renewcommand{\IBSlen}{8}
	\renewcommand{\IBSResultCSVacc}{data/IBS_results_POM_manual_vinyl_a_Time.csv}
	\renewcommand{\IBSResultCSVAa}{data/POM_manual_vinyl_DS/IBS_results_POM_manual_vinyl_a_DS_2.csv}
	\renewcommand{\IBSResultCSVBa}{data/POM_manual_vinyl_LPDS/IBS_results_POM_manual_vinyl_a_LPDS_2.csv}
	\renewcommand{\IBSResultCSVAb}{data/POM_manual_vinyl_DS/IBS_results_POM_manual_vinyl_a_DS_3.csv}
	\renewcommand{\IBSResultCSVBb}{data/POM_manual_vinyl_LPDS/IBS_results_POM_manual_vinyl_a_LPDS_3.csv}
	\renewcommand{\IBSResultCSVAc}{data/POM_manual_vinyl_DS/IBS_results_POM_manual_vinyl_a_DS_4.csv}
	\renewcommand{\IBSResultCSVBc}{data/POM_manual_vinyl_LPDS/IBS_results_POM_manual_vinyl_a_LPDS_4.csv}
	\renewcommand{\IBSResultCSVAd}{data/POM_manual_vinyl_DS/IBS_results_POM_manual_vinyl_a_DS_5.csv}
	\renewcommand{\IBSResultCSVBd}{data/POM_manual_vinyl_LPDS/IBS_results_POM_manual_vinyl_a_LPDS_5.csv}
	\renewcommand{\IBSResultCSVAe}{data/POM_manual_vinyl_DS/IBS_results_POM_manual_vinyl_a_DS_7.csv}
	\renewcommand{\IBSResultCSVBe}{data/POM_manual_vinyl_LPDS/IBS_results_POM_manual_vinyl_a_LPDS_7.csv}
	\renewcommand{\IBSResultCSVAf}{data/POM_manual_vinyl_DS/IBS_results_POM_manual_vinyl_a_DS_10.csv}
	\renewcommand{\IBSResultCSVBf}{data/POM_manual_vinyl_LPDS/IBS_results_POM_manual_vinyl_a_LPDS_10.csv}
	\renewcommand{\IBSResultCSVAg}{data/POM_manual_vinyl_DS/IBS_results_POM_manual_vinyl_a_DS_15.csv}
	\renewcommand{\IBSResultCSVBg}{data/POM_manual_vinyl_LPDS/IBS_results_POM_manual_vinyl_a_LPDS_15.csv}
	\renewcommand{\yAmin}{-670.4906805419922}
	\renewcommand{\yAmax}{802.0985375976562}
	\renewcommand{\LMmin}{-265.38158669608407}
	\renewcommand{\LMmax}{265.38158669608407}
	\begin{minipage}{0.5\textwidth}
		\centering
		\hspace{2.8cm}\textbf{Low-Pass Filter + Downsampling}
	\end{minipage}
	\begin{minipage}{0.5\textwidth}
		\centering
		\hspace{1.4cm}\textbf{Only Low-Pass Filter}
	\end{minipage}\\\vspace{-0.6cm}
	\begin{flushright}
		\input{figures/experimental/IBS_results_DS_vs_LP_POM_acc.tikz}
	\end{flushright}\vspace{-0.4cm}
	\caption{Comparison of experimental results using a low-pass filter with downsampling versus only low-pass filtering, applied to \textbf{accelerations} of \textbf{POM} rods using manual hammer with \textbf{vinyl tip}, \legendary}
	\label{fig:IBS_results_DS_vs_LP_POM_vinyl_a}
\end{figure}

\begin{figure}[ht!]
	\small
	\renewcommand{\IBSlen}{2}
	\renewcommand{\IBSResultCSVdisp}{data/IBS_results_ENAW7075_manual_vinyl_d_Time.csv}
	\renewcommand{\IBSResultCSVAa}{data/ENAW7075_manual_vinyl_DS/IBS_results_ENAW7075_manual_vinyl_d_DS_2.csv}
	\renewcommand{\IBSResultCSVBa}{data/ENAW7075_manual_vinyl_LPDS/IBS_results_ENAW7075_manual_vinyl_d_LPDS_2.csv}
	\renewcommand{\IBSResultCSVAb}{data/ENAW7075_manual_vinyl_DS/IBS_results_ENAW7075_manual_vinyl_d_DS_3.csv}
	\renewcommand{\IBSResultCSVBb}{data/ENAW7075_manual_vinyl_LPDS/IBS_results_ENAW7075_manual_vinyl_d_LPDS_3.csv}
	\renewcommand{\IBSResultCSVAc}{data/ENAW7075_manual_vinyl_DS/IBS_results_ENAW7075_manual_vinyl_d_DS_4.csv}
	\renewcommand{\IBSResultCSVBc}{data/ENAW7075_manual_vinyl_LPDS/IBS_results_ENAW7075_manual_vinyl_d_LPDS_4.csv}
	\renewcommand{\IBSResultCSVAd}{data/ENAW7075_manual_vinyl_DS/IBS_results_ENAW7075_manual_vinyl_d_DS_5.csv}
	\renewcommand{\IBSResultCSVBd}{data/ENAW7075_manual_vinyl_LPDS/IBS_results_ENAW7075_manual_vinyl_d_LPDS_5.csv}
	\renewcommand{\yAmin}{-7.017584765625001e-06}
	\renewcommand{\yAmax}{4.182299296875e-05}
	\renewcommand{\LMmin}{-304.67055783962905}
	\renewcommand{\LMmax}{304.67055783962905}
	\begin{minipage}{0.5\textwidth}
		\centering
		\hspace{2.8cm}\textbf{Low-Pass Filter + Downsampling}
	\end{minipage}
	\begin{minipage}{0.5\textwidth}
		\centering
		\hspace{1.4cm}\textbf{Only Low-Pass Filter}
	\end{minipage}\\\vspace{-0.6cm}
	\begin{flushright}
		% This file was created with tikzplotlib v0.10.1.
\begin{tikzpicture}

\definecolor{darkgray176}{RGB}{176,176,176}
\definecolor{gray}{RGB}{128,128,128}
\definecolor{lightgray204}{RGB}{204,204,204}

\begin{groupplot}[group style={group size=4 by 5, horizontal sep=1.5cm, vertical sep=0.6cm}]

% DS 1 - A
\nextgroupplot[
height=3.6cm,
scaled x ticks=manual:{}{\pgfmathparse{#1}},
yticklabel style={
        /pgf/number format/fixed,
        /pgf/number format/precision=3
},
tick align=outside,
tick pos=left,
title={Displacement \\\(\displaystyle d_1^{(0)}\) / \si[per-mode=symbol]{\meter}},
title style = {align = center},
title style={yshift=0.7ex},
width=0.24\textwidth,
x grid style={darkgray176},
xmin=0, xmax=\IBSlen,
ylabel={Original},
xtick style={color=black},
xticklabels={},
y grid style={darkgray176},
ymin=\yAmin, ymax=\yAmax,
ytick style={color=black}
]
\addplot [very thick, TUMBlue]
table[x=t, y=y1IBS, col sep=comma]{\IBSResultCSVdisp};

\addplot [very thick, TUMOrange]
table[x=t, y=y1Meas, col sep=comma]{\IBSResultCSVdisp};

\nextgroupplot[
height=3.6cm,
tick align=outside,
tick pos=left,
title={Lagrange multiplier \\\(\displaystyle \lambda\) / N},
title style = {align = center},
title style={yshift=0.7ex},
width=0.24\textwidth,
x grid style={darkgray176},
xmin=0, xmax=\IBSlen,
xtick style={color=black},
xticklabels={},
y grid style={darkgray176},
ymin=\LMmin, ymax=\LMmax,
ytick style={color=black}
]
\addplot [very thick, TUMBlue]
table[x=t, y=lambda, col sep=comma]{\IBSResultCSVdisp};

% DS 1 - B
\nextgroupplot[
height=3.6cm,
width=0.24\textwidth,
x grid style=none,
xticklabels={},
yticklabels={},
xtick style={draw=none},
ytick style={draw=none},
hide x axis,
hide y axis
]

\nextgroupplot[
height=3.6cm,
width=0.24\textwidth,
x grid style=none,
xticklabels={},
yticklabels={},
xtick style={draw=none},
ytick style={draw=none},
hide x axis,
hide y axis
]

% DS 2 - A
\nextgroupplot[
height=3.6cm,
scaled x ticks=manual:{}{\pgfmathparse{#1}},
yticklabel style={
        /pgf/number format/fixed,
        /pgf/number format/precision=3
},
tick align=outside,
tick pos=left,
width=0.24\textwidth,
x grid style={darkgray176},
xmin=0, xmax=\IBSlen,
ylabel={DS 2x},
xtick style={color=black},
xticklabels={},
y grid style={darkgray176},
ymin=\yAmin, ymax=\yAmax,
ytick style={color=black}
]
\addplot [very thick, TUMBlue]
table[x=t, y=y1IBS, col sep=comma]{\IBSResultCSVAa};

\addplot [very thick, TUMOrange]
table[x=t, y=y1Meas, col sep=comma]{\IBSResultCSVAa};

\nextgroupplot[
height=3.6cm,
tick align=outside,
tick pos=left,
width=0.24\textwidth,
x grid style={darkgray176},
xmin=0, xmax=\IBSlen,
xtick style={color=black},
xticklabels={},
y grid style={darkgray176},
ymin=\LMmin, ymax=\LMmax,
ytick style={color=black}
]
\addplot [very thick, TUMBlue]
table[x=t, y=lambda, col sep=comma]{\IBSResultCSVAa};

% DS 2 - B
\nextgroupplot[
height=3.6cm,
scaled x ticks=manual:{}{\pgfmathparse{#1}},
yticklabel style={
        /pgf/number format/fixed,
        /pgf/number format/precision=3
},
tick align=outside,
tick pos=left,
title={Displacement \\\(\displaystyle d_1^{(0)}\) / \si[per-mode=symbol]{\meter}},
title style = {align = center},
title style={yshift=0.7ex},
width=0.24\textwidth,
x grid style={darkgray176},
xmin=0, xmax=\IBSlen,
xtick style={color=black},
xticklabels={},
y grid style={darkgray176},
ymin=\yAmin, ymax=\yAmax,
ytick style={color=black}
]
\addplot [very thick, TUMBlue]
table[x=t, y=y1IBS, col sep=comma]{\IBSResultCSVBa};

\addplot [very thick, TUMOrange]
table[x=t, y=y1Meas, col sep=comma]{\IBSResultCSVBa};

\nextgroupplot[
height=3.6cm,
tick align=outside,
tick pos=left,
title={Lagrange multiplier \\\(\displaystyle \lambda\) / N},
title style = {align = center},
title style={yshift=0.7ex},
width=0.24\textwidth,
x grid style={darkgray176},
xmin=0, xmax=\IBSlen,
xtick style={color=black},
xticklabels={},
y grid style={darkgray176},
ymin=\LMmin, ymax=\LMmax,
ytick style={color=black}
]
\addplot [very thick, TUMBlue]
table[x=t, y=lambda, col sep=comma]{\IBSResultCSVBa};

% DS 3 - A
\nextgroupplot[
height=3.6cm,
scaled x ticks=manual:{}{\pgfmathparse{#1}},
yticklabel style={
        /pgf/number format/fixed,
        /pgf/number format/precision=3
},
tick align=outside,
tick pos=left,
width=0.24\textwidth,
x grid style={darkgray176},
xmin=0, xmax=\IBSlen,
ylabel={DS 3x},
xtick style={color=black},
xticklabels={},
y grid style={darkgray176},
ymin=\yAmin, ymax=\yAmax,
ytick style={color=black}
]
\addplot [very thick, TUMBlue]
table[x=t, y=y1IBS, col sep=comma]{\IBSResultCSVAb};

\addplot [very thick, TUMOrange]
table[x=t, y=y1Meas, col sep=comma]{\IBSResultCSVAb};

\nextgroupplot[
height=3.6cm,
tick align=outside,
tick pos=left,
width=0.24\textwidth,
x grid style={darkgray176},
xmin=0, xmax=\IBSlen,
xtick style={color=black},
xticklabels={},
y grid style={darkgray176},
ymin=\LMmin, ymax=\LMmax,
ytick style={color=black}
]
\addplot [very thick, TUMBlue]
table[x=t, y=lambda, col sep=comma]{\IBSResultCSVAb};

% DS 3 - B
\nextgroupplot[
height=3.6cm,
scaled x ticks=manual:{}{\pgfmathparse{#1}},
yticklabel style={
        /pgf/number format/fixed,
        /pgf/number format/precision=3
},
tick align=outside,
tick pos=left,
width=0.24\textwidth,
x grid style={darkgray176},
xmin=0, xmax=\IBSlen,
xtick style={color=black},
xticklabels={},
y grid style={darkgray176},
ymin=\yAmin, ymax=\yAmax,
ytick style={color=black}
]
\addplot [very thick, TUMBlue]
table[x=t, y=y1IBS, col sep=comma]{\IBSResultCSVBb};

\addplot [very thick, TUMOrange]
table[x=t, y=y1Meas, col sep=comma]{\IBSResultCSVBb};

\nextgroupplot[
height=3.6cm,
tick align=outside,
tick pos=left,
width=0.24\textwidth,
x grid style={darkgray176},
xmin=0, xmax=\IBSlen,
xtick style={color=black},
xticklabels={},
y grid style={darkgray176},
ymin=\LMmin, ymax=\LMmax,
ytick style={color=black}
]
\addplot [very thick, TUMBlue]
table[x=t, y=lambda, col sep=comma]{\IBSResultCSVBb};

% DS 4 - A
\nextgroupplot[
height=3.6cm,
scaled x ticks=manual:{}{\pgfmathparse{#1}},
yticklabel style={
        /pgf/number format/fixed,
        /pgf/number format/precision=3
},
tick align=outside,
tick pos=left,
width=0.24\textwidth,
x grid style={darkgray176},
xmin=0, xmax=\IBSlen,
ylabel={DS 4x},
xtick style={color=black},
xticklabels={},
y grid style={darkgray176},
ymin=\yAmin, ymax=\yAmax,
ytick style={color=black}
]
\addplot [very thick, TUMBlue]
table[x=t, y=y1IBS, col sep=comma]{\IBSResultCSVAc};

\addplot [very thick, TUMOrange]
table[x=t, y=y1Meas, col sep=comma]{\IBSResultCSVAc};

\nextgroupplot[
height=3.6cm,
tick align=outside,
tick pos=left,
width=0.24\textwidth,
x grid style={darkgray176},
xmin=0, xmax=\IBSlen,
xtick style={color=black},
xticklabels={},
y grid style={darkgray176},
ymin=\LMmin, ymax=\LMmax,
ytick style={color=black}
]
\addplot [very thick, TUMBlue]
table[x=t, y=lambda, col sep=comma]{\IBSResultCSVAc};

% DS 4 - B
\nextgroupplot[
height=3.6cm,
scaled x ticks=manual:{}{\pgfmathparse{#1}},
yticklabel style={
        /pgf/number format/fixed,
        /pgf/number format/precision=3
},
tick align=outside,
tick pos=left,
width=0.24\textwidth,
x grid style={darkgray176},
xmin=0, xmax=\IBSlen,
xtick style={color=black},
xticklabels={},
y grid style={darkgray176},
ymin=\yAmin, ymax=\yAmax,
ytick style={color=black}
]
\addplot [very thick, TUMBlue]
table[x=t, y=y1IBS, col sep=comma]{\IBSResultCSVBc};

\addplot [very thick, TUMOrange]
table[x=t, y=y1Meas, col sep=comma]{\IBSResultCSVBc};

\nextgroupplot[
height=3.6cm,
tick align=outside,
tick pos=left,
width=0.24\textwidth,
x grid style={darkgray176},
xmin=0, xmax=\IBSlen,
xtick style={color=black},
xticklabels={},
y grid style={darkgray176},
ymin=\LMmin, ymax=\LMmax,
ytick style={color=black}
]
\addplot [very thick, TUMBlue]
table[x=t, y=lambda, col sep=comma]{\IBSResultCSVBc};

% DS 5 - A
\nextgroupplot[
height=3.6cm,
scaled x ticks=manual:{}{\pgfmathparse{#1}},
yticklabel style={
        /pgf/number format/fixed,
        /pgf/number format/precision=3
},
tick align=outside,
tick pos=left,
width=0.24\textwidth,
x grid style={darkgray176},
xmin=0, xmax=\IBSlen,
ylabel={DS 5x},
xtick style={color=black},
xlabel={Time / ms},
y grid style={darkgray176},
ymin=\yAmin, ymax=\yAmax,
ytick style={color=black}
]
\addplot [very thick, TUMBlue]
table[x=t, y=y1IBS, col sep=comma]{\IBSResultCSVAd};

\addplot [very thick, TUMOrange]
table[x=t, y=y1Meas, col sep=comma]{\IBSResultCSVAd};

\nextgroupplot[
height=3.6cm,
tick align=outside,
tick pos=left,
width=0.24\textwidth,
x grid style={darkgray176},
xmin=0, xmax=\IBSlen,
xtick style={color=black},
xlabel={Time / ms},
y grid style={darkgray176},
ymin=\LMmin, ymax=\LMmax,
ytick style={color=black}
]
\addplot [very thick, TUMBlue]
table[x=t, y=lambda, col sep=comma]{\IBSResultCSVAd};

% DS 5 - B
\nextgroupplot[
height=3.6cm,
scaled x ticks=manual:{}{\pgfmathparse{#1}},
yticklabel style={
        /pgf/number format/fixed,
        /pgf/number format/precision=3
},
tick align=outside,
tick pos=left,
width=0.24\textwidth,
x grid style={darkgray176},
xmin=0, xmax=\IBSlen,
xtick style={color=black},
xlabel={Time / ms},
y grid style={darkgray176},
ymin=\yAmin, ymax=\yAmax,
ytick style={color=black}
]
\addplot [very thick, TUMBlue]
table[x=t, y=y1IBS, col sep=comma]{\IBSResultCSVBd};

\addplot [very thick, TUMOrange]
table[x=t, y=y1Meas, col sep=comma]{\IBSResultCSVBd};

\nextgroupplot[
height=3.6cm,
tick align=outside,
tick pos=left,
width=0.24\textwidth,
x grid style={darkgray176},
xmin=0, xmax=\IBSlen,
xtick style={color=black},
xlabel={Time / ms},
y grid style={darkgray176},
ymin=\LMmin, ymax=\LMmax,
ytick style={color=black}
]
\addplot [very thick, TUMBlue]
table[x=t, y=lambda, col sep=comma]{\IBSResultCSVBd};

\end{groupplot}
\end{tikzpicture}
	\end{flushright}\vspace{-0.4cm}
	\caption{Comparison of experimental results using a low-pass filter with downsampling versus only low-pass filtering, applied to \textbf{displacements} of \textbf{aluminum} rods using manual hammer with \textbf{vinyl tip}, \legendary}
	\label{fig:IBS_results_DS_vs_LP_ENAW7075_vinyl_d}
\end{figure}
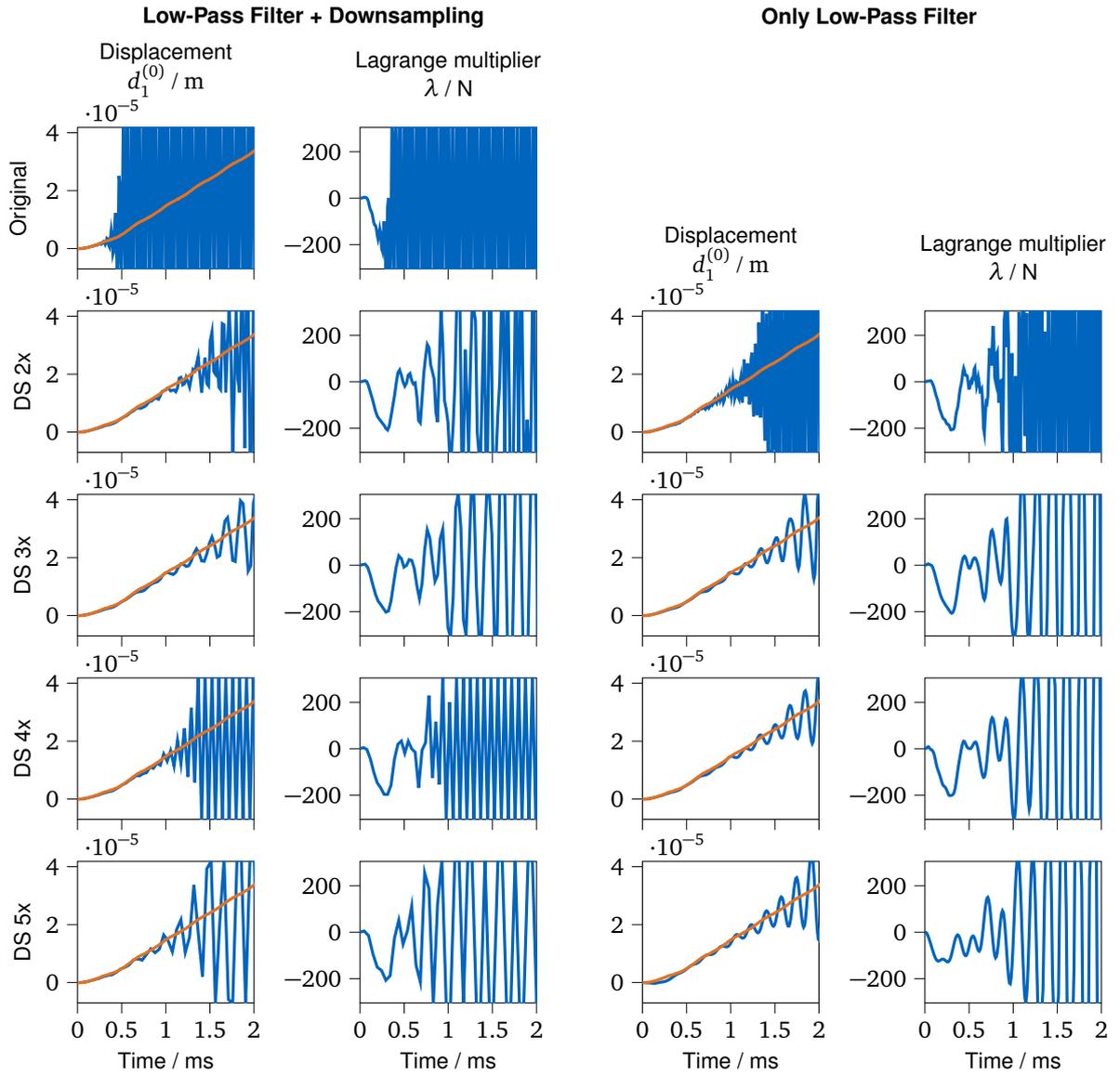

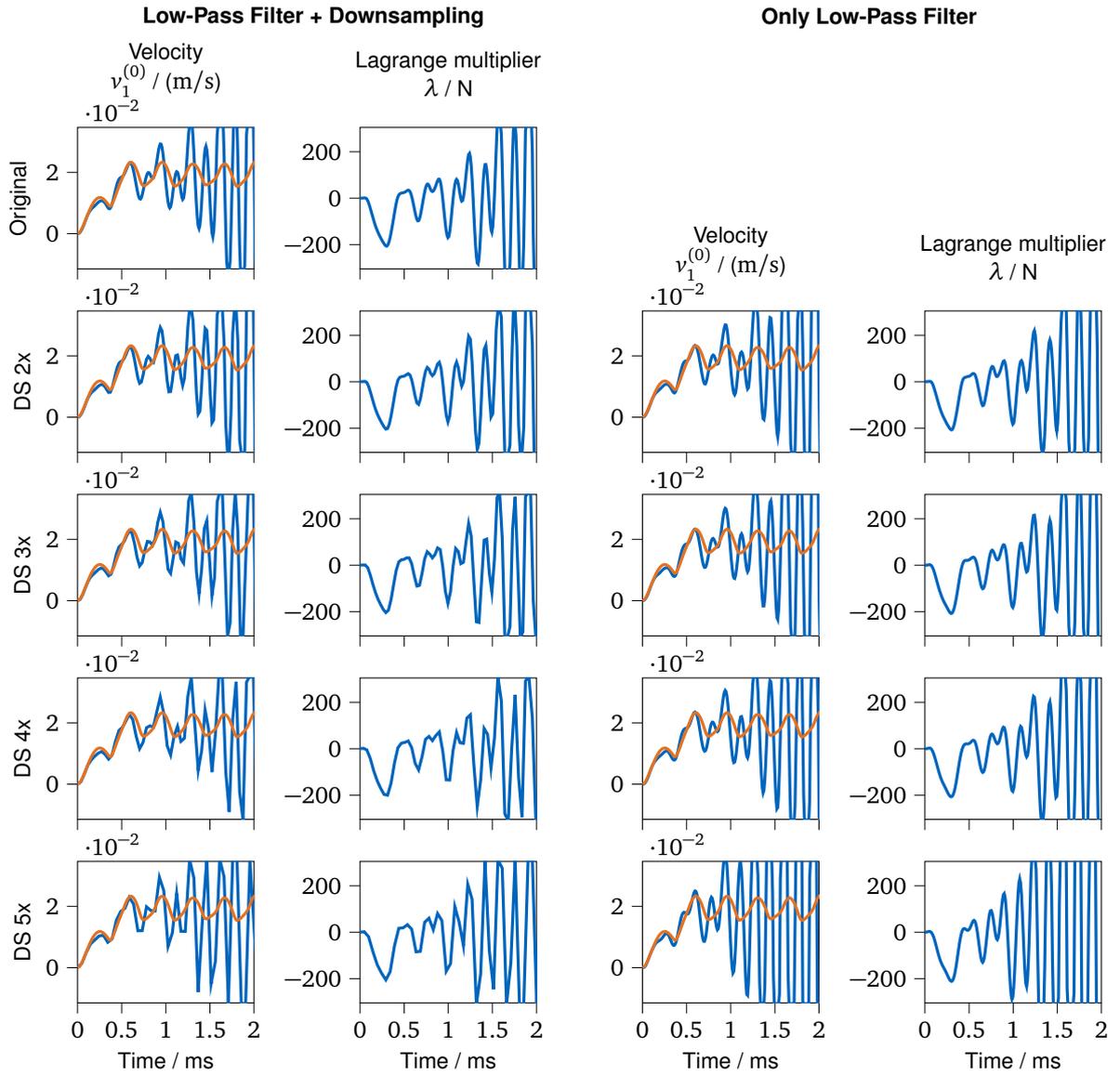
\begin{figure}[ht!]
	\small
	\renewcommand{\IBSlen}{2}
	\renewcommand{\IBSResultCSVvelo}{data/IBS_results_ENAW7075_manual_vinyl_v_Time.csv}
	\renewcommand{\IBSResultCSVAa}{data/ENAW7075_manual_vinyl_DS/IBS_results_ENAW7075_manual_vinyl_v_DS_2.csv}
	\renewcommand{\IBSResultCSVBa}{data/ENAW7075_manual_vinyl_LPDS/IBS_results_ENAW7075_manual_vinyl_v_LPDS_2.csv}
	\renewcommand{\IBSResultCSVAb}{data/ENAW7075_manual_vinyl_DS/IBS_results_ENAW7075_manual_vinyl_v_DS_3.csv}
	\renewcommand{\IBSResultCSVBb}{data/ENAW7075_manual_vinyl_LPDS/IBS_results_ENAW7075_manual_vinyl_v_LPDS_3.csv}
	\renewcommand{\IBSResultCSVAc}{data/ENAW7075_manual_vinyl_DS/IBS_results_ENAW7075_manual_vinyl_v_DS_4.csv}
	\renewcommand{\IBSResultCSVBc}{data/ENAW7075_manual_vinyl_LPDS/IBS_results_ENAW7075_manual_vinyl_v_LPDS_4.csv}
	\renewcommand{\IBSResultCSVAd}{data/ENAW7075_manual_vinyl_DS/IBS_results_ENAW7075_manual_vinyl_v_DS_5.csv}
	\renewcommand{\IBSResultCSVBd}{data/ENAW7075_manual_vinyl_LPDS/IBS_results_ENAW7075_manual_vinyl_v_LPDS_5.csv}
	\renewcommand{\yAmin}{-0.0114898013671875}
	\renewcommand{\yAmax}{0.034632754980468757}
	\renewcommand{\LMmin}{-304.67055783962905}
	\renewcommand{\LMmax}{304.67055783962905}
	\begin{minipage}{0.5\textwidth}
		\centering
		\hspace{2.8cm}\textbf{Low-Pass Filter + Downsampling}
	\end{minipage}
	\begin{minipage}{0.5\textwidth}
		\centering
		\hspace{1.4cm}\textbf{Only Low-Pass Filter}
	\end{minipage}\\\vspace{-0.6cm}
	\begin{flushright}
		% This file was created with tikzplotlib v0.10.1.
\begin{tikzpicture}

\definecolor{darkgray176}{RGB}{176,176,176}
\definecolor{gray}{RGB}{128,128,128}
\definecolor{lightgray204}{RGB}{204,204,204}

\begin{groupplot}[group style={group size=4 by 5, horizontal sep=1.5cm, vertical sep=0.6cm}]

% DS 1 - A
\nextgroupplot[
height=3.6cm,
scaled x ticks=manual:{}{\pgfmathparse{#1}},
yticklabel style={
        /pgf/number format/fixed,
        /pgf/number format/precision=3
},
tick align=outside,
tick pos=left,
title={Velocity \\\(\displaystyle v_1^{(0)}\) / (\si[per-mode=symbol]{\meter\per\second})},
title style = {align = center},
title style={yshift=0.7ex},
width=0.24\textwidth,
x grid style={darkgray176},
xmin=0, xmax=\IBSlen,
ylabel={Original},
xtick style={color=black},
xticklabels={},
y grid style={darkgray176},
ymin=\yAmin, ymax=\yAmax,
ytick style={color=black}
]
\addplot [very thick, TUMBlue]
table[x=t, y=y1IBS, col sep=comma]{\IBSResultCSVvelo};

\addplot [very thick, TUMOrange]
table[x=t, y=y1Meas, col sep=comma]{\IBSResultCSVvelo};

\nextgroupplot[
height=3.6cm,
tick align=outside,
tick pos=left,
title={Lagrange multiplier \\\(\displaystyle \lambda\) / N},
title style = {align = center},
title style={yshift=0.7ex},
width=0.24\textwidth,
x grid style={darkgray176},
xmin=0, xmax=\IBSlen,
xtick style={color=black},
xticklabels={},
y grid style={darkgray176},
ymin=\LMmin, ymax=\LMmax,
ytick style={color=black}
]
\addplot [very thick, TUMBlue]
table[x=t, y=lambda, col sep=comma]{\IBSResultCSVvelo};

% DS 1 - B
\nextgroupplot[
height=3.6cm,
width=0.24\textwidth,
x grid style=none,
xticklabels={},
yticklabels={},
xtick style={draw=none},
ytick style={draw=none},
hide x axis,
hide y axis
]

\nextgroupplot[
height=3.6cm,
width=0.24\textwidth,
x grid style=none,
xticklabels={},
yticklabels={},
xtick style={draw=none},
ytick style={draw=none},
hide x axis,
hide y axis
]

% DS 2 - A
\nextgroupplot[
height=3.6cm,
scaled x ticks=manual:{}{\pgfmathparse{#1}},
yticklabel style={
        /pgf/number format/fixed,
        /pgf/number format/precision=3
},
tick align=outside,
tick pos=left,
width=0.24\textwidth,
x grid style={darkgray176},
xmin=0, xmax=\IBSlen,
ylabel={DS 2x},
xtick style={color=black},
xticklabels={},
y grid style={darkgray176},
ymin=\yAmin, ymax=\yAmax,
ytick style={color=black}
]
\addplot [very thick, TUMBlue]
table[x=t, y=y1IBS, col sep=comma]{\IBSResultCSVAa};

\addplot [very thick, TUMOrange]
table[x=t, y=y1Meas, col sep=comma]{\IBSResultCSVAa};

\nextgroupplot[
height=3.6cm,
tick align=outside,
tick pos=left,
width=0.24\textwidth,
x grid style={darkgray176},
xmin=0, xmax=\IBSlen,
xtick style={color=black},
xticklabels={},
y grid style={darkgray176},
ymin=\LMmin, ymax=\LMmax,
ytick style={color=black}
]
\addplot [very thick, TUMBlue]
table[x=t, y=lambda, col sep=comma]{\IBSResultCSVAa};

% DS 2 - B
\nextgroupplot[
height=3.6cm,
scaled x ticks=manual:{}{\pgfmathparse{#1}},
yticklabel style={
        /pgf/number format/fixed,
        /pgf/number format/precision=3
},
tick align=outside,
tick pos=left,
title={Velocity \\\(\displaystyle v_1^{(0)}\) / (\si[per-mode=symbol]{\meter\per\second})},
title style = {align = center},
title style={yshift=0.7ex},
width=0.24\textwidth,
x grid style={darkgray176},
xmin=0, xmax=\IBSlen,
xtick style={color=black},
xticklabels={},
y grid style={darkgray176},
ymin=\yAmin, ymax=\yAmax,
ytick style={color=black}
]
\addplot [very thick, TUMBlue]
table[x=t, y=y1IBS, col sep=comma]{\IBSResultCSVBa};

\addplot [very thick, TUMOrange]
table[x=t, y=y1Meas, col sep=comma]{\IBSResultCSVBa};

\nextgroupplot[
height=3.6cm,
tick align=outside,
tick pos=left,
title={Lagrange multiplier \\\(\displaystyle \lambda\) / N},
title style = {align = center},
title style={yshift=0.7ex},
width=0.24\textwidth,
x grid style={darkgray176},
xmin=0, xmax=\IBSlen,
xtick style={color=black},
xticklabels={},
y grid style={darkgray176},
ymin=\LMmin, ymax=\LMmax,
ytick style={color=black}
]
\addplot [very thick, TUMBlue]
table[x=t, y=lambda, col sep=comma]{\IBSResultCSVBa};

% DS 3 - A
\nextgroupplot[
height=3.6cm,
scaled x ticks=manual:{}{\pgfmathparse{#1}},
yticklabel style={
        /pgf/number format/fixed,
        /pgf/number format/precision=3
},
tick align=outside,
tick pos=left,
width=0.24\textwidth,
x grid style={darkgray176},
xmin=0, xmax=\IBSlen,
ylabel={DS 3x},
xtick style={color=black},
xticklabels={},
y grid style={darkgray176},
ymin=\yAmin, ymax=\yAmax,
ytick style={color=black}
]
\addplot [very thick, TUMBlue]
table[x=t, y=y1IBS, col sep=comma]{\IBSResultCSVAb};

\addplot [very thick, TUMOrange]
table[x=t, y=y1Meas, col sep=comma]{\IBSResultCSVAb};

\nextgroupplot[
height=3.6cm,
tick align=outside,
tick pos=left,
width=0.24\textwidth,
x grid style={darkgray176},
xmin=0, xmax=\IBSlen,
xtick style={color=black},
xticklabels={},
y grid style={darkgray176},
ymin=\LMmin, ymax=\LMmax,
ytick style={color=black}
]
\addplot [very thick, TUMBlue]
table[x=t, y=lambda, col sep=comma]{\IBSResultCSVAb};

% DS 3 - B
\nextgroupplot[
height=3.6cm,
scaled x ticks=manual:{}{\pgfmathparse{#1}},
yticklabel style={
        /pgf/number format/fixed,
        /pgf/number format/precision=3
},
tick align=outside,
tick pos=left,
width=0.24\textwidth,
x grid style={darkgray176},
xmin=0, xmax=\IBSlen,
xtick style={color=black},
xticklabels={},
y grid style={darkgray176},
ymin=\yAmin, ymax=\yAmax,
ytick style={color=black}
]
\addplot [very thick, TUMBlue]
table[x=t, y=y1IBS, col sep=comma]{\IBSResultCSVBb};

\addplot [very thick, TUMOrange]
table[x=t, y=y1Meas, col sep=comma]{\IBSResultCSVBb};

\nextgroupplot[
height=3.6cm,
tick align=outside,
tick pos=left,
width=0.24\textwidth,
x grid style={darkgray176},
xmin=0, xmax=\IBSlen,
xtick style={color=black},
xticklabels={},
y grid style={darkgray176},
ymin=\LMmin, ymax=\LMmax,
ytick style={color=black}
]
\addplot [very thick, TUMBlue]
table[x=t, y=lambda, col sep=comma]{\IBSResultCSVBb};

% DS 4 - A
\nextgroupplot[
height=3.6cm,
scaled x ticks=manual:{}{\pgfmathparse{#1}},
yticklabel style={
        /pgf/number format/fixed,
        /pgf/number format/precision=3
},
tick align=outside,
tick pos=left,
width=0.24\textwidth,
x grid style={darkgray176},
xmin=0, xmax=\IBSlen,
ylabel={DS 4x},
xtick style={color=black},
xticklabels={},
y grid style={darkgray176},
ymin=\yAmin, ymax=\yAmax,
ytick style={color=black}
]
\addplot [very thick, TUMBlue]
table[x=t, y=y1IBS, col sep=comma]{\IBSResultCSVAc};

\addplot [very thick, TUMOrange]
table[x=t, y=y1Meas, col sep=comma]{\IBSResultCSVAc};

\nextgroupplot[
height=3.6cm,
tick align=outside,
tick pos=left,
width=0.24\textwidth,
x grid style={darkgray176},
xmin=0, xmax=\IBSlen,
xtick style={color=black},
xticklabels={},
y grid style={darkgray176},
ymin=\LMmin, ymax=\LMmax,
ytick style={color=black}
]
\addplot [very thick, TUMBlue]
table[x=t, y=lambda, col sep=comma]{\IBSResultCSVAc};

% DS 4 - B
\nextgroupplot[
height=3.6cm,
scaled x ticks=manual:{}{\pgfmathparse{#1}},
yticklabel style={
        /pgf/number format/fixed,
        /pgf/number format/precision=3
},
tick align=outside,
tick pos=left,
width=0.24\textwidth,
x grid style={darkgray176},
xmin=0, xmax=\IBSlen,
xtick style={color=black},
xticklabels={},
y grid style={darkgray176},
ymin=\yAmin, ymax=\yAmax,
ytick style={color=black}
]
\addplot [very thick, TUMBlue]
table[x=t, y=y1IBS, col sep=comma]{\IBSResultCSVBc};

\addplot [very thick, TUMOrange]
table[x=t, y=y1Meas, col sep=comma]{\IBSResultCSVBc};

\nextgroupplot[
height=3.6cm,
tick align=outside,
tick pos=left,
width=0.24\textwidth,
x grid style={darkgray176},
xmin=0, xmax=\IBSlen,
xtick style={color=black},
xticklabels={},
y grid style={darkgray176},
ymin=\LMmin, ymax=\LMmax,
ytick style={color=black}
]
\addplot [very thick, TUMBlue]
table[x=t, y=lambda, col sep=comma]{\IBSResultCSVBc};

% DS 5 - A
\nextgroupplot[
height=3.6cm,
scaled x ticks=manual:{}{\pgfmathparse{#1}},
yticklabel style={
        /pgf/number format/fixed,
        /pgf/number format/precision=3
},
tick align=outside,
tick pos=left,
width=0.24\textwidth,
x grid style={darkgray176},
xmin=0, xmax=\IBSlen,
ylabel={DS 5x},
xtick style={color=black},
xlabel={Time / ms},
y grid style={darkgray176},
ymin=\yAmin, ymax=\yAmax,
ytick style={color=black}
]
\addplot [very thick, TUMBlue]
table[x=t, y=y1IBS, col sep=comma]{\IBSResultCSVAd};

\addplot [very thick, TUMOrange]
table[x=t, y=y1Meas, col sep=comma]{\IBSResultCSVAd};

\nextgroupplot[
height=3.6cm,
tick align=outside,
tick pos=left,
width=0.24\textwidth,
x grid style={darkgray176},
xmin=0, xmax=\IBSlen,
xtick style={color=black},
xlabel={Time / ms},
y grid style={darkgray176},
ymin=\LMmin, ymax=\LMmax,
ytick style={color=black}
]
\addplot [very thick, TUMBlue]
table[x=t, y=lambda, col sep=comma]{\IBSResultCSVAd};

% DS 5 - B
\nextgroupplot[
height=3.6cm,
scaled x ticks=manual:{}{\pgfmathparse{#1}},
yticklabel style={
        /pgf/number format/fixed,
        /pgf/number format/precision=3
},
tick align=outside,
tick pos=left,
width=0.24\textwidth,
x grid style={darkgray176},
xmin=0, xmax=\IBSlen,
xtick style={color=black},
xlabel={Time / ms},
y grid style={darkgray176},
ymin=\yAmin, ymax=\yAmax,
ytick style={color=black}
]
\addplot [very thick, TUMBlue]
table[x=t, y=y1IBS, col sep=comma]{\IBSResultCSVBd};

\addplot [very thick, TUMOrange]
table[x=t, y=y1Meas, col sep=comma]{\IBSResultCSVBd};

\nextgroupplot[
height=3.6cm,
tick align=outside,
tick pos=left,
width=0.24\textwidth,
x grid style={darkgray176},
xmin=0, xmax=\IBSlen,
xtick style={color=black},
xlabel={Time / ms},
y grid style={darkgray176},
ymin=\LMmin, ymax=\LMmax,
ytick style={color=black}
]
\addplot [very thick, TUMBlue]
table[x=t, y=lambda, col sep=comma]{\IBSResultCSVBd};

\end{groupplot}
\end{tikzpicture}
	\end{flushright}\vspace{-0.4cm}
	\caption{Comparison of experimental results using a low-pass filter with downsampling versus only low-pass filtering, applied to \textbf{velocities} of \textbf{aluminum} rods using manual hammer with \textbf{vinyl tip}, \legendary}
	\label{fig:IBS_results_DS_vs_LP_ENAW7075_vinyl_v}
\end{figure}

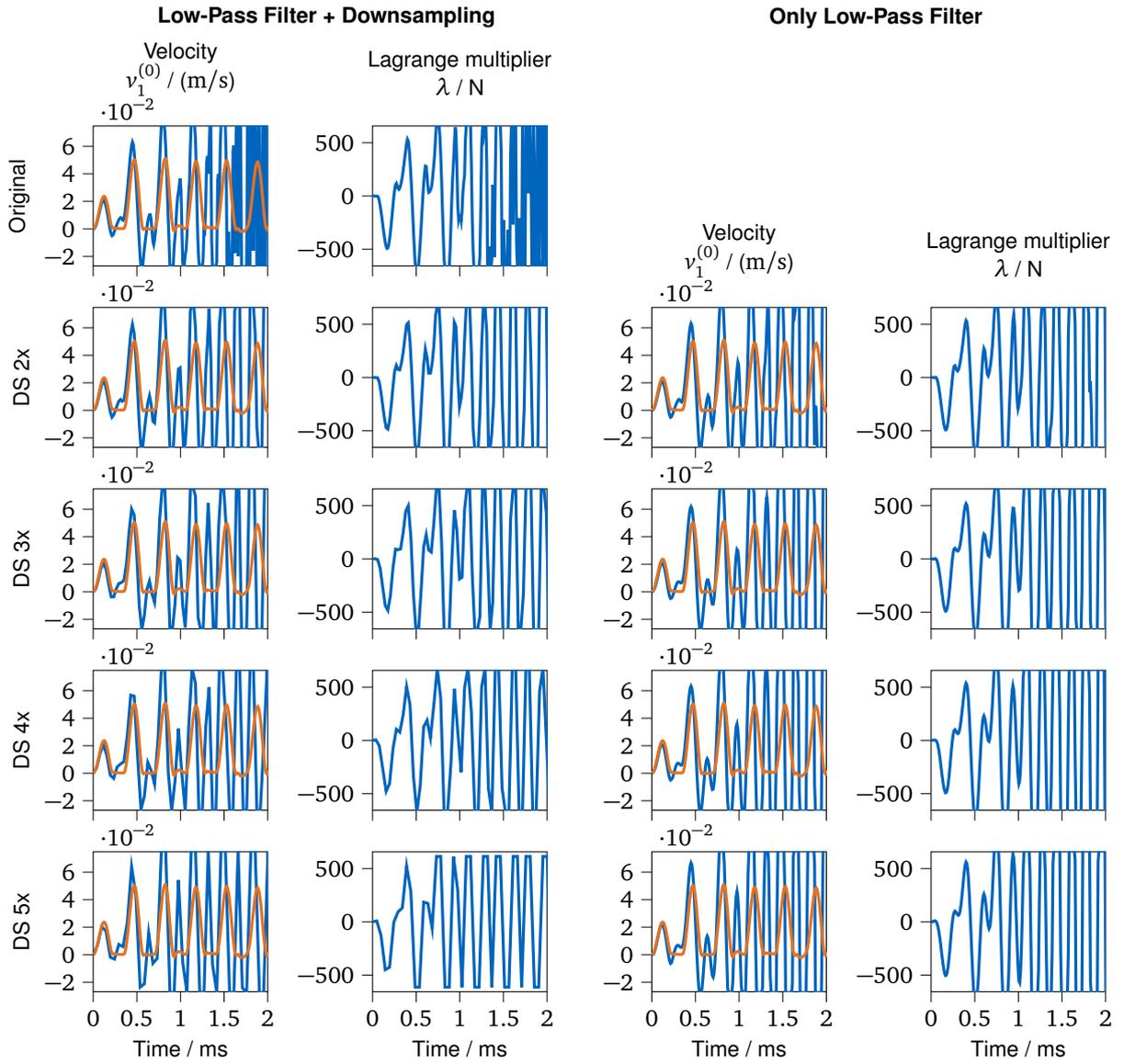
\begin{figure}[ht!]
	\small
	\renewcommand{\IBSlen}{2}
	\renewcommand{\IBSResultCSVvelo}{data/IBS_results_ENAW7075_manual_steel_v_Time.csv}
	\renewcommand{\IBSResultCSVAa}{data/ENAW7075_manual_steel_DS/IBS_results_ENAW7075_manual_steel_v_DS_2.csv}
	\renewcommand{\IBSResultCSVBa}{data/ENAW7075_manual_steel_LPDS/IBS_results_ENAW7075_manual_steel_v_LPDS_2.csv}
	\renewcommand{\IBSResultCSVAb}{data/ENAW7075_manual_steel_DS/IBS_results_ENAW7075_manual_steel_v_DS_3.csv}
	\renewcommand{\IBSResultCSVBb}{data/ENAW7075_manual_steel_LPDS/IBS_results_ENAW7075_manual_steel_v_LPDS_3.csv}
	\renewcommand{\IBSResultCSVAc}{data/ENAW7075_manual_steel_DS/IBS_results_ENAW7075_manual_steel_v_DS_4.csv}
	\renewcommand{\IBSResultCSVBc}{data/ENAW7075_manual_steel_LPDS/IBS_results_ENAW7075_manual_steel_v_LPDS_4.csv}
	\renewcommand{\IBSResultCSVAd}{data/ENAW7075_manual_steel_DS/IBS_results_ENAW7075_manual_steel_v_DS_5.csv}
	\renewcommand{\IBSResultCSVBd}{data/ENAW7075_manual_steel_LPDS/IBS_results_ENAW7075_manual_steel_v_LPDS_5.csv}
	\renewcommand{\yAmin}{-0.026929833833128212}
	\renewcommand{\yAmax}{0.07476644910871982}
	\renewcommand{\LMmin}{-656.911739006243}
	\renewcommand{\LMmax}{656.911739006243}
	\begin{minipage}{0.5\textwidth}
		\centering
		\hspace{2.8cm}\textbf{Low-Pass Filter + Downsampling}
	\end{minipage}
	\begin{minipage}{0.5\textwidth}
		\centering
		\hspace{1.4cm}\textbf{Only Low-Pass Filter}
	\end{minipage}\\\vspace{-0.6cm}
	\begin{flushright}
		% This file was created with tikzplotlib v0.10.1.
\begin{tikzpicture}

\definecolor{darkgray176}{RGB}{176,176,176}
\definecolor{gray}{RGB}{128,128,128}
\definecolor{lightgray204}{RGB}{204,204,204}

\begin{groupplot}[group style={group size=4 by 5, horizontal sep=1.5cm, vertical sep=0.6cm}]

% DS 1 - A
\nextgroupplot[
height=3.6cm,
scaled x ticks=manual:{}{\pgfmathparse{#1}},
yticklabel style={
        /pgf/number format/fixed,
        /pgf/number format/precision=3
},
tick align=outside,
tick pos=left,
title={Velocity \\\(\displaystyle v_1^{(0)}\) / (\si[per-mode=symbol]{\meter\per\second})},
title style = {align = center},
title style={yshift=0.7ex},
width=0.24\textwidth,
x grid style={darkgray176},
xmin=0, xmax=\IBSlen,
ylabel={Original},
xtick style={color=black},
xticklabels={},
y grid style={darkgray176},
ymin=\yAmin, ymax=\yAmax,
ytick style={color=black}
]
\addplot [very thick, TUMBlue]
table[x=t, y=y1IBS, col sep=comma]{\IBSResultCSVvelo};

\addplot [very thick, TUMOrange]
table[x=t, y=y1Meas, col sep=comma]{\IBSResultCSVvelo};

\nextgroupplot[
height=3.6cm,
tick align=outside,
tick pos=left,
title={Lagrange multiplier \\\(\displaystyle \lambda\) / N},
title style = {align = center},
title style={yshift=0.7ex},
width=0.24\textwidth,
x grid style={darkgray176},
xmin=0, xmax=\IBSlen,
xtick style={color=black},
xticklabels={},
y grid style={darkgray176},
ymin=\LMmin, ymax=\LMmax,
ytick style={color=black}
]
\addplot [very thick, TUMBlue]
table[x=t, y=lambda, col sep=comma]{\IBSResultCSVvelo};

% DS 1 - B
\nextgroupplot[
height=3.6cm,
width=0.24\textwidth,
x grid style=none,
xticklabels={},
yticklabels={},
xtick style={draw=none},
ytick style={draw=none},
hide x axis,
hide y axis
]

\nextgroupplot[
height=3.6cm,
width=0.24\textwidth,
x grid style=none,
xticklabels={},
yticklabels={},
xtick style={draw=none},
ytick style={draw=none},
hide x axis,
hide y axis
]

% DS 2 - A
\nextgroupplot[
height=3.6cm,
scaled x ticks=manual:{}{\pgfmathparse{#1}},
yticklabel style={
        /pgf/number format/fixed,
        /pgf/number format/precision=3
},
tick align=outside,
tick pos=left,
width=0.24\textwidth,
x grid style={darkgray176},
xmin=0, xmax=\IBSlen,
ylabel={DS 2x},
xtick style={color=black},
xticklabels={},
y grid style={darkgray176},
ymin=\yAmin, ymax=\yAmax,
ytick style={color=black}
]
\addplot [very thick, TUMBlue]
table[x=t, y=y1IBS, col sep=comma]{\IBSResultCSVAa};

\addplot [very thick, TUMOrange]
table[x=t, y=y1Meas, col sep=comma]{\IBSResultCSVAa};

\nextgroupplot[
height=3.6cm,
tick align=outside,
tick pos=left,
width=0.24\textwidth,
x grid style={darkgray176},
xmin=0, xmax=\IBSlen,
xtick style={color=black},
xticklabels={},
y grid style={darkgray176},
ymin=\LMmin, ymax=\LMmax,
ytick style={color=black}
]
\addplot [very thick, TUMBlue]
table[x=t, y=lambda, col sep=comma]{\IBSResultCSVAa};

% DS 2 - B
\nextgroupplot[
height=3.6cm,
scaled x ticks=manual:{}{\pgfmathparse{#1}},
yticklabel style={
        /pgf/number format/fixed,
        /pgf/number format/precision=3
},
tick align=outside,
tick pos=left,
title={Velocity \\\(\displaystyle v_1^{(0)}\) / (\si[per-mode=symbol]{\meter\per\second})},
title style = {align = center},
title style={yshift=0.7ex},
width=0.24\textwidth,
x grid style={darkgray176},
xmin=0, xmax=\IBSlen,
xtick style={color=black},
xticklabels={},
y grid style={darkgray176},
ymin=\yAmin, ymax=\yAmax,
ytick style={color=black}
]
\addplot [very thick, TUMBlue]
table[x=t, y=y1IBS, col sep=comma]{\IBSResultCSVBa};

\addplot [very thick, TUMOrange]
table[x=t, y=y1Meas, col sep=comma]{\IBSResultCSVBa};

\nextgroupplot[
height=3.6cm,
tick align=outside,
tick pos=left,
title={Lagrange multiplier \\\(\displaystyle \lambda\) / N},
title style = {align = center},
title style={yshift=0.7ex},
width=0.24\textwidth,
x grid style={darkgray176},
xmin=0, xmax=\IBSlen,
xtick style={color=black},
xticklabels={},
y grid style={darkgray176},
ymin=\LMmin, ymax=\LMmax,
ytick style={color=black}
]
\addplot [very thick, TUMBlue]
table[x=t, y=lambda, col sep=comma]{\IBSResultCSVBa};

% DS 3 - A
\nextgroupplot[
height=3.6cm,
scaled x ticks=manual:{}{\pgfmathparse{#1}},
yticklabel style={
        /pgf/number format/fixed,
        /pgf/number format/precision=3
},
tick align=outside,
tick pos=left,
width=0.24\textwidth,
x grid style={darkgray176},
xmin=0, xmax=\IBSlen,
ylabel={DS 3x},
xtick style={color=black},
xticklabels={},
y grid style={darkgray176},
ymin=\yAmin, ymax=\yAmax,
ytick style={color=black}
]
\addplot [very thick, TUMBlue]
table[x=t, y=y1IBS, col sep=comma]{\IBSResultCSVAb};

\addplot [very thick, TUMOrange]
table[x=t, y=y1Meas, col sep=comma]{\IBSResultCSVAb};

\nextgroupplot[
height=3.6cm,
tick align=outside,
tick pos=left,
width=0.24\textwidth,
x grid style={darkgray176},
xmin=0, xmax=\IBSlen,
xtick style={color=black},
xticklabels={},
y grid style={darkgray176},
ymin=\LMmin, ymax=\LMmax,
ytick style={color=black}
]
\addplot [very thick, TUMBlue]
table[x=t, y=lambda, col sep=comma]{\IBSResultCSVAb};

% DS 3 - B
\nextgroupplot[
height=3.6cm,
scaled x ticks=manual:{}{\pgfmathparse{#1}},
yticklabel style={
        /pgf/number format/fixed,
        /pgf/number format/precision=3
},
tick align=outside,
tick pos=left,
width=0.24\textwidth,
x grid style={darkgray176},
xmin=0, xmax=\IBSlen,
xtick style={color=black},
xticklabels={},
y grid style={darkgray176},
ymin=\yAmin, ymax=\yAmax,
ytick style={color=black}
]
\addplot [very thick, TUMBlue]
table[x=t, y=y1IBS, col sep=comma]{\IBSResultCSVBb};

\addplot [very thick, TUMOrange]
table[x=t, y=y1Meas, col sep=comma]{\IBSResultCSVBb};

\nextgroupplot[
height=3.6cm,
tick align=outside,
tick pos=left,
width=0.24\textwidth,
x grid style={darkgray176},
xmin=0, xmax=\IBSlen,
xtick style={color=black},
xticklabels={},
y grid style={darkgray176},
ymin=\LMmin, ymax=\LMmax,
ytick style={color=black}
]
\addplot [very thick, TUMBlue]
table[x=t, y=lambda, col sep=comma]{\IBSResultCSVBb};

% DS 4 - A
\nextgroupplot[
height=3.6cm,
scaled x ticks=manual:{}{\pgfmathparse{#1}},
yticklabel style={
        /pgf/number format/fixed,
        /pgf/number format/precision=3
},
tick align=outside,
tick pos=left,
width=0.24\textwidth,
x grid style={darkgray176},
xmin=0, xmax=\IBSlen,
ylabel={DS 4x},
xtick style={color=black},
xticklabels={},
y grid style={darkgray176},
ymin=\yAmin, ymax=\yAmax,
ytick style={color=black}
]
\addplot [very thick, TUMBlue]
table[x=t, y=y1IBS, col sep=comma]{\IBSResultCSVAc};

\addplot [very thick, TUMOrange]
table[x=t, y=y1Meas, col sep=comma]{\IBSResultCSVAc};

\nextgroupplot[
height=3.6cm,
tick align=outside,
tick pos=left,
width=0.24\textwidth,
x grid style={darkgray176},
xmin=0, xmax=\IBSlen,
xtick style={color=black},
xticklabels={},
y grid style={darkgray176},
ymin=\LMmin, ymax=\LMmax,
ytick style={color=black}
]
\addplot [very thick, TUMBlue]
table[x=t, y=lambda, col sep=comma]{\IBSResultCSVAc};

% DS 4 - B
\nextgroupplot[
height=3.6cm,
scaled x ticks=manual:{}{\pgfmathparse{#1}},
yticklabel style={
        /pgf/number format/fixed,
        /pgf/number format/precision=3
},
tick align=outside,
tick pos=left,
width=0.24\textwidth,
x grid style={darkgray176},
xmin=0, xmax=\IBSlen,
xtick style={color=black},
xticklabels={},
y grid style={darkgray176},
ymin=\yAmin, ymax=\yAmax,
ytick style={color=black}
]
\addplot [very thick, TUMBlue]
table[x=t, y=y1IBS, col sep=comma]{\IBSResultCSVBc};

\addplot [very thick, TUMOrange]
table[x=t, y=y1Meas, col sep=comma]{\IBSResultCSVBc};

\nextgroupplot[
height=3.6cm,
tick align=outside,
tick pos=left,
width=0.24\textwidth,
x grid style={darkgray176},
xmin=0, xmax=\IBSlen,
xtick style={color=black},
xticklabels={},
y grid style={darkgray176},
ymin=\LMmin, ymax=\LMmax,
ytick style={color=black}
]
\addplot [very thick, TUMBlue]
table[x=t, y=lambda, col sep=comma]{\IBSResultCSVBc};

% DS 5 - A
\nextgroupplot[
height=3.6cm,
scaled x ticks=manual:{}{\pgfmathparse{#1}},
yticklabel style={
        /pgf/number format/fixed,
        /pgf/number format/precision=3
},
tick align=outside,
tick pos=left,
width=0.24\textwidth,
x grid style={darkgray176},
xmin=0, xmax=\IBSlen,
ylabel={DS 5x},
xtick style={color=black},
xlabel={Time / ms},
y grid style={darkgray176},
ymin=\yAmin, ymax=\yAmax,
ytick style={color=black}
]
\addplot [very thick, TUMBlue]
table[x=t, y=y1IBS, col sep=comma]{\IBSResultCSVAd};

\addplot [very thick, TUMOrange]
table[x=t, y=y1Meas, col sep=comma]{\IBSResultCSVAd};

\nextgroupplot[
height=3.6cm,
tick align=outside,
tick pos=left,
width=0.24\textwidth,
x grid style={darkgray176},
xmin=0, xmax=\IBSlen,
xtick style={color=black},
xlabel={Time / ms},
y grid style={darkgray176},
ymin=\LMmin, ymax=\LMmax,
ytick style={color=black}
]
\addplot [very thick, TUMBlue]
table[x=t, y=lambda, col sep=comma]{\IBSResultCSVAd};

% DS 5 - B
\nextgroupplot[
height=3.6cm,
scaled x ticks=manual:{}{\pgfmathparse{#1}},
yticklabel style={
        /pgf/number format/fixed,
        /pgf/number format/precision=3
},
tick align=outside,
tick pos=left,
width=0.24\textwidth,
x grid style={darkgray176},
xmin=0, xmax=\IBSlen,
xtick style={color=black},
xlabel={Time / ms},
y grid style={darkgray176},
ymin=\yAmin, ymax=\yAmax,
ytick style={color=black}
]
\addplot [very thick, TUMBlue]
table[x=t, y=y1IBS, col sep=comma]{\IBSResultCSVBd};

\addplot [very thick, TUMOrange]
table[x=t, y=y1Meas, col sep=comma]{\IBSResultCSVBd};

\nextgroupplot[
height=3.6cm,
tick align=outside,
tick pos=left,
width=0.24\textwidth,
x grid style={darkgray176},
xmin=0, xmax=\IBSlen,
xtick style={color=black},
xlabel={Time / ms},
y grid style={darkgray176},
ymin=\LMmin, ymax=\LMmax,
ytick style={color=black}
]
\addplot [very thick, TUMBlue]
table[x=t, y=lambda, col sep=comma]{\IBSResultCSVBd};

\end{groupplot}
\end{tikzpicture}
	\end{flushright}\vspace{-0.4cm}
	\caption{Comparison of experimental results using a low-pass filter with downsampling versus only low-pass filtering, applied to \textbf{velocities} of \textbf{aluminum} rods using manual hammer with \textbf{steel tip}, \legendary}
	\label{fig:IBS_results_DS_vs_LP_ENAW7075_steel_v}
\end{figure}

\clearpage

\end{document}